\documentclass[aps,prd,a4paper,showpacs,nofootinbib,twocolumn]{revtex4-1}
\usepackage{hyperref}
\usepackage{ifpdf}
\usepackage{subfigure}
\usepackage{amsmath}
\usepackage{amssymb}
\usepackage{amsfonts}
\usepackage{epsf}
\usepackage{rotating}
\usepackage{graphicx}
\usepackage{amsmath}
\usepackage{fancyhdr}
\usepackage{lineno}
\usepackage{braket}
\usepackage{stfloats}
\usepackage{nicefrac}
\usepackage{graphics}
\usepackage{pstricks}
\usepackage{color}
\usepackage{multirow}
\usepackage{url}
\usepackage{empheq}
\usepackage[title]{appendix}
\def\sss{\scriptscriptstyle}

\allowdisplaybreaks
\begin{document}
	
\title{Beauty baryon non-leptonic decays into decuplet baryons and $CP$-asymmetries based on 
$SU(3)$-Flavor analysis}
\author{Shibasis Roy} 
\email{shibasisr@imsc.res.in}
\affiliation{The Institute of Mathematical Sciences, Taramani,
	Chennai 600113, India}
\affiliation{Homi Bhabha National Institute, Training School 
	Complex, Anushaktinagar, Mumbai  400094, India} 

\author{N.~G.~Deshpande}
\email{desh@uoregon.edu}
\affiliation{Institute for Fundamental Science, University of Oregon,
		Eugene, Oregon 94703, USA}

\author{Rahul Sinha} 
\email{sinha@imsc.res.in}
\affiliation{The Institute of Mathematical Sciences, Taramani,
		Chennai 600113, India}
\affiliation{Homi Bhabha National Institute, Training School Complex, 
		Anushaktinagar, Mumbai  400094, India}

	\date{\today}
	
\begin{abstract}
We consider charmless weak decays of beauty-baryons into decuplet baryons and 
pseudoscalar mesons in a general framework based on $SU(3)$-flavor decomposition of 
the decay amplitudes. The dynamical assumption independent $SU(3)$ analysis accounts for the effects of an arbitrarily broken $SU(3)$ symmetry in these decays. An alternative approach in terms of quark diagrams is also provided and compared with the $SU(3)$ decomposition in the limit of exact $SU(3)$-flavor symmetry. Furthermore, the symmetries of the effective Hamiltonian is used to relate or neglect reduced 
$SU(3)$ amplitudes to derive several sum rule relations between amplitudes and relations between $CP$ asymmetries in 
these decays and identify those that hold even if $SU(3)$ is broken.
\end{abstract}
\maketitle
\section{Introduction}
In recent times, bottom baryon decays have come under increased scrutiny chiefly due to the large number of $b$-baryon production at LHCb~\cite{Aaij:2019pqz}. Among several other interesting rare decays of $b$-baryons, LHCb has analyzed $CP$-asymmetries in bottom baryon charmless decays to two-body and multibody hadronic final states~\cite{Aaltonen:2011qt,Aaij:2012as,Aaltonen:2014vra,Aaij:2016cla,Aaij:2017pgy,Aaij:2018lsx,Aaij:2018tlk,Aaij:2019rkf,Aaij:2019mmy}. The theoretical foundation laid out to study $B$ meson decays\cite{Zeppenfeld:1980ex,Savage:1989ub,Gronau:1990ka,Chau:1990ay,
Gronau:1994rj,Deshpande:1994pw,Gronau:1995hn,Gronau:1995hm,Grinstein:1996us,
Deshpande:1997rr,Deshpande:1997ar,Gronau:1998fn,Neubert:1997wb,Atwood:1997iw,Buras:1997cv,Buras:1998ra,Falk:1998wc,Beneke:2000ry,
He:2000dg,Deshpande:2000jp,Gronau:2000pk,Fu:2002nr,Paz:2002ev,Dariescu:2002hw,
Wu:2002nz,Beneke:2003zv,Grossman:2003qp,Chiang:2003rb,He:2004ck,Chiang:2004nm,
Buras:2004ub,Gronau:2006eb,Ali:2007ff,Cheng:2011qh,He:2013vta,Cheng:2014rfa,
Grinstein:2014aza,Hsiao:2015iiu,He:2018php,He:2018joe,Wang:2020gmn} is also useful to study  bottom and charmed baryon decay amplitude sum-rules and $CP$ asymmetry relations for various decay modes. In particular, the general framework of $SU(3)$ analysis in beauty mesons 
as well as charm meson decays~\cite{Hinchliffe:1995hz, 
Bhattacharya:2008ss,Pirtskhalava:2011va,Hiller:2012xm,Feldmann:2012js,Bhattacharya:2012ah, 
Grossman:2012ry,Gronau:2015rda,Muller:2015lua,Muller:2015rna,Cheng:2016ejf,Grossman:2019xcj} 
into two pseudoscalars ($PP$), pseudoscalar-vector 
boson ($PV$), and two vector mesons ($VV$) can be easily extended to study bottom and charmed baryon decaying into an octet baryon and a pseudo-scalar meson~\cite{Lu:2009cm,Gronau:2013mza,Hsiao:2014mua,He:2015fwa,He:2015fsa,Zhu:2016bra,Hsiao:2017tif,Savage:1989qr,Pakvasa:1990if,Savage:1991wu,Lu:2016ogy,Geng:2017mxn,Geng:2018upx,Geng:2018bow,Geng:2018plk,Geng:2018rse,Grossman:2018ptn,Hsiao:2019yur,Geng:2019xbo,Wang:2019alu,Wang:2019dls,Jia:2019zxi,Dery:2020lbc}.

On the other hand, experimental evidences suggest a $b$ baryon to a decuplet baryon transition alongside the pseudo-scalar(vector) meson is also possible. In this paper, we consider such hadronic beauty-baryon decays featuring a decuplet 
of light baryons and a pseudoscalar meson based on the $SU(3)$ 
decomposition of the decay amplitudes in terms of $SU(3)$-reduced amplitudes. An alternate approach based on flavor flow along quark lines is also presented which is equivalent to the $SU(3)$ decomposition in the limit of exact $SU(3)$-flavor symmetry. In contrast to previous studies, our approach~\cite{Grinstein:1996us,Grinstein:2004kn,Roy:2019cky} facilitates an $SU(3)$ decomposition of the decays in terms of $SU(3)$-reduced amplitudes without any 
particular set of assumptions about the underlying dynamics.

The formalism of $SU(3)$ decomposition of the decay amplitudes in terms of 
reduced $SU(3)$ amplitudes is presented in Sec.~\ref{sec:t}. The results are 
summarized in Appendix~\ref{sec:App1} and \ref{sec:App2}. The quark flow 
diagrams are indicated in Appendix~\ref{sec:App3}. The relations between the amplitudes 
for beauty baryon decays into decuplet of light baryons and pseudoscalar mesons 
are derived in Sec.~\ref{sec:amp rel}. The effects of $SU(3)$ breaking on 
account of $s$-quark mass are considered in Sec.~\ref{sec:brk SU3}. The 
corresponding relations between $CP$ asymmetries are derived in 
Sec.~\ref{sec:CP rel}. We finally conclude in Sec.~\ref{sec:Conc}.

\section{Formalism}
\label{sec:t}
In this paper we study the charmless decay of an anti-triplet  
($\overline{\textbf{3}}$) beauty-baryon into a decuplet baryon ($B$) and an 
octet pseudoscalar meson ($P$), i.e. $B_b(\overline{\textbf{3}})\to 
B(\textbf{10})\, 
P(\textbf{8})$. This decay is described by a Hamiltonian with $\Delta Q=0$ and 
$\Delta S=-1,0$ (equivalent to $\Delta I_{3}$ and $\Delta Y$ representation).
The possible decays can be divided into two sub classes, namely the $\Delta 
S=0$ 
and $\Delta S=-1$ transitions. The allowed final state $SU(3)$ representations 
($\mathbf{f}$) are; $\mathbf{8}$, $\mathbf{10}$, 
$\mathbf{27}$,  $\mathbf{35}$. There are twenty physical 
process possible for $\Delta S=0$ and another twenty for $\Delta S=-1$. In 
Appendix~\ref{sec:App1}, each of these 
decay modes are decomposed in terms of the $SU(3)$ reduced amplitudes that add upto 
forty. Since the physical $\eta$ and $\eta^{'}$ mesons are admixtures of octet 
$\eta_{8}$ and singlet $\eta_{1}$ mesons, a study of 
$B_{b}(\overline{\textbf{3}})\to B(\textbf{10})\, 
P(\textbf{1})$ is also necessary. Therefore one has to take into account four (two 
each for $\Delta S=-1$ and $\Delta S=0$) additional independent $SU(3)$ amplitudes. Since no assumption about the particular form of effective Hamilton has been made yet, these forty four reduced amplitudes are all independent of each other and no amplitude relation exist between the decay modes.
The expression of the amplitudes in terms of reduced $SU(3)$ amplitudes 
is concisely given as,
\begin{widetext}
\vspace{-0.2cm}
	\begin{eqnarray}
	\label{Master Formula}
	\mathcal{A}(i \to f_{b} 
	f_{m})=(-1)^{I_{3}-\frac{Y}{2}-\frac{T}{3}}\quad\sum_{\mathclap 
	{\substack{\{f,	\, R\} \\
				Y^{b}+Y^{m}=Y^{f}, \, Y^{f}-Y^{i}=Y^{H}\\
				I_{3}^{b}+I_{3}^{m}=I_{3}^{f},\, 
				I_{3}^{f}-I_{3}^{i}=I_{3}^{H}}}} 
	C^{I_{3}^{b}\, I_{3}^{m}\, I_{3}^{f} }_{I^{b}\, I^{m}\, I^{f}} 
	\quad \begin{pmatrix}
	\mathbf{f_{b}}       & \mathbf{f_{m}}  & \mathbf{f} \\
	\noalign{\smallskip}    
	(Y^{b},I^{b},I^{b}_{3}) & (Y^{m},I^{m},I^{m}_{3}) 	& 
	(Y^{f},I^{f},I^{f}_{3}) \\
	\end{pmatrix} \nonumber\\
	\quad \begin{pmatrix}
	\mathbf{f}       & \mathbf{\bar{i}}  & \mathbf{R} \\
	\noalign{\smallskip}    
	(Y^{f},I^{f},I^{f}_{3}) & (-Y^{i},I^{i},-I^{i}_{3}) 	& 
	(Y^{H},I^{H},I^{H}_{3}) \\
	\end{pmatrix}  C^{I_{3}^{f}\, -I_{3}^{i}\, I_{3}^{H} }_{I^{f}\, I^{i}\, 
	I^{H}} 
	\langle \mathbf{f}\parallel \mathbf{R_{I}} \parallel 
	\mathbf{i}\rangle,
	\end{eqnarray}     
\end{widetext}
where, $C^{a,b,c}_{A,B,C}$ are the $SU(2)$ Clebsch-Gordon coefficients and 
\begin{equation}
\begin{pmatrix}
\mathbf{R_{a}}       & \mathbf{R_{b}}  & \mathbf{R_{c}} \\
\noalign{\smallskip}    
(Y^{a},I^{a},I^{a}_{3}) & (Y^{b},I^{b},I^{b}_{3}) 	& 
(Y^{c},I^{c},I^{c}_{3}) 
\end{pmatrix} .
\end{equation}
are the $SU(3)$ isoscalar coefficients~\cite{Shortley,deSwart:1963pdg,Samios:1974tw,Kaeding:1995re,Kaeding:1995vq} obtained by coupling the 
representations $\mathbf{R_{a}}\otimes\mathbf{R_{b}} \to \mathbf{R_{c}}$. $T$ 
is the triality of a $SU(3)$ representation of the initial state conjugate $\overline{\mathbf{i}}$ and the phase factor appearing in front ensures that correct signs are assigned to the individual initial $b$-baryon anti-triplet. Given a form of effective Hamiltonian ($\mathcal{H}_{\text{eff}}$), it 
can be $SU(3)$ decomposed, 
\begin{align}
\label{eff Ham definite}
\mathcal{H}_{\text{eff}}=\sum_{\substack{ \{Y,I,I_{3}\}\\ 
\mathbf{R}}}\mathcal{F}^{\{Y,I,I_{3}\}}_{\mathbf{R}} \mathbf{R_{I}},
\end{align}
where $\mathcal{F}^{\{Y,I,I_{3}\}}_{\mathbf{R}}$ depends on the $SU(3)$ CG 
coefficients appearing in front of the $SU(3)$ representations 
($\mathbf{R_{I}}$). Moreover $\mathcal{F}^{\{Y,I,I_{3}\}}_{\mathbf{R}}$ also 
contains additional factors entering Eq.~\eqref{eff Ham definite} in form of 
Wilson coefficients and CKM elements. It is also important to note that by 
knowing the dynamical coefficients for different isospin values in a given 
$SU(3)$ representation, one can drop the isospin Casimir label ($I$) and 
express the Wigner-Eckart reduced matrix element $\langle \mathbf{f}\parallel 
\mathbf{R} \parallel \mathbf{i}\rangle$, in its usual form, independent of the 
isospin $I$ 
label. By using completeness of $SU(3)$ CG coefficients up to a phase factor, 
\begin{align}
\langle \mathbf{f}\parallel \mathbf{R_{I}} \parallel 
\mathbf{i}\rangle=\underbrace{\mathcal{F}^{\{Y,I,I_{3}\}}_{\mathbf{R}} 
\sqrt{\frac{\text{dim f }}{\text{dim R }}}}_{\text{dynamical Coeff.
of}~\mathcal{H}} \langle \mathbf{f}\parallel \mathbf{R} \parallel 
\mathbf{i}\rangle.
\end{align}
\subsection*{}
The lowest order effective Hamiltonian~\cite{Ciuchini:1993vr,Buchalla:1995vs,Buras:1998raa}
for charmless $b$-baryon decays consists of $\Delta S=-1$ and $ \Delta S=0$ parts. Each part is composed from the 
operators $ O_{1}$,\ldots, $ O_{10}$. The complete Hamiltonian can be written 
as:
\begin{widetext}
\begin{eqnarray}
\label{eff H}
\mathcal{H}_{\text{eff}}=\frac{4G_{F}}{\sqrt{2}}\Big[\lambda^{(s)}_{u}
\Big(C_{1}(O^{(u)}_{1}-O^{(c)}_{1})+C_{2}(O^{(u)}_{2}-O^{(c)}_{2})\Big)-
\lambda^{(s)}_{t} \sum_{i=1,2}C_{i}O^{(c)}_{i}-
\lambda^{(s)}_{t}
\sum_{i=3}^{10}C_{i}O_{i}^{(s)}\nonumber \\ +\lambda^{(d)}_{u}
\Big(C_{1}(O^{(u)}_{1}-O^{(c)}_{1})+C_{2}(O^{(u)}_{2}-O^{(c)}_{2})\Big)-\lambda^{(d)}_{t} \sum_{i=1,2}C_{i}O^{(c)}_{i}-
\lambda^{(d)}_{t}
\sum_{i=3}^{10}C_{i}O_{i}^{(d)} \Big],
\end{eqnarray} 
\end{widetext}
where $V_{ub}V_{us}^{*}=\lambda_{u}^{s}$, $V_{ub}V_{ud}^{*}=\lambda_{u}^{d}$, 
$V_{tb}V_{ts}^{*}=\lambda_{t}^{s}$, $V_{tb}V_{td}^{*}=\lambda_{t}^{d}$ are the 
CKM elements and $C_{i}$ s are the Wilson coefficients. $O_{1}$ and $ O_{2}$ 
are the ``tree'' operators:
\begin{align}
O_{1}^{(u)} & =  (\overline{u}^{i}_{L}\gamma^{\mu}b^{j}_{L})
(\overline{s}^{j}_{L}\gamma_{\mu}u^{i}_{L}) \nonumber \\
O_{1}^{(c)} & =  (\overline{c}^{i}_{L}\gamma^{\mu}b^{j}_{L})
(\overline{s}^{j}_{L}\gamma_{\mu}c^{i}_{L}) \nonumber\\
O_{2}^{(u)}&=(\overline{u}^{i}_{L}\gamma^{\mu}b^{i}_{L})
(\overline{s}^{j}_{L}\gamma_{\mu}u^{j}_{L})  \nonumber  \\
O_{2}^{(c)}&=(\overline{c}^{i}_{L}\gamma^{\mu}b^{i}_{L})
(\overline{s}^{j}_{L}\gamma_{\mu}c^{j}_{L}). 
\end{align}
$ O_{3}$, \ldots, $ O_{6}$ are the ``gluonic penguin'' operators:
\begin{align}
O_{3}^{(s)} & = (\overline{s}^{i}_{L}\gamma^{\mu}b^{i}_{L})
\sum_{q=u,d,s}(\overline{q}^{j}_{L}\gamma_{\mu}q^{j}_{L})\nonumber \\
O_{4}^{(s)} & = (\overline{s}^{i}_{L}\gamma^{\mu}b^{j}_{L}) 
\sum_{q=u,d,s}(\overline{q}^{j}_{L}\gamma_{\mu}q^{i}_{L})\nonumber \\
O_{5}^{(s)} & = (\overline{s}^{i}_{L}\gamma^{\mu}b^{i}_{L})
\sum_{q=u,d,s}(\overline{q}^{j}_{R}\gamma_{\mu}q^{j}_{R})\nonumber \\
O_{6}^{(s)} & = (\overline{s}^{i}_{L}\gamma^{\mu}b^{j}_{L})
\sum_{q=u,d,s}(\overline{q}^{j}_{R}\gamma_{\mu}q^{i}_{R}). 
\end{align}
and finally $ O_{7}$, \ldots, $ O_{10}$ are the four ``Electroweak penguin" (EWP) operators:  
\begin{align}
O_{7}^{(s)} & = \frac{3}{2} (\overline{s}^{i}_{L}\gamma^{\mu}b^{i}_{L})
\sum_{q=u,d,s}e_{q}(\overline{q}^{j}_{R}\gamma_{\mu}q^{j}_{R}),\nonumber \\ 
O_{8}^{(s)} & =\frac{3}{2} (\overline{s}^{i}_{L}\gamma^{\mu}b^{j}_{L})
\sum_{q=u,d,s}e_{q}(\overline{q}^{j}_{R}\gamma_{\mu}q^{i}_{R}),\\
O_{9}^{(s)} & = \frac{3}{2} (\overline{s}^{i}_{L}\gamma^{\mu}b^{i}_{L})
\sum_{q=u,d,s}e_{q}(\overline{q}^{j}_{L}\gamma_{\mu}q^{j}_{L}),\nonumber \\
O_{10}^{(s)} & =\frac{3}{2} (\overline{s}^{i}_{L}\gamma^{\mu}b^{j}_{L})
\sum_{q=u,d,s}e_{q}(\overline{q}^{j}_{L}\gamma_{\mu}q^{i}_{L}).
\end{align}

\noindent ${\cal H}_{\rm eff}$  is a  linear combinations of four quark operators of the form $(\overline{q}_{1}b)(\overline{q_{2}}q_{3})$.
These operators transform as $3\bigotimes 3 \bigotimes \overline{3}$
under $SU(3)$-flavor and can be decomposed into sums of irreducible  operators corresponding to irreducible $SU(3)$ representations:
$\textbf{15}\,$,$\,\mathbf{\overline{6}}\,$,$\,\mathbf{3^{(6)}}\,$,$\,\mathbf{3^{(\bar{3})}}\,$ where  the superscript index: `6' (`$\overline{3}$') indicates the origin of $\mathbf{3}$ out of the two possible representations arising from the tensor product of $q_{1}$ and $q_{2}$. The $SU(3)$ triplet representation of quarks ($q_{i}$) and its conjugate denoting the anti-quarks ($\overline{q_{i}}$) consist of the flavor states;
\begin{align}
\label{eqn:q-antiq}
q_{i}=
\begin{pmatrix}
& u &\\& d &\\& s &\\ 
\end{pmatrix}
\qquad
\overline{q_{i}}=
\begin{pmatrix}
&\overline{d}& \\ &-\overline{u}& \\ &\overline{s}& \\ 
\end{pmatrix}
\end{align}
The unbroken dim-6 effective Hamiltonian contains parts that transform as $\bf{3},\,\overline{\bf{6}},\,\bf{15}$ under $SU(3)$-flavor. The tree part of the effective Hamiltonian responsible for $b\to u$ transition has the following $SU(3)$ decomposition ,\\
\begin{widetext}
\begin{align}
\label{T Ham}
\frac{\sqrt{2}\mathcal{H}_{\sss{\rm T}}}{4G_{F}}=&\left\{\lambda_{u}^{s}
\left[\frac{(C_{1}+C_{2})}{2}\left(-\textbf{15}_{1}-\frac{1}{\sqrt{2}}
\textbf{15}_{0}-\frac{1}{\sqrt{2}}\textbf{3}^{(\textbf{6})}_{0}\right)
+\frac{(C_{1}-C_{2})}{2}\left(\overline{\textbf{6}}_{1}+\textbf{3}^{(\overline
	{\textbf{3}})}_{0}\right)\right]\right.\nonumber
\\&+\left.\lambda_{u}^{d}\left[\frac{(C_{1}+C_{2})}{2}\left(-\frac{2}{\sqrt{3}}
\textbf{15}_{3/2}-\frac{1}{\sqrt{6}}\textbf{15}_{1/2}-\frac{1}{\sqrt{2}}
\textbf{3}^{(\textbf{6})}_{1/2}\right)+\frac{(C_{1}-C_{2})}{2}
\left(-\overline{\textbf{6}}_{1/2}+\textbf{3}^{(\overline{\textbf{3}})}_{1/2}\right)\right]
\right\}.
\end{align}
\end{widetext}
The penguin part of the effective Hamiltonian is also $SU(3)$ decomposed below,
\begin{widetext} 
\begin{align}
\label{P g Ham}
\frac{\sqrt{2}\mathcal{H}_{\sss{\rm g}}}{4G_{F}}=&\left\{-\lambda_{t}^{s}
\left[-\sqrt{2}(C_{3}+C_{4})\textbf{3}^{(\textbf{6})}_{0}+(C_{3}-C_{4})
\textbf{3}^{(\overline{\textbf{3}})}_{0}\right]\right. 
-\left.\lambda_{t}^{d}\left[-\sqrt{2}(C_{3}+C_{4})\textbf{3}^{(\textbf{6})}_{1/2}
+(C_{3}-C_{4})\textbf{3}^{(\overline{\textbf{3}})}_{1/2}\right]\right.\nonumber
\\&-\left.\lambda_{t}^{s}
\left[-\sqrt{2}(C_{5}+C_{6})\textbf{3}^{(\textbf{6})}_{0}+(C_{5}-C_{6})
\textbf{3}^{(\overline{\textbf{3}})}_{0}\right]\right. 
-\left.\lambda_{t}^{d}\left[-\sqrt{2}(C_{5}+C_{6})\textbf{3}^{(\textbf{6})}_{1/2}
+(C_{5}-C_{6})\textbf{3}^{(\overline{\textbf{3}})}_{1/2}\right]\right\},
\end{align} 
\begin{align}
\label{P EW Ham}
\frac{\sqrt{2}\mathcal{H}_{\sss{\rm EWP}}}{4G_{F}}=&\left\{-\lambda_{t}^{s}
\left[\frac{(C_{9}+C_{10})}{2}\left(-\frac{3}{2}\textbf{15}_{1}
-\frac{3}{2\sqrt{2}}\textbf{15}_{0}+\frac{1}{2\sqrt{2}}
\textbf{3}^{(\textbf{6})}_{0}\right)+\frac{(C_{9}-C_{10})}{2}
\left(\frac{3}{2}\overline{\textbf{6}}_{1}+\frac{1}{2}
\textbf{3}^{(\overline{\textbf{3}})}_{0}\right)\right]\right.\nonumber
 \\ &\left.-\lambda_{t}^{d}\left[\frac{(C_{9}+C_{10})}{2}\left(-\sqrt{3}\, 
\textbf{15}_{3/2}-\frac{1}{2}\sqrt{\frac{3}{2}}\textbf{15}_{1/2}
+\frac{1}{2\sqrt{2}}\textbf{3}^{(\textbf{6})}_{1/2}\right)
+\frac{(C_{9}-C_{10})}{2}\left(-\frac{3}{2}\overline{\textbf{6}}_{1/2}
+\frac{1}{2}\textbf{3}^{(\overline{\textbf{3}})}_{1/2}\right)\right]\right\},
\end{align}
\end{widetext}
where we have ignored the contributions from $O_{7}$ and $O_{8}$ Electroweak 
penguin operators due to the smallness of the Wilson coefficients $C_{7}$ and 
$C_{8}$ appearing in front of them. For charmless $b$-decays, $b\to 
qc\overline{c}$, $q=d,s$ contributes as an independent $\mathbf{3}$ of $SU(3)$ 
without affecting the $SU(3)$ structure of Eq.~\eqref{T Ham}-\eqref{P EW Ham}. 
With this particular choice of effective Hamiltonian and allowed final state 
$SU(3)$ representations, there are five independent $SU(3)$-reduced matrix 
elements:
\vspace{-0.1cm}
\begin{align}
\langle 8 \parallel\textbf{3}\parallel\overline{3}\rangle,\qquad 
\langle 8\parallel\overline{\textbf{6}}\parallel\overline{3}\rangle,\qquad \langle 8\parallel\textbf{15}\parallel\overline{3}\rangle\,\nonumber\\ \langle 10\parallel\textbf{15}\parallel\overline{3}\rangle,\qquad\langle 27 \parallel\textbf{15}\parallel\overline{3}\rangle\qquad
\end{align}
and the $SU(3)$-reduced amplitudes  arising from $\mathcal{H}_{\sss{\rm T}}$ 
are conveniently expressed as follows;
\begin{align}
\label{alg amp}
c_{8}&=-(C_{1}+C_{2})\,\langle\text{8}\parallel\textbf{3}^{(6)}\parallel\overline{\text{3}}\rangle
 \nonumber\\&\quad +\sqrt{2}(C_{1}-C_2)\,\langle 
\text{8}\parallel\textbf{3}^{(\overline{3})}\parallel\overline{\text{3}}\rangle\nonumber\\
b_{8}&=(C_{1}-C_2)\,\langle 
\text{8}\parallel\overline{\textbf{6}}\parallel\overline{\text{3}}\rangle\qquad\nonumber\\
a_{8}&=(C_{1}+C_{2})\,\langle\text{8}\parallel\textbf{15}\parallel\overline{\text{3}}\rangle\qquad\\
a_{10}&=(C_{1}+C_{2})\,\langle\text{10}\parallel\textbf{15}\parallel\overline{\text{3}}\rangle\qquad\nonumber\\
a_{27}&=(C_{1}+C_{2})\,\langle\text{27}\parallel\textbf{15}\parallel\overline{\text{3}}\rangle.\qquad\nonumber
\end{align}  
The decay amplitudes for all possible $\Delta S=-1$ and $\Delta S=0$ processes are expressed using Eq~\eqref{alg amp} and are given in Table~\ref{Tab S1} and Table~\ref{Tab S0} respectively. \\
An alternate description of decay amplitudes is obtained in terms of topological quark diagrams. The symmetry properties of the final state decuplet baryons allow five possible diagrams starting from a flavor anti-triplet $b$-baryon whose light quarks are in a flavor anti-symmetric states. Those five independent topologies
consist of three $W$-exchanges ($E_1$, $E_1$, $E_3$), one tree ($T$) and a 
penguin-like ($P_{q}$) 
($q$ being the flavor of the quark going in the loop) amplitude where the marked quarks are anti-symmetrized in the initial state baryon. In addition, a sixth diagram ($S$) appears after including the decay modes containing the singlet $\eta_{1}$. The mapping between the topological amplitudes and the $SU(3)$-reduced amplitudes is given below,

\onecolumngrid
\begin{figure*}[t]
  \centering
  \includegraphics[ width=\textwidth]{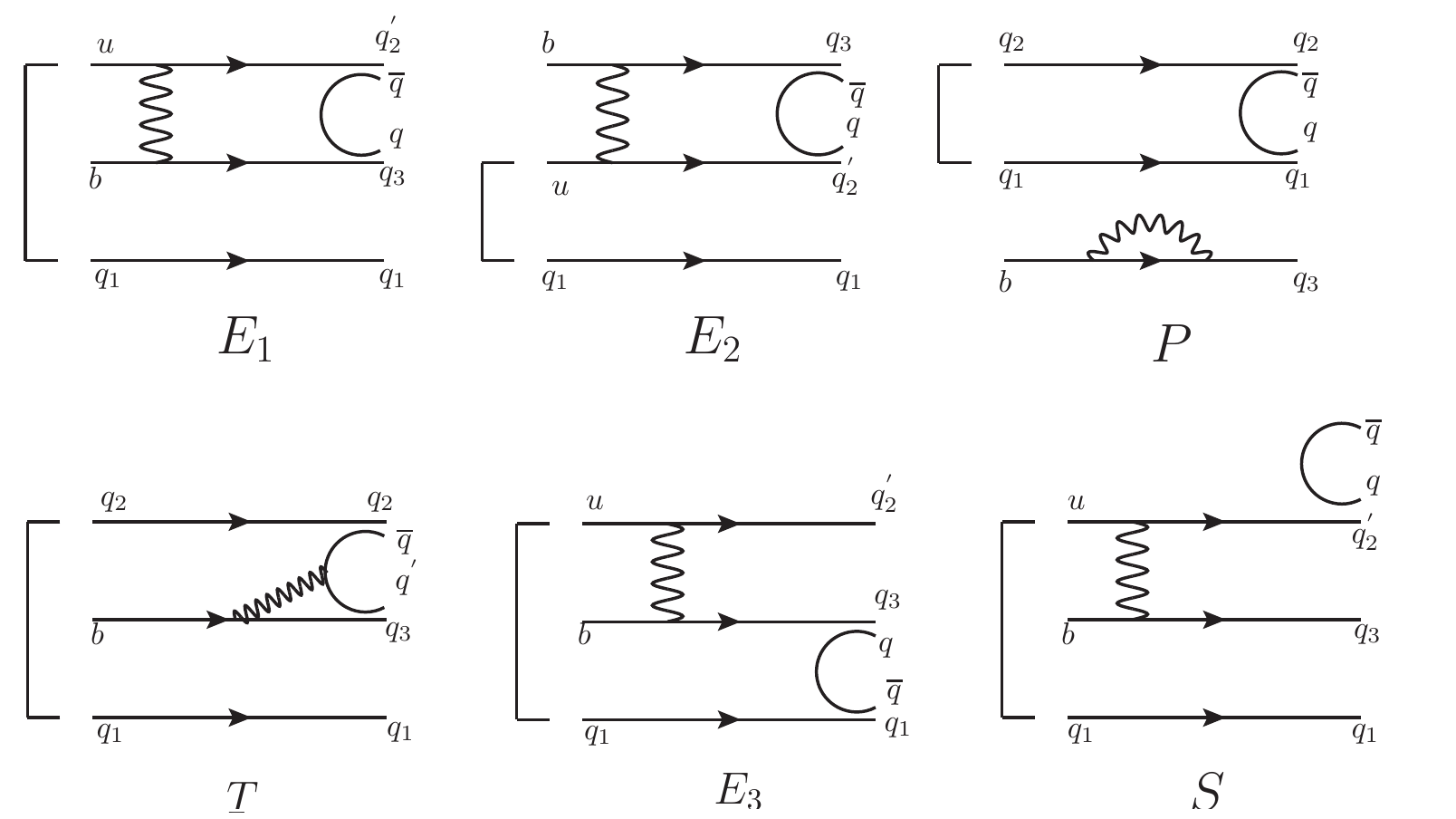}
\caption{Topological diagrams contributing to tree amplitude for anti-triplet $b$-baryon hadronic decays. The first five diagrams contribute to $B_b(\overline{\textbf{3}})\to B(\textbf{10})\, P(\textbf{8})$ processes. The last diagram contributes specifically to $B_b(\overline{\textbf{3}})\to B(\textbf{10})\, P(\textbf{1})$ processes.}  
\label{diagram1}
\end{figure*}
\twocolumngrid
\begin{widetext}
\begin{align}
&c_{8}=-\frac{1}{8} \Big(-\sqrt{10}
   \,E_{1}+3 \sqrt{10}\, E_{2}-2 \sqrt{10}\, E_{3} \nonumber\\& \qquad\qquad\qquad+8 \sqrt{10}\, P_{u}+2 \sqrt{10}\, T\Big)\\
&b_{8}=\frac{1}{4} \left(\sqrt{15}\,
   E_{2}-\sqrt{15}\, E_{1}\right)\nonumber\\
&a_{8}=\frac{1}{24} \left(5 \sqrt{6}\, E_{1}+5 \sqrt{6}\, E_{2}-10 \sqrt{6}\, E_{3}+2 \sqrt{6}\, T\right)\nonumber\\
&a_{10}=\frac{1}{3} \left(2 \sqrt{3}\, E_{1}+2 \sqrt{3}\, E_{2}+2 \sqrt{3}\, E_{3}-\sqrt{3}\, T\right)\nonumber\\
&a_{27}=\sqrt{\frac{2}{3}}\, T.\nonumber
\end{align}
\end{widetext}

The topologies $E_3$ and $T$ are sometimes ignored in accordance with		
Korner-Pati-Woo theorem~\cite{Korner:1970xq,Pati:1970fg,Stech:1987fa,
Neubert:1991zd,Kohara:1991ug,Xu:1992sw,Chau:1995gk,Neubert:1997uc,Geng:2019awr,
Liang:2018rkl,Gutsche:2018utw,Gutsche:2018msz} based on $``\Delta I=1/2"$ 
rule which suggests that the quark pair produced in weak interaction ending up in the same baryon must be anti-symmetric in flavour space. However, the accuracy of 
such a statement is dependent on the modeling of the baryon-baryon transition and 
exact $SU(3)$ flavor symmetry. The details of the arguments made to ignore 
$E_3$ and $T$ are reproduced in Appendix~\ref{sec:App4} for completeness.  We 
derive several interesting tests (Eqs.~\eqref{USpin-1}--\eqref{USpin-3} of 
Appendix~\ref{sec:App4}) to verify if the contribution from the topological 
amplitudes $E_3$ and $T$  are indeed suppressed. We 
have chosen to include the $E_{3}$ and $T$ diagrams throughout the rest of this 
paper for generality.

The two approaches are equivalent and imply that individual topological amplitudes cannot be expressed in terms of a single $SU(3)$-reduced amplitude and vice-versa in these two basis. It is also important to note that the topology, $P_{u}$, originate 
purely from the tree operators i.e. $O_{1}$ and $O_{2}$ as emphasized in 
\cite{Neubert:1997wb,Atwood:1997iw,Buras:1997cv,Buras:1998ra,Falk:1998wc}, 
even though it is denoted by $P_{u}$ and referred to as a penguin 
topology. Moreover the $W$-exchange topologies being a 
four-quark tree-like structure also contribute to the tree amplitudes.
 
The transition induced by QCD penguin operators, given in Eq~\eqref{P g Ham} is only a $\mathbf{3}$ under $SU(3)$ and corresponding $SU(3)$-reduced amplitude is identified with the `penguin' amplitude $P_{t}$,
		
%
\begin{align}
P_{t}=
-\sqrt{2}(C_{3}+C_{4}+C_{5}+C_{6})\langle\text{8}\parallel\textbf{3}^{(\textbf{6})}\parallel\overline{\text{3}}\rangle\nonumber\\+(C_{3}-C_{4}+C_{5}-C_{6})
\langle\text{8}\parallel\textbf{3}^{(\overline{\textbf{3}})}_{0}\parallel\overline{\text{3}}\rangle.
\end{align}
%


\subsection{Relating EWP and Tree amplitudes}

We begin this section by noting that the $\mathbf{15}$ and 
$\overline{\mathbf{6}}$ part of the Hamiltonian described in Eqs.~\eqref{T 
Ham} and \eqref{P EW Ham} relate the contributions to the decay from the tree 
and EWP operators~\cite{Deshpande:1994pw,Gronau:1998fn,Deshpande:2000jp} respectively, so as to effectively obey the following 
relations, 
\begin{align}
\mathcal{H}^{\text{EWP}}_{\mathbf{15}}(\Delta 
S)&=-\frac{3}{2}\frac{\lambda^{s(d)}_{t}(C_{9}+C_{10})}
{\lambda^{s(d)}_{u}(C_{1}+C_{2})}\mathcal{H}^{\text{T}}_{\mathbf{15}}(\Delta
S)\nonumber\\
\mathcal{H}^{\text{EWP}}_{\overline{\mathbf{6}}}(\Delta 
S)&=-\frac{3}{2}\frac{\lambda^{s(d)}_{t}(C_{9}-C_{10})}
{\lambda^{s(d)}_{u}(C_{1}-C_{2})}\mathcal{H}^{\text{T}}_{\overline{\mathbf{6}}}
(\Delta S)
\end{align}
These relations are valid independently for both $\Delta S=\{0,-1\}$ decays and 
remain unaffected by the QCD penguin operators since $O_{3}$ and $O_{4}$ 
transforms entirely as a $\mathbf{3}$ under $SU(3)$.
Apriori, the $\mathbf{3}^{(6)}$ and $\mathbf{3}^{(\overline{3})}$ operators do 
not follow a simple relation for arbitrary values of $C_{1},\,C_{2},\, C_{9}$ 
and $C_{10}$. Since the set of $SU(3)$-reduced elements remains the same, in 
analogy to Eq.~\eqref{alg amp} one can define,
\begin{align}
c_{8}^{EW}&=(C_{9}+C_{10})\,\langle\text{8}\parallel\textbf{3}^{(6)}\parallel
	\overline{\text{3}}\rangle\quad\nonumber\\
	&\qquad+\sqrt{2}(C_{9}-C_{10})\,\langle 
	\text{8}\parallel\textbf{3}^{(\overline{3})}\parallel\overline{\text{3}}
	\rangle\nonumber\\
b_{8}^{EW}&=-\frac{3}{2}(C_{9}-C_{10})\,\langle 
\text{8}\parallel\overline{\textbf{6}}\parallel\overline{\text{3}}
	\rangle\quad\nonumber\\
a_{8}^{EW}&=-\frac{3}{2}(C_{9}+C_{10})\,\langle\text{8}\parallel\textbf{15}
	\parallel\overline{\text{3}}\rangle\quad\\
a_{10}^{EW}&=-\frac{3}{2}(C_{9}+C_{10})\,\langle\text{10}\parallel\textbf{15}
	\parallel\overline{\text{3}}\rangle\quad\nonumber\\
a_{27}^{EW}&=-\frac{3}{2}(C_{9}+C_{10})\,\langle\text{27}\parallel\textbf{15}
	\parallel\overline{\text{3}}\rangle.\quad\nonumber
\end{align}
The EWP reduced amplitudes $b_{8}^{EW},a_{8}^{EW},a_{10}^{EW}$ and 
$a_{27}^{EW}$ are expressed in terms of $b_{8},a_{8},a_{10}$ and $a_{27}$ 
defined in Eq.~\eqref{alg amp};
\begin{align}
\label{PEW rel}
b_{8}^{EW}=-\frac{3}{2}\frac{C_{9}-C_{10}}{C_{1}-C_{2}}b_{8},\quad
a_{8}^{EW}=-\frac{3}{2}\frac{C_{9}+C_{10}}{C_{1}+C_{2}}a_{8},\\
a_{10}^{EW}=-\frac{3}{2}\frac{C_{9}+C_{10}}{C_{1}+C_{2}}a_{10},\quad
a_{27}^{EW}=-\frac{3}{2}\frac{C_{9}+C_{10}}{C_{1}+C_{2}}a_{27}\nonumber
\end{align}

Using numerical values of the Wilson coefficients to leading logarithmic order 
one obtains, \\
\begin{align}
\frac{C_{9}+C_{10}}{C_{1}+C_{2}}=-1.139\,\alpha, \qquad  \frac{C_{9}-C_{10}}{C_{1}-C_{2}}=-1.107\,\alpha.
\end{align}
To a good approximation~\cite{Gronau:1998fn} these two ratios of Wilson coefficients can be taken to be a common value~\cite{Buchalla:1995vs,Buras:1998raa} given by $\kappa$;\\
\begin{align}
\kappa=\frac{C_{9}+C_{10}}{C_{1}+C_{2}}=\frac{C_{9}-C_{10}}{C_{1}-C_{2}}\simeq-1.12\,\alpha.
\end{align}
With this additional assumption, Eq.~\eqref{PEW rel} implies that the 
 $b_{8}^{EW},a_{8}^{EW},a_{10}^{EW}$ and $a_{27}^{EW}$ are proportional to the 
tree reduced amplitudes $b_{8},a_{8},a_{10}$ and $a_{27}$ 
\begin{align}
\label{PEW and T}
b_{8}^{EW}=-\frac{3}{2}\kappa\, b_{8}, \qquad a_{8}^{EW}=-\frac{3}{2}\kappa\,a_{8},\quad \nonumber\\
a_{10}^{EW}=-\frac{3}{2}\kappa\,a_{10},\qquad a_{27}^{EW}=-\frac{3}{2}\kappa\,a_{27}.\quad 
\end{align}
The equivalence of $SU(3)$-reduced amplitudes to topological diagrams, allows one to interpret $P_{i}^{EW}$ as Electroweak quark diagrams with one insertion of the Electroweak penguin operator;
\begin{align}
P_{E_i}^{EW}&=-\frac{3}{2}\kappa E_i\\
P_{T}^{EW}&=-\frac{3}{2}\kappa T
\end{align}
where, $i=\{1,2,3\}$. 
As mentioned earlier, there is no simple relation between the $\mathbf{3}$ part of the EWP Hamiltonian to the tree part. There are, however, decays where the $\mathbf{3}$ part of the Hamiltonian cannot contribute to the formation of final states which require a pure $\Delta I= 1$ or $\Delta I=3/2$ transition. For example, the decay $\Lambda_{b}^{0}\to\Delta^{+}K^{-}$ receives contribution from $\overline{\mathbf{6}}$ and $\mathbf{15}$ part of the effective Hamiltonian and in this particular case the ratio of EWP and tree contributions is given entirely by the simple ratio $-3/2\kappa(\lambda^{s}_{t}/\lambda^{s}_{u})$. The $SU(3)$-reduced amplitudes for the penguin operators are provided in Eq.~\ref{PEW S0 tab} and Eq.~\ref{PEW S1 Tab}.

\onecolumngrid
\vspace{-0.1cm}
\begin{table*}[ht!]
\vspace{-0.3cm}
\begin{tabular}{|c|c|c|}	
\hline
$\Delta\,S=-1$&$SU(3)-\text{reduced amplitude}$&$\text{Topological diagrams}$\\ 
\hline
$\mathcal{A}(\Lambda_{b}^{0}\to  \Delta^{+}K^{-})$&$\frac{1}{60} \left(-5 \sqrt{2} a_{10}-3 a_{27}-8 a_{8}+4 \sqrt{10} b_{8}\right)$&$-\frac{1}{\sqrt{6}}E_{1}$\\
$\mathcal{A}(\Lambda_{b}^{0}\to  \Delta^{0}\bar{K}^{0})$&$\frac{1}{60} \left(5 \sqrt{2} a_{10}+3 a_{27}+8 a_{8}-4 \sqrt{10} b_{8}\right)$&$\frac{1}{\sqrt{6}}E_{1}$\\
$\mathcal{A}(\Lambda_{b}^{0}\to  \Sigma^{'+}\pi^{-})$&$\frac{1}{60} \left(-5 \sqrt{2} a_{10}+15 a_{27}+10 a_{8}+2 \sqrt{15} c_{8}-2 \sqrt{10} b_{8}\right)$&$\frac{1}{\sqrt{6}}\left(-E_{3}+P_{u}+T\right)$\\
$\mathcal{A}(\Lambda_{b}^{0}\to  \Sigma^{'0}\pi^{0})$&$\frac{1}{30} \left(-3 a_{27}-3 a_{8}-\sqrt{15} c_{8}\right)$&$-\frac{1}{2\sqrt{6}}\left(E_{2}-E_{3}+2P_{u}+T\right)$\\
$\mathcal{A}(\Lambda_{b}^{0}\to  \Sigma^{'0}\eta_{8})$&$-\frac{1}{10 \sqrt{3}} \left(3a_{27}-2 a_{8}+\sqrt{10} b_{8}\right)$&$\frac{1}{6\sqrt{2}}\left(2E_{1}-E_{2}-E_{3}-T\right)$\\
$\mathcal{A}(\Lambda_{b}^{0}\to  \Sigma^{'-}\pi^{+})$&$\frac{1}{60} \left(5 \sqrt{2} a_{10}-3 a_{27}+2 \left(a_{8}+\sqrt{15} c_{8}+\sqrt{10} b_{8}\right)\right) $&$\frac{1}{\sqrt{6}}\left(E_{2}+P_{u}\right)$\\
$\mathcal{A}(\Lambda_{b}^{0}\to  \Xi^{'0}K^{0})$&$\frac{1}{60} \left(5 \sqrt{2} a_{10}+15 a_{27}-10 a_{8}-2 \sqrt{15} c_{8}+2 \sqrt{10} b_{8}\right)$&$\frac{1}{\sqrt{6}}\left(E_{3}-P_{u}\right)$\\
$\mathcal{A}(\Lambda_{b}^{0}\to  \Xi^{'-}K^{+})$&$\frac{1}{60} \left(5 \sqrt{2} a_{10}-3 a_{27}+2 \left(a_{8}+\sqrt{15} c_{8}+\sqrt{10} b_{8}\right)\right)$&$\frac{1}{\sqrt{6}}\left(E_{2}+P_{u}\right)$\\
$\mathcal{A}(\Xi_{b}^{0}\to  \Sigma^{'+}K^{-})$&$\frac{1}{30} \left(-5 \sqrt{2} a_{10}+6 a_{27}+a_{8}+\sqrt{15} c_{8}+\sqrt{10} b_{8}\right)$&$\frac{1}{\sqrt{6}}\left(-E_{1}-E_{3}+P_{u}+T\right)$\\
$\mathcal{A}(\Xi_{b}^{0}\to  \Sigma^{'0} \bar{K}^{0})$&$\frac{1}{60} \left(10 a_{10}+9 \sqrt{2} a_{27}-\sqrt{2} a_{8}-\sqrt{30} c_{8}-2 \sqrt{5} b_{8}\right)$&$\frac{1}{2\sqrt{3}}\left(E_{1}+E_{3}-P_{u}\right)$\\
$\mathcal{A}(\Xi_{b}^{0}\to  \Xi^{'0} \pi^{0})$&$\frac{1}{120} \left(-10 a_{10}-27 \sqrt{2} a_{27}-2 \left(\sqrt{2} a_{8}+\sqrt{30} c_{8}+2 \sqrt{5} b_{8}\right)\right)$&$-\frac{1}{2\sqrt{3}}\left(E_{2}+P_{u}+T\right)$\\
$\mathcal{A}(\Xi_{b}^{0}\to  \Xi^{'0}\eta_{8})$&$\frac{1}{120} \left(10 \sqrt{3} a_{10}+3 \sqrt{6} a_{27}-2 \left(\sqrt{6} a_{8}+3 \sqrt{10} c_{8}+2 \sqrt{15} b_{8}\right)\right)$&$\frac{1}{6}\left(2E_{1}-E_{2}+2E_{3}-3P_{u}-T\right)$\\
$\mathcal{A}(\Xi_{b}^{0}\to  \Xi^{'-}\pi^{+})$&$\frac{1}{60} \left(5 \sqrt{2} a_{10}-3 a_{27}+2 \left(a_{8}+\sqrt{15} c_{8}+\sqrt{10} b_{8}\right)\right)$&$\frac{1}{\sqrt{6}}\left(E_{2}+P_{u}\right)$\\
$\mathcal{A}(\Xi_{b}^{0}\to  \Omega^{-}K^{+})$&$\frac{1}{60} \left(5 \sqrt{6} a_{10}-3 \sqrt{3} a_{27}+2 \sqrt{3} a_{8}+6 \sqrt{5} c_{8}+2 \sqrt{30} b_{8}\right)$&$\frac{1}{\sqrt{2}}\left(E_{2}+P_{u}\right)$\\
$\mathcal{A}(\Xi_{b}^{-}\to  \Sigma^{'0}K^{-})$&$\frac{1}{60} \left(12 \sqrt{2} a_{27}-3 \sqrt{2} a_{8}+\sqrt{30} c_{8}-2 \sqrt{5} b_{8}\right)$&$\frac{1}{2\sqrt{3}}\left(P_{u}+T\right)$\\
$\mathcal{A}(\Xi_{b}^{-}\to  \Sigma^{'-}\bar{K}^{0})$&$\frac{1}{30} \left(3 a_{27}+3 a_{8}-\sqrt{15} c_{8}+\sqrt{10} b_{8}\right)$&$-\frac{1}{\sqrt{6}}P_{u}$\\
$\mathcal{A}(\Xi_{b}^{-}\to  \Xi^{'0}\pi^{-})$&$\frac{1}{30} \left(-12 a_{27}+3 a_{8}-\sqrt{15} c_{8}+\sqrt{10} b_{8}\right)$&$-\frac{1}{\sqrt{6}}\left(P_{u}+T\right)$\\
$\mathcal{A}(\Xi_{b}^{-}\to  \Xi^{'-} \pi^{0})$&$\frac{1}{60} \left(-3 \sqrt{2} a_{27}-3 \sqrt{2} a_{8}+\sqrt{30} c_{8}-2 \sqrt{5} b_{8}\right)$&$\frac{1}{2\sqrt{3}}P_{u}$\\
$\mathcal{A}(\Xi_{b}^{-}\to  \Xi^{'-}\eta_{8})$&$\frac{1}{60} \left(3 \sqrt{6} a_{27}+3 \sqrt{6} a_{8}-3 \sqrt{10} c_{8}+2 \sqrt{15} b_{8}\right)$&-$\frac{1}{2}P_{u}$\\
$\mathcal{A}(\Xi_{b}^{-}\to  \Omega^{-} K^{0})$&$\frac{1}{30} \left(-3 \sqrt{3} a_{27}-3 \sqrt{3} a_{8}+3 \sqrt{5} c_{8}-\sqrt{30} b_{8}\right)$&$\frac{1}{\sqrt{2}}P_{u}$\\
\hline
\end{tabular}
\caption{$B_b(\overline{\textbf{3}})\to B(\textbf{10})\, P(\textbf{8})$ decay $\Delta S=-1$ transitions}
\label{Tab S1}
\vspace{0.65cm}
\begin{tabular}{|c|c|c|}	
\hline
$\Delta\,S=0$&$SU(3)-\text{reduced amplitude}$&$\text{Topological Diagram}$\\ 
\hline
$\mathcal{A}(\Lambda_{b}^{0}\to  \Delta^{+} \pi^{-})$&$\frac{1}{30} \left(-5 \sqrt{2} a_{10}+6 a_{27}+a_{8}+\sqrt{15} c_{8}+\sqrt{10} b_{8}\right)$&$\frac{1}{\sqrt{6}}\left(-E_{1}-E_{3}+P_{u}+T\right)$ \\
$\mathcal{A}(\Lambda_{b}^{0}\to  \Delta^{0} \pi^{0})$&$ \frac{1}{120} \left(10 a_{10}-9 \sqrt{2} a_{27}-4 \left(\sqrt{2} a_{8}+\sqrt{30} c_{8}+2 \sqrt{5} b_{8}\right)\right)$&$\frac{1}{2\sqrt{3}}\left(E_{1}-E_{2}+E_{3}-2P_{u}-T\right)$ \\
$\mathcal{A}(\Lambda_{b}^{0}\to  \Delta^{0} \eta_{8})$&$ \frac{1}{24} \left(-2 \sqrt{3} a_{10}-3 \sqrt{6} a_{27}\right)$&$-\frac{1}{6}\left(E_{1}+E_{2}+E_{3}+T\right)$ \\
$\mathcal{A}(\Lambda_{b}^{0}\to  \Delta^{-}\pi^{+})$&$ \frac{1}{60} \left(5 \sqrt{6} a_{10}-3 \sqrt{3} a_{27}+2 \sqrt{3} a_{8}+6 \sqrt{5} c_{8}+2 \sqrt{30} b_{8}\right)$&$\frac{1}{\sqrt{2}}\left(E_{2}+P_{u}\right)$ \\
$\mathcal{A}(\Lambda_{b}^{0}\to  \Sigma^{'0}K^{0})$&$ \frac{1}{60} \left(10 a_{10}+9 \sqrt{2} a_{27}-\sqrt{2} a_{8}-\sqrt{30} c_{8}-2 \sqrt{5} b_{8}\right)$&$\frac{1}{2\sqrt{3}}\left(E_{1}+E_{3}-P_{u}\right)$ \\
$\mathcal{A}(\Lambda_{b}^{0}\to  \Sigma^{'-}K^{+})$&$ \frac{1}{60} \left(5 \sqrt{2} a_{10}-3 a_{27}+2 \left(a_{8}+\sqrt{15} c_{8}+\sqrt{10} b_{8}\right)\right)$&$\frac{1}{\sqrt{6}}\left(E_{2}+P_{u}\right)$ \\
$\mathcal{A}(\Xi_{b}^{0}\to  \Delta^{+} K^{-})$&$ \frac{1}{60} \left(-5 \sqrt{2} a_{10}+15 a_{27}+10 a_{8}+2 \sqrt{15} c_{8}-2 \sqrt{10} b_{8}\right)$& $\frac{1}{\sqrt{6}}\left(-E_{3}+P_{u}+T\right)$\\
$\mathcal{A}(\Xi_{b}^{0}\to  \Delta^{0} \overline{K}^{0})$&$ \frac{1}{60} \left(5 \sqrt{2} a_{10}+15 a_{27}-10 a_{8}-2 \sqrt{15} c_{8}+2 \sqrt{10} b_{8}\right)$& $\frac{1}{\sqrt{6}}\left(E_{3}-P_{u}\right)$\\
$\mathcal{A}(\Xi_{b}^{0}\to  \Sigma^{'+}\pi^{-})$&$ \frac{1}{60} \left(-5 \sqrt{2} a_{10}-3 a_{27}-8 a_{8}+4 \sqrt{10} b_{8}\right)$&$-\frac{1}{\sqrt{6}}E_{1}$ \\
$\mathcal{A}(\Xi_{b}^{0}\to  \Sigma^{'0}\pi^{0})$&$\frac{1}{60} \left(-12 a_{27}+3 a_{8}-\sqrt{15} c_{8}-3 \sqrt{10} b_{8}\right)$&$\frac{1}{2\sqrt{6}}\left(E_{1}-E_{2}-P_{u}-T\right)$ \\
$\mathcal{A}(\Xi_{b}^{0}\to  \Sigma^{'0} \eta_{8})$&$\frac{1}{60} \left(-5 \sqrt{3} a_{8}-3 \sqrt{5} c_{8}+\sqrt{30} b_{8}\right)$& $-\frac{1}{6\sqrt{2}}\left(E_{1}+E_{2}-2E_{3}+3P_{u}+T\right)$\\
$\mathcal{A}(\Xi_{b}^{0}\to  \Sigma^{'-}\pi^{+})$&$\frac{1}{60} \left(5 \sqrt{2} a_{10}-3 a_{27}+2 \left(a_{8}+\sqrt{15} c_{8}+\sqrt{10} b_{8}\right)\right)$&$\frac{1}{\sqrt{6}}\left(E_{2}+P_{u}\right)$ \\
$\mathcal{A}(\Xi_{b}^{0}\to \Xi^{'0} K^{0})$&$ \frac{1}{60} \left(5 \sqrt{2} a_{10}+3 a_{27}+8 a_{8}-4 \sqrt{10} b_{8}\right)$&$\frac{1}{\sqrt{6}}E_{1}$ \\
$\mathcal{A}(\Xi_{b}^{0}\to  \Xi^{'-}K^{+})$&$ \frac{1}{60} \left(5 \sqrt{2} a_{10}-3 a_{27}+2 \left(a_{8}+\sqrt{15} c_{8}+\sqrt{10} b_{8}\right)\right)$&$\frac{1}{\sqrt{6}}\left(E_{2}+P_{u}\right)$ \\
$\mathcal{A}(\Xi_{b}^{-}\to  \Delta^{'0}K^{-})$&$ \frac{1}{30} \left(12 a_{27}-3 a_{8}+\sqrt{15} c_{8}-\sqrt{10} b_{8}\right)$& $\frac{1}{\sqrt{6}}\left(P_{u}+T\right)$\\
$\mathcal{A}(\Xi_{b}^{-}\to \Delta^{'-}\overline{K}^{0})$&$ \frac{1}{30} \left(3 \sqrt{3} a_{27}+3 \sqrt{3} a_{8}-3 \sqrt{5} c_{8}+\sqrt{30} b_{8}\right)$&$-\frac{1}{\sqrt{2}}P_{u}$\\
$\mathcal{A}(\Xi_{b}^{-}\to \Sigma^{'0} \pi^{-})$&$ \frac{1}{60} \left(-12 \sqrt{2} a_{27}+3 \sqrt{2} a_{8}-\sqrt{30} c_{8}+2 \sqrt{5} b_{8}\right)$& $-\frac{1}{2\sqrt{3}}\left(P_{u}+T\right)$\\
$\mathcal{A}(\Xi_{b}^{-}\to \Sigma^{'-} \pi^{0})$&$ \frac{1}{60} \left(-3 \sqrt{2} a_{27}-3 \sqrt{2} a_{8}+\sqrt{30} c_{8}-2 \sqrt{5} b_{8}\right)$& $\frac{1}{2\sqrt{3}}P_{u}$\\
$\mathcal{A}(\Xi_{b}^{-}\to \Sigma^{'-} \eta_{8})$&$ \frac{1}{60} \left(3 \sqrt{6} a_{27}+3 \sqrt{6} a_{8}-3 \sqrt{10} c_{8}+2 \sqrt{15} b_{8}\right)$& $-\frac{1}{2}P_{u}$\\
$\mathcal{A}(\Xi_{b}^{-}\to  \Xi^{'-} K^{0})$&$ \frac{1}{30} \left(-3 a_{27}-3 a_{8}+\sqrt{15} c_{8}-\sqrt{10} b_{8}\right)$&$\frac{1}{\sqrt{6}}P_{u}$ \\
\hline
\end{tabular}
\caption{$B_b(\overline{\textbf{3}})\to B(\textbf{10})\, P(\textbf{8})$ decay $\Delta S=0$ transitions}
\label{Tab S0}
\end{table*}
\pagebreak
\twocolumngrid
\section{Amplitude relations}
\label{sec:amp rel}
The complete decay amplitude is given in terms of tree and penguin $SU(3)$-reduced amplitudes and the CKM elements, 
\begin{align}
\label{tp decomp}
\mathcal{A}=&\lambda_{u}^{q}\mathcal{A}_{\text{tree}}+
\lambda_{t}^{q}\mathcal{A}_{\text{penguin}},
\end{align}
where $q=s,\,d$ denote the $\Delta S=-1,\,0$ process. Since the $SU(3)$ operators appear in EW and tree part of the Hamiltonian in a particular combination, the same amplitude relations between $\Delta S=-1$ and $\Delta S=0$ processes are satisfied by the EWP part and the tree part. Therefore, the following decay amplitude relations are obtained; 
\begin{align}
\label{USpin}
&\mathcal{A}(\Lambda_{b}^{0}\to \Sigma^{'+}\pi^{-})=\mathcal{A}(\Xi_{b}^{0}\to 
\Delta^{+}K^{-}),\\
&\mathcal{A}(\Lambda_{b}^{0}\to \Xi^{'0}K^{0})=
	\mathcal{A}(\Xi_{b}^{0}\to \Delta^{0}\overline{K^{0}}),\\
&\mathcal{A}(\Lambda_{b}^{0}\to \Delta^{+}\pi^{-})=
	\mathcal{A}(\Xi_{b}^{0}\to \Sigma^{'+} K^{-}),\\ 
&\mathcal{A}(\Lambda_{b}^{0}\to\Sigma^{'0}K^{0})=
	\mathcal{A}(\Xi^{b}_{0}\to\Sigma^{'0}\overline{K^{0}}),\\
&\mathcal{A}(\Lambda_{b}^{0}\to\Delta^{+}K^{-})=
\mathcal{A}(\Xi_{b}^{0}\to\Sigma^{'+}\pi^{-})\nonumber\\
&\quad=
-\mathcal{A}(\Lambda^{0}_{b}\to\Delta^{0}\overline{K^{0}})=
-\mathcal{A}(\Xi_{b}^{0}\to\Xi^{'0}K^{0}),\\ 
&\mathcal{A}(\Xi_{b}^{-}\to\Sigma^{'0}K^{-})=
-\frac{1}{\sqrt{2}}\mathcal{A}(\Xi_{b}^{-}\to\Xi^{'0}\pi^{-})\nonumber\\
&\quad=
\frac{1}{\sqrt{2}}\mathcal{A}(\Xi_{b}^{-}\to\Delta^{0}K^{-})=
-\mathcal{A}(\Xi_{b}^{-}\to\Sigma^{'0}\pi^{-}),\\
&\mathcal{A}(\Lambda_{b}^{0}\to\Sigma^{'-}\pi^{+})=
\mathcal{A}(\Lambda_{b}^{0}\to\Xi^{'-}K^{+})\nonumber\\
&\quad=\mathcal{A}(\Xi_{b}^{0}\to\Xi^{'-}\pi^{+})=
\frac{1}{\sqrt{3}}\mathcal{A}(\Xi_{b}^{0}\to\Omega^{-}K^{+})\nonumber\\
&\quad=\frac{1}{\sqrt{3}}\mathcal{A}(\Lambda_{b}^{0}\to\Delta^{+}\pi^{-})=
\mathcal{A}(\Lambda_{b}^{0}\to\Sigma^{'-}K^{+})\nonumber\\
&\quad=\mathcal{A}(\Xi_{b}^{0}\to\Sigma^{'-}\pi^{+})=
\mathcal{A}(\Xi_{b}^{0}\to\Xi^{'-}K^{+}),\\[3ex]
&\mathcal{A}(\Xi_{b}^{-}\to\Sigma^{'-}\overline{K^{0}})=
-\sqrt{2}\mathcal{A}(\Xi_{b}^{-}\to\Xi^{'-}\pi^{0})\nonumber\\
&\quad=\sqrt{\frac{2}{3}}\mathcal{A}(\Xi_{b}^{-}\to\Xi^{'-}\eta_{8})=
-\frac{1}{\sqrt{3}}\mathcal{A}(\Xi_{b}^{-}\to\Omega^{-}K^{0})\nonumber\\
&\quad=\frac{1}{\sqrt{3}}\mathcal{A}(\Xi_{b}^{-}\to\Delta^{'-}\overline{K^{0}})=
\sqrt{\frac{2}{3}}\mathcal{A}(\Xi_{b}^{-}\to\Sigma^{'-}\eta_{8})\nonumber\\&
\quad=-\mathcal{A}(\Xi_{b}^{-}\to\Xi^{'-}K^{0})=
-\sqrt{2}\mathcal{A}(\Xi_{b}^{-}\to\Sigma^{'-}\pi^{0}).
\end{align}
\begin{widetext}
The $\Delta S=-1$ triangle relations are given below,
\begin{gather}
2\mathcal{A}(\Lambda_{b}^{0}\to\Sigma^{'0}\pi^{0})+
\mathcal{A}(\Lambda_{b}^{0}\to\Sigma^{'-}\pi^{+})+
\mathcal{A}(\Lambda_{b}^{0}\to\Sigma^{'+}\pi^{-})=0,\\
\mathcal{A}(\Lambda_{b}^{0}\to\Delta^{+}K^{-})+
\mathcal{A}(\Lambda_{b}^{0}\to\Sigma^{'+}\pi^{-})- 
\mathcal{A}(\Xi_{b}^{0}\to\Sigma^{'+}K^{-})=0,\\
\mathcal{A}(\Xi_{b}^{0}\to\Sigma^{'0}\overline{K^{0}})
-\frac{1}{\sqrt{2}}\mathcal{A}(\Lambda_{b}^{0}\to\Delta^{+}K^{-})
-\frac{1}{\sqrt{2}}\mathcal{A}(\Lambda_{b}^{0}\to\Xi^{'0}K^{0})=0,\\
\sqrt{2}\mathcal{A}(\Lambda_{b}^{0}\to\Sigma^{'0}\pi^{0})
-\frac{1}{\sqrt{2}}\mathcal{A}(\Lambda_{b}^{0}\to\Xi^{'0}K^{0})- 
\mathcal{A}(\Xi_{b}^{0}\to\Xi^{'0}\pi^{0})=0,\\
\sqrt{2} \mathcal{A}(\Lambda_{b}^{0} \to \Sigma^{'0}\eta_{8})+ 
\sqrt{\frac{3}{2}} \mathcal{A}(\Lambda_{b}^{0}\to\Xi^{'0}K^{0})- 
\mathcal{A}(\Xi_{b}^{0}\to\Xi^{'0}\eta_{8})=0,\\
\frac{2}{\sqrt{3}} 
\mathcal{A}(\Xi_{b}^{0}\to\Sigma^{'0}\overline{K^{0}})- 		
\mathcal{A}(\Xi_{b}^{0}\to\Xi^{'0}\eta_{8})+
\frac{1}{\sqrt{3}}\mathcal{A}(\Xi_{b}^{0}\to\Xi^{'0}\pi^{0})=0,\\
2\mathcal{A}(\Lambda_{b}^{0}\to\Sigma^{'0}\pi^{0})+
\mathcal{A}(\Lambda_{b}^{0}\to\Sigma^{'+}\pi^{-})+
\mathcal{A}(\Xi_{b}^{0}\to\Xi^{'-}\pi^{+})=0,\\
2\mathcal{A}(\Lambda_{b}^{0}\to\Sigma^{'0}\pi^{0})+
\mathcal{A}(\Lambda_{b}^{0}\to\Sigma^{'+}\pi^{-})+
\mathcal{A}(\Lambda_{b}^{0}\to\Xi^{'-}K^{+})=0,\\
\mathcal{A}(\Lambda_{b}^{0}\to\Delta^{0}\overline{K^{0}})-
\mathcal{A}(\Lambda_{b}^{0}\to\Sigma^{'+}\pi^{-})+
\mathcal{A}(\Xi_{b}^{0}\to\Sigma^{'+}K^{-})=0,\\
\mathcal{A}(\Lambda_{b}^{0}\to\Delta^{+}K^{-})-
\mathcal{A}(\Lambda_{b}^{0}\to\Xi^{'0}K^{0})+
\sqrt{2}\mathcal{A}(\Xi_{b}^{0}\to\Sigma^{'0}\overline{K^{0}})=0,\\
\mathcal{A}(\Lambda_{b}^{0}\to\Sigma^{'0}\pi^{0})+
\frac{1}{2}\mathcal{A}(\Lambda_{b}^{0}\to\Sigma^{+}\pi^{-})+
\frac{1}{2\sqrt{3}}\mathcal{A}(\Xi_{b}^{0}\to\Omega^{-}K^{+})=0.
\end{gather}

The $\Delta S=0$ triangle relations are also obtained,
\begin{gather}
\sqrt{6} \mathcal{A}(\Lambda_{b}^{0}\to\Delta^{0}\pi^{0})+
	\mathcal{A}(\Lambda_{b}^{0}\to\Delta^{-}\pi^{+})+
	\sqrt{3}\mathcal{A}(\Lambda_{b}^{0}\to\Delta^{+}\pi^{-})=0,\\
\sqrt{3} \mathcal{A}(\Lambda_{b}^{0}\to\Delta^{0}\eta_{8})-
	\mathcal{A}(\Lambda_{b}^{0}\to\Delta^{0}\pi^{0})+2 
	\mathcal{A}(\Lambda_{b}^{0}\to\Sigma^{'0}K^{0})=0,\\
\mathcal{A}(\Lambda_{b}^{0}\to\Delta^{+}\pi^{-})-
	\mathcal{A}(\Xi_{b}^{0}\to\Delta^{+}K^{-})+
	\mathcal{A}(\Xi_{b}^{0}\to\Sigma^{'+}\pi^{-})=0,\\
\sqrt{2}\mathcal{A}(\Lambda_{b}^{0}\to\Sigma^{'0}K^{0})-
	\mathcal{A}(\Xi^{b}_{0}\to\Delta^{0}\overline{K^{0}})+
	\mathcal{A}(\Xi_{b}^{0}\to\Sigma^{'+}\pi^{-})=0,\\
\sqrt{2}\mathcal{A}(\Lambda_{b}^{0}\to\Delta^{0}\pi^{0})-
	\mathcal{A}(\Xi_{b}^{0}\to\Delta^{0}\overline{K^{0}})-2 
	\mathcal{A}(\Xi_{b}^{0}\to\Sigma^{'0}\pi^{0})=0,\\
\sqrt{2}\mathcal{A}(\Lambda_{b}^{0}\to\Delta^{0}\eta_{8})+
	\sqrt{3}\mathcal{A}(\Xi_{b}^{0}\to\Delta^{0}\overline{K^{0}})-
	2\mathcal{A}(\Xi_{b}^{0}\to\Sigma^{'0}\eta_{8})=0,\\
\mathcal{A}(\Xi_{b}^{0}\to\Sigma^{'-}\pi^{+})+
	\sqrt{2}\mathcal{A}(\Lambda_{b}^{0}\to\Delta^{0}\pi^{0})+
	\mathcal{A}(\Lambda_{b}^{0}\to\Delta^{+}\pi^{-})=0,\\ 
\mathcal{A}(\Lambda_{b}^{0}\to\Delta^{+}\pi^{-})-
	\mathcal{A}(\Xi_{b}^{0}\to\Delta^{+}K^{-})+ 
	\mathcal{A}(\Xi_{b}^{0}\to\Xi^{'0}K^{0})=0,\\
\sqrt{2}\mathcal{A}(\Lambda_{b}^{0}\to\Sigma^{'0}K^{0})-
	\mathcal{A}(\Xi_{b}^{0}\to\Delta^{0}\overline{K^{0}})-
	\mathcal{A}(\Xi_b^{0}\to\Xi^{'0}K^{0})=0,\\
\mathcal{A}(\Xi_{b}^{0}\to\Xi^{'-}K^{+})+
	\sqrt{2}\mathcal{A}(\Lambda_{b}^{0}\to\Delta^{0}\pi^{0})+
	\mathcal{A}(\Lambda_{b}^{0}\to\Delta^{+}\pi^{-})=0,\\
\sqrt{2}\mathcal{A}(\Lambda_{b}^{0}\to\Delta^{0}\pi^{0})+
	\mathcal{A}(\Lambda_{b}^{0}\to\Delta^{+}\pi^{-})+
	\mathcal{A}(\Lambda_{b}^{0}\to\Sigma^{'-}K^{+})=0
\end{gather}
\end{widetext}
\section{$SU(3)$ breaking effects}
\label{sec:brk SU3}
The $SU(3)$ structure of the 
unbroken Hamiltonian is modified by this term and to the first order in strange 
quark mass, the broken Hamiltonian is made of the following $SU(3)$ 
representations~\cite{Hinchliffe:1995hz,Grinstein:1996us,Pirtskhalava:2011va,Hiller:2012xm,Xu:2013dta,Muller:2015rna,Egolf:2002nk,Grossman:2013lya,Xu:2013dta}, 
\begin{multline}
(\mathbf{3}\oplus \overline{\mathbf{6}} \oplus 
\mathbf{15})\otimes(\mathbf{1}+\epsilon 
\, \mathbf{8})=(\mathbf{3}\oplus \overline{\mathbf{6}} \oplus 
\mathbf{15})\nonumber\\+\epsilon (\mathbf{3}_{i}\oplus 
\overline{\mathbf{6}}_{i} \oplus \mathbf{15}_{1} \oplus \mathbf{15}_{2} \oplus 
\mathbf{15}_{3}^{1} \nonumber\\ \oplus \mathbf{15}_{3}^{2}\oplus 
\mathbf{15^{'}}\oplus \mathbf{24}\oplus \mathbf{42}),
\end{multline} 
where the subscript $i=1,2,3$ indicates the origin of that representation from $\mathbf{3}$ ,$\overline{\mathbf{6}}$, $ \mathbf{15}$ respectively and $\epsilon$ is the $SU(3)$-breaking parameter. $SU(3)$ breaking effects will induce higher $SU(3)$ representations and some of the amplitude relations will cease to hold as there are now thirteen independent $SU(3)$-reduced matrix elements. The isospin relation,
\begin{equation}
\mathcal{A}(\Lambda_{b}^{0}\to\Delta^{0}\overline{K^{0}})=
-\mathcal{A}(\Lambda_{b}^{0}\to\Delta^{+}K^{-})
\end{equation}
and isospin triangle relations, for $\Delta S=-1$,
\begin{align}
\label{iso1}
2\mathcal{A}(\Lambda_{b}^{0}\to\Sigma^{'0}\pi^{0})&+
	\mathcal{A}(\Lambda_{b}^{0}\to\Sigma^{'-}\pi^{+})\nonumber\\
	&\qquad+\mathcal{A}(\Lambda_{b}^{0}\to\Sigma^{'+}\pi^{-})=0,
\end{align}
and for $\Delta S=0$,\\
\begin{align}
\sqrt{6}\mathcal{A}(\Lambda_{b}^{0}\to\Delta^{0}\pi^{0})&+
	\mathcal{A}(\Lambda_{b}^{0}\to\Delta^{-}\pi^{+})\nonumber\\
	&\qquad+\sqrt{3}\mathcal{A}(\Lambda_{b}^{0}\to\Delta^{+}\pi^{-})=0,
\end{align}
continue to hold. In addition, arbitrary $SU(3)$-breaking but isospin 
conserving effects still forbid $\Delta I=2$ and $\Delta I=\frac{5}{2}$ 
transitions which results in general amplitude sum rules,  
\begin{widetext}
	\begin{align}
	\mathcal{A}(\Lambda_{b}^{0}\to\Delta^{0}\overline{K}^{0})&+ 
	\mathcal{A}(\Lambda_{b}^{0}\to\Delta^{+}K^{-})+ 
	2\, \mathcal{A}(\Lambda_{b}^{0}\to\Sigma^{'0}\pi^{0}) + 
	\mathcal{A}(\Lambda_{b}^{0}\to\Sigma^{'-}\pi^{+})+ 
	\mathcal{A}(\Lambda_{b}^{0}\to\Sigma^{'+}\pi^{-})
	\nonumber\\&-\sqrt{2} 
	\mathcal{A}(\Xi_{b}^{0}\to\Sigma^{'0}\overline{K}^{0})-
	\mathcal{A}(\Xi_{b}^{0}\to\Sigma^{'+}K^{-})-\sqrt{2} 
	\mathcal{A}(\Xi_{b}^{0}\to\Xi^{'0}\pi^{0}) 
	-\mathcal{A}(\Xi_{b}^{0}\to\Xi^{'-}\pi^{+})
	\nonumber\\& + \sqrt{2} 
	\mathcal{A}(\Xi_{b}^{-}\to\Sigma^{'0}K^{-})+
	\mathcal{A}(\Xi_{b}^{-}\to\Sigma^{'-}\overline{K}^{0})
	+ \mathcal{A}(\Xi_{b}^{-}\to\Xi^{'0}\pi^{-}) 
	+ \sqrt{2} \mathcal{A}(\Xi_{b}^{-}\to\Xi^{'-}\pi^{0})=0 \\[15pt]
2 \sqrt{3} \mathcal{A}(\Lambda_{b}^{0}\to\Delta^{0}\pi^{0})
&+\sqrt{2} \mathcal{A}(\Lambda_{b}^{0}\to\Delta^{-}\pi^{+})
+\sqrt{6} \mathcal{A}(\Lambda_{b}^{0}\to\Delta^{+}\pi^{-})
-\sqrt{6} \mathcal{A}(\Xi_{b}^{0}\to\Delta^{0}\overline{K}^{0}) 
-\sqrt{6} \mathcal{A}(\Xi_{b}^{0}\to\Delta^{+}K^{-})
\nonumber\\
&-2 \sqrt{6} \mathcal{A}(\Xi_{b}^{0}\to\Sigma^{'0}\pi^{0}) 
-\sqrt{6} \mathcal{A}(\Xi_{b}^{0}\to\Sigma^{'-}\pi^{+}) 
-\sqrt{6} \mathcal{A}(\Xi_{b}^{0}\to\Sigma^{'+}\pi^{-})  
+ \sqrt{6} \mathcal{A}(\Xi_{b}^{-}\to\Delta^{0}K^{-})
\nonumber\\
&+ \sqrt{2}\mathcal{A}(\Xi_{b}^{-}\to\Delta^{-}\overline{K}^{0}) 
+ 2 \sqrt{3} \mathcal{A}(\Xi_{b}^{-}\to\Sigma^{'0}\pi^{-}) 
+ \mathcal{A}(\Xi_{b}^{-}\to\Sigma^{'-}\pi^{0})=0.
%
%
	\end{align} 
\end{widetext}
\section{CP relations}
\label{sec:CP rel}
The general decay amplitude for a spin-$1/2$
$b$-baryon ($\mathcal{B}$) to a spin 0 pseudo-scalar ($\mathcal{P}$) and a 
spin-$3/2$ ($\mathcal{D}$) is given by,
\begin{align}
\mathcal{M}=-iq_{\mu}\overline{u}^{\mu}_{\mathcal{D}}(a+b\gamma_{5})u_{\mathcal{B}},
\end{align}
where $\overline{u}^{\mu}_{\mathcal{D}}$ is the Rarita-Schwinger spinor for the spin-3/2 decuplet baryon, $q_{\mu}$ is the momentum of the pseudo-scalar meson and $u_{\mathcal{B}}$ is the spinor for the initial spin-1/2 $b$-baryon. The two coefficients $a$ and $b$ contain the CKM elements as well as the same flavor structure as $\mathcal{A}_{\text{tree}}$ and $\mathcal{A}_{\text{penguin}}$. The total decay rate for an unpolarized $b$-baryon is given by,
\begin{align}
\Gamma(\mathcal{B}\to \mathcal{D}\,\mathcal{P})&=\frac{\vert 
	\mathbf{p}_{\mathcal{D}}\vert}{8\pi m_{\mathcal{B}}^{2}}\Big(\frac{2}{3}
\frac{m_{B}^{2}}{m_{D}^{2}}2m_{B}(E_{D}+m_{D})\vert 
	\mathbf{p}_{\mathcal{D}}\vert^{2}\Big)\nonumber\\
& \quad\Big(\vert a \vert^{2}+\frac{\vert 
\mathbf{p}_{\mathcal{D}}\vert^{2}}{(E_{D}+m_{D})^{2}}\vert b \vert^{2}\Big)
\end{align}
where $\vert\mathbf{p}_{\mathcal{D}}\vert$ is the magnitude of the 3-momentum 
of the decuplet baryon in the rest frame of the initial $b$-baryon. The decay 
products can be in any one of the two possible relative 
angular momentum states, $l=1$  and $l=2$ identified as $\mathsf{P}$-wave and 
$\mathsf{D}$-wave respectively. It is straightforward to connect the 
$\mathsf{P}$-wave and $\mathsf{D}$-wave amplitude to the $SU(3)$-amplitudes
\begin{align}
\label{eq:Gamma-SP}
\Gamma(\mathcal{B}\to \mathcal{D}\,\mathcal{P})&
=\frac{m_{\mathcal{B}}(E_{D}+m_{D})\vert 
	\mathbf{p}_{\mathcal{D}}\vert^{3}}{6\pi m_{\mathcal{D}}^{2}}\nonumber \\&
	\times\Big( \vert \mathcal{A}^{\mathsf{P}} \vert^{2} +\frac{\vert 
\mathbf{p}_{\mathcal{D}}\vert^{2}}{(E_{D}+m_{D})^{2}}\vert \mathcal{A}^{\mathsf{D}} \vert^{2}\Big)
\end{align}
where $\mathcal{A}^{\mathsf{P}}$ and $\mathcal{A}^{\mathsf{D}}$ are 
\begin{align}
\label{tp0 decomp}
\mathcal{A}^{\mathsf{P}}=a=\,&\lambda_{u}^{q}\mathcal{A}_{\text{tree}}^{\mathsf{P}}+
\lambda_{t}^{q}\mathcal{A}^{\mathsf{P}}_{\text{penguin}},\nonumber\\
\mathcal{A}^{\mathsf{D}}=b=\,&\lambda_{u}^{q}\mathcal{A}_{\text{tree}}^{\mathsf{D}}+
\lambda_{t}^{q}\mathcal{A}^{\mathsf{D}}_{\text{penguin}},
\end{align}


$A_{CP}$ is defined~\cite{Grossman:2013lya} subsequently as,
\begin{align}
A_{CP}=&\frac{\Gamma(\mathcal{B}\to 
	\mathcal{D}\,\mathcal{P})-\Gamma(\overline{\mathcal{B}}
	\to\overline{\mathcal{D}}\,\overline{\mathcal{P}})}
	{\Gamma(\mathcal{B}\to 
	\mathcal{D}\,\mathcal{P})+\Gamma(\overline{\mathcal{B}}
	\to\overline{\mathcal{D}}\,\overline{\mathcal{P}})}\nonumber\\ 			    
	=&\frac{\Delta_{CP}(\mathcal{B}\to 
	\mathcal{D}\,\mathcal{P})}{2\tilde{\Gamma}(\mathcal{B}\to 
	\mathcal{D}\,\mathcal{P})},\label{eq:ACP}
\end{align}
where,
$$\tilde{\Gamma}(\mathcal{B}\to 
	\mathcal{D}\,\mathcal{P})=\frac{1}{2}\big(\Gamma(\mathcal{B}\to 
	\mathcal{D}\,\mathcal{P})+\Gamma(\overline{\mathcal{B}}
	\to\overline{\mathcal{D}}\,\overline{\mathcal{P}})\big).$$ 
$A_{CP}$ is the sum of $CP$ violating contribution from the 
$\delta^{\mathsf{P}}_{CP}$ and $\delta^{\mathsf{D}}_{CP}$ with appropriate 
phase-space factor multiplied:
\begin{align}
A_{CP}(\mathcal{B}\to\mathcal{D}\,\mathcal{P})&=
\frac{\tau_{\mathcal{B}}}{\mathcal{BR}(\mathcal{B}\to
	\mathcal{D}\,\mathcal{P})} 
\frac{m_{\mathcal{B}}(E_{D}+m_{D})\vert 
	\mathbf{p}_{\mathcal{D}}\vert^{3}}{6\pi m_{\mathcal{D}}^{2}}\nonumber 
	\\&\qquad
	\Big( \delta^{\mathsf{P}}_{CP}  +\frac{\vert 
\mathbf{p}_{\mathcal{D}}\vert^{2}}{(E_{D}+m_{D})^{2}}\delta^{\mathsf{D}}_{CP}\Big)
\end{align} 
where $\tau_{\mathcal{B}}$ is the lifetime of the beauty-baryon.

By definition, 
\begin{multline}
\delta_{CP}^{l}(\mathcal{B}\to 
\mathcal{D}\,\mathcal{P})=-\,4\textbf{J}\\
\times\text{Im}\Big[\mathcal{A}^{l*}_{\text{tree}}(\mathcal{B}\to\mathcal{D}\,\mathcal{P})\mathcal{A}^{l}_{\text{penguin}}(\mathcal{B}\to\mathcal{D}\,\mathcal{P})\Big],
\end{multline}
$\textbf{J}$ being the well known Jarlskog invariant.
Based on amplitude relations for the tree and penguin parts obtained in 
Eq.~\eqref{USpin} following $\delta_{CP}^{a}$ 
relations~\cite{Brown:1983wd,Donoghue:1985ww,Dunietz:1992ti} are obtained,
\begin{align}
\label{Delta CP USpin}
&\delta^{l}(\Lambda_{b}^{0}\to \Sigma^{'+}\pi^{-})=
	-\delta^{l}(\Xi_{b}^{0}\to \Delta^{+}K^{-}),\\
&\delta^{l}(\Lambda_{b}^{0}\to \Xi^{'0}K^{0})=
	-\delta^{l}(\Xi_{b}^{0}\to \Delta^{0}\overline{K^{0}}),\\
&\delta^{l}(\Lambda_{b}^{0}\to \Delta^{+}\pi^{-})=
	-\delta^{l}(\Xi_{b}^{0}\to \Sigma^{'+} K^{-}),\\ 
&\delta^{l}(\Lambda_{b}^{0}\to\Sigma^{'0}K^{0})=
	-\delta^{l}(\Xi^{b}_{0}\to\Sigma^{'0}\overline{K^{0}}),\\
&\delta^{l}(\Lambda_{b}^{0}\to\Delta^{+}K^{-})=
	-\delta^{l}(\Xi_{b}^{0}\to\Sigma^{'+}\pi^{-})\nonumber\\
&\quad=-\delta^{l}(\Lambda^{0}_{b}\to\Delta^{0}\overline{K^{0}})=
	-\delta^{l}(\Xi_{b}^{0}\to\Xi^{'0}K^{0}),\\ 
&\delta^{l}(\Xi_{b}^{-}\to\Sigma^{'0}K^{-})=
	-\frac{1}{2}\delta^{l}(\Xi_{b}^{-}\to\Xi^{'0}\pi^{-})\nonumber\\
&\quad=-\frac{1}{2}\delta^{l}(\Xi_{b}^{-}\to\Delta^{0}K^{-})=
	-\delta^{l}(\Xi_{b}^{-}\to\Sigma^{'0}\pi^{-}),\\
&\delta^{l}(\Lambda_{b}^{0}\to\Sigma^{'-}\pi^{+})=
	-\delta^{l}(\Lambda_{b}^{0}\to\Xi^{'-}K^{+})\nonumber\\
&\quad=-\delta^{l}(\Xi_{b}^{0}\to\Xi^{'-}\pi^{+})=
	-\frac{1}{3}\delta^{l}(\Xi_{b}^{0}\to\Omega^{-}K^{+})\nonumber\\
&\quad=-\frac{1}{3}\delta^{l}(\Lambda_{b}^{0}\to\Delta^{+}\pi^{-})=
-\delta^{l}(\Lambda_{b}^{0}\to\Sigma^{'-}K^{+})\nonumber\\
&\quad=-\delta^{l}(\Xi_{b}^{0}\to\Sigma^{'-}\pi^{+})=
-\delta^{l}(\Xi_{b}^{0}\to\Xi^{'-}K^{+}),\\
&\delta^{l}(\Xi_{b}^{-}\to\Sigma^{'-}\overline{K^{0}})=-2 
\delta^{l}(\Xi_{b}^{-}\to\Xi^{'-}\pi^{0})\nonumber\\
&\quad=-\frac{2}{3}\delta^{l}(\Xi_{b}^{-}\to\Xi^{'-}\eta_{8})
=-\frac{1}{3}\delta^{l}(\Xi_{b}^{-}\to\Omega^{-}K^{0})\nonumber\\
&\quad=-\frac{1}{3}\delta^{l}(\Xi_{b}^{-}\to\Delta^{'-}\overline{K^{0}})=
-\frac{2}{3}\delta^{l}(\Xi_{b}^{-}\to\Sigma^{'-}\eta_{8})\nonumber\\
&\quad=-\delta^{l}(\Xi_{b}^{-}\to\Xi^{'-}K^{0})=
-2\delta^{l}(\Xi_{b}^{-}\to\Sigma^{'-}\pi^{0}).
\end{align}
for both $l=\mathsf{P}$ and $l=\mathsf{D}$.
Since, $A_{CP}$ depends on the masses of the initial and final baryons as 
well as the final state meson \cite{Savage:1989qr,Dunietz:1992ti}, some approximation is needed to obtain $A_{CP}$ 
relations between various modes.
Ignoring $\mathbf{p_{\mathcal{D}}}$ and $m_{\mathcal{D}}$ differences between 
such modes, $CP$ violation relations can be 
experimentally verified using the relation, 
\begin{align}
\frac{A_{CP}(\mathcal{B}_{i}\to\,\mathcal{D}_{j}\mathcal{P}_{k})}
{A_{CP}(\mathcal{B}_{l}\to \mathcal{D}_{m}\mathcal{P}_{n})}\!
\simeq-\frac{\tau_{\mathcal{B}_{i}}}
{\tau_{\mathcal{B}_{l}}}
\frac{\mathcal{BR}(\mathcal{B}_{l}\to\mathcal{D}_{m}\mathcal{P}_{n})} 
{\mathcal{BR}(\mathcal{B}_{i}\to\mathcal{D}_{j}\mathcal{P}_{k})},
\end{align}
where $i$, $j$, $k$ and $l$, $m$, $n$ are indices corresponding to the various 
baryons belonging to the above mentioned $\delta_{CP}$ relations. There is a 
further simplification in case $i=l$, resulting in
\begin{align}
\frac{A_{CP}(\mathcal{B}_{i}\to \mathcal{D}_{j}\,\mathcal{P}_{k})}
{A_{CP}(\mathcal{B}_{l} \to \mathcal{D}_{m}\,\mathcal{P}_{n})}
\simeq-
\frac{\mathcal{BR}(\mathcal{B}_{l}\to\mathcal{D}_{m}\mathcal{P}_{n})} 
{\mathcal{BR}(\mathcal{B}_{i}\to\mathcal{D}_{j}\mathcal{D}_{k})},
\end{align}
where the uncertainties due to lifetime measurement also cancel out~\cite{He:2015fwa}. 
The decay asymmetry parameter ($\alpha$) can be measured from an angular distribution study of the final states provided that the subsequent decay of the decuplet baryon is parity violating. The relative strength of the $\mathsf{D}$-wave contribution~\cite{Pakvasa:1990if,Dunietz:1992ti} is extracted from $\alpha$; 
\begin{align}
\alpha=\frac{2\text{Re}(\mathcal{A}^{\mathsf{P}*}\mathcal{A}^{\mathsf{D}})\vert
 \mathbf{p_{\mathcal{D}}} \vert/E_{\mathcal{D}}+m_{\mathcal{D}}}{\vert 
\mathcal{A}^{\mathsf{P}}\vert^{2}+\vert 
\mathcal{A}^{\mathsf{D}}\vert^{2}(\vert \mathbf{p_{\mathcal{D}}} 
\vert/E_{\mathcal{D}}+m_{\mathcal{D}})^{2}}
\end{align}
By systematically taking into account both partial wave contributions, a reliable prediction for $A_{CP}$ relations is possible.
\section{Conclusion}
\label{sec:Conc}

We have explored hadronic anti-triplet 
($\overline{\textbf{3}}$) beauty-baryon into a decuplet baryon ($B$) and an 
octet pseudoscalar meson ($P$), i.e. $B_b(\overline{\textbf{3}})\to 
B(\textbf{10})\,P(\textbf{8})$, based on $SU(3)$ decomposition
of the decay amplitudes in a  general framework. This extends our previous 
analysis~\cite{Roy:2019cky} of the anti-triplet beauty
baryon decays into the octet or singlet of a light baryon and
a pseudoscalar meson and completes the application of the method to decays 
involving any non-charmed baryon. We have shown that in the most general
case, the forty distinct decay modes require forty independent
reduced $SU(3)$ amplitudes to describe all possible $\Delta S=-1$ and $\Delta 
S=0$ processes. The dimension-6 effective Hamiltonian and allowed final 
state $SU(3)$ representations constrain the number of independent $SU(3)$-reduced 
matrix elements to five. An alternative approach in terms of quark diagrams is also provided and compared with the $SU(3)$ decomposition in the limit of exact 
$SU(3)$-flavor symmetry. We explicitly demonstrate a one to one correspondence 
between the quark-diagrams and $SU(3)$-reduced matrix elements. Both the approaches indicate that there exist several amplitude relations between different decay modes. We explicitly derive those sum rules relations between decay amplitudes as well as relations between $CP$ asymmetries. We further probe the $SU(3)$-breaking effects in the decay amplitudes at leading order in the $SU(3)$-breaking parameter and identify those amplitude relations that survive even when the $SU(3)$ flavor symmetry is no longer exact.

\onecolumngrid 
\begin{appendices}
\section{}
\label{sec:App1}
$SU(3)$-decomposition of $\Delta S=-1$ processes for a generic Hamiltonian without any dynamical assumptions:
\begin{equation}
	\centering
\rotatebox{90}{
$
\resizebox{0.9\vsize}{!}{%
$
\left(
\begin{array}{c}
\mathcal{A}(\Lambda_{b}^{0}\to  \Delta^{+}K^{-})\\
\mathcal{A}(\Lambda_{b}^{0}\to  \Delta^{0}\bar{K}^{0})\\
\mathcal{A}(\Lambda_{b}^{0}\to  \Sigma^{'+}\pi^{-})\\
\mathcal{A}(\Lambda_{b}^{0}\to  \Sigma^{'0}\pi^{0})\\
\mathcal{A}(\Lambda_{b}^{0}\to  \Sigma^{'0}\eta_{8})\\
\mathcal{A}(\Lambda_{b}^{0}\to  \Sigma^{'-}\pi^{+})\\
\mathcal{A}(\Lambda_{b}^{0}\to  \Xi^{'0}K^{0})\\
\mathcal{A}(\Lambda_{b}^{0}\to  \Xi^{'-}K^{+})\\
\mathcal{A}(\Xi_{b}^{0}\to  \Sigma^{'+}K^{-})\\
\mathcal{A}(\Xi_{b}^{0}\to  \Sigma^{'0} \bar{K}^{0})\\
\mathcal{A}(\Xi_{b}^{0}\to  \Xi^{'0} \pi^{0})\\
\mathcal{A}(\Xi_{b}^{0}\to  \Xi^{'0}\eta_{8})\\
\mathcal{A}(\Xi_{b}^{0}\to  \Xi^{'-}\pi^{+})\\
\mathcal{A}(\Xi_{b}^{0}\to  \Omega^{-}K^{+})\\
\mathcal{A}(\Xi_{b}^{-}\to  \Sigma^{'0}K^{-})\\
\mathcal{A}(\Xi_{b}^{-}\to  \Sigma^{'-}\bar{K}^{0})\\
\mathcal{A}(\Xi_{b}^{-}\to  \Xi^{'0}\pi^{-})\\
\mathcal{A}(\Xi_{b}^{-}\to  \Xi^{'-} \pi^{0})\\
\mathcal{A}(\Xi_{b}^{-}\to  \Xi^{'-}\eta_{8})\\
\mathcal{A}(\Xi_{b}^{-}\to  \Omega^{-} K^{0})\\
\end{array}
\right)
=\left(
\begin{array}{cccccccccccccccccccc}
\sqrt{\frac{2}{15}} & \sqrt{\frac{2}{15}} & \frac{1}{2 \sqrt{3}} & \frac{1}{2 \sqrt{3}} & \frac{1}{6 \sqrt{5}} & \frac{1}{6 \sqrt{5}} & \frac{1}{2} & \frac{1}{6 \sqrt{2}} & \frac{1}{2 \sqrt{2}} &
   \frac{1}{2 \sqrt{30}} & \frac{\sqrt{\frac{3}{10}}}{2} & \frac{\sqrt{\frac{3}{35}}}{2} & \frac{1}{2 \sqrt{5}} & \frac{1}{2 \sqrt{21}} & 0 & 0 & 0 & 0 & 0 & 0 \\
 -\sqrt{\frac{2}{15}} & -\sqrt{\frac{2}{15}} & -\frac{1}{2 \sqrt{3}} & -\frac{1}{2 \sqrt{3}} & -\frac{1}{6 \sqrt{5}} & -\frac{1}{6 \sqrt{5}} & \frac{1}{2} & -\frac{1}{6 \sqrt{2}} & \frac{1}{2 \sqrt{2}} &
   -\frac{1}{2 \sqrt{30}} & \frac{\sqrt{\frac{3}{10}}}{2} & -\frac{\sqrt{\frac{3}{35}}}{2} & \frac{1}{2 \sqrt{5}} & -\frac{1}{2 \sqrt{21}} & 0 & 0 & 0 & 0 & 0 & 0 \\
 -\frac{1}{\sqrt{30}} & -\frac{1}{\sqrt{30}} & \frac{1}{2 \sqrt{3}} & \frac{1}{2 \sqrt{3}} & -\frac{1}{2 \sqrt{5}} & -\frac{1}{2 \sqrt{5}} & -\frac{1}{6} & -\frac{1}{2 \sqrt{2}} & -\frac{1}{6 \sqrt{2}} &
   \frac{1}{2 \sqrt{30}} & \frac{\sqrt{\frac{3}{10}}}{2} & \frac{\sqrt{\frac{3}{35}}}{2} & \frac{1}{2 \sqrt{5}} & \frac{1}{2 \sqrt{21}} & -\frac{1}{2 \sqrt{5}} & -\frac{\sqrt{\frac{3}{5}}}{2} &
   -\frac{\sqrt{\frac{2}{5}}}{3} & -\frac{2}{3 \sqrt{5}} & 0 & 0 \\
 0 & 0 & 0 & 0 & 0 & 0 & -\frac{1}{3} & 0 & -\frac{1}{3 \sqrt{2}} & 0 & \sqrt{\frac{3}{10}} & 0 & \frac{1}{\sqrt{5}} & 0 & \frac{1}{2 \sqrt{5}} & \frac{\sqrt{\frac{3}{5}}}{2} & \frac{\sqrt{\frac{2}{5}}}{3}
   & \frac{2}{3 \sqrt{5}} & 0 & 0 \\
 -\frac{1}{\sqrt{10}} & -\frac{1}{\sqrt{10}} & 0 & 0 & \frac{1}{\sqrt{15}} & \frac{1}{\sqrt{15}} & 0 & \frac{1}{\sqrt{6}} & 0 & \frac{1}{\sqrt{10}} & 0 & \frac{3}{\sqrt{35}} & 0 & \frac{1}{\sqrt{7}} & 0 &
   0 & 0 & 0 & 0 & 0 \\
 \frac{1}{\sqrt{30}} & \frac{1}{\sqrt{30}} & -\frac{1}{2 \sqrt{3}} & -\frac{1}{2 \sqrt{3}} & \frac{1}{2 \sqrt{5}} & \frac{1}{2 \sqrt{5}} & -\frac{1}{6} & \frac{1}{2 \sqrt{2}} & -\frac{1}{6 \sqrt{2}} &
   -\frac{1}{2 \sqrt{30}} & \frac{\sqrt{\frac{3}{10}}}{2} & -\frac{\sqrt{\frac{3}{35}}}{2} & \frac{1}{2 \sqrt{5}} & -\frac{1}{2 \sqrt{21}} & -\frac{1}{2 \sqrt{5}} & -\frac{\sqrt{\frac{3}{5}}}{2} &
   -\frac{\sqrt{\frac{2}{5}}}{3} & -\frac{2}{3 \sqrt{5}} & 0 & 0 \\
 \frac{1}{\sqrt{30}} & \frac{1}{\sqrt{30}} & -\frac{1}{2 \sqrt{3}} & -\frac{1}{2 \sqrt{3}} & -\frac{1}{3 \sqrt{5}} & -\frac{1}{3 \sqrt{5}} & 0 & -\frac{1}{3 \sqrt{2}} & 0 & \frac{1}{\sqrt{30}} & 0 &
   \sqrt{\frac{3}{35}} & 0 & \frac{1}{\sqrt{21}} & \frac{1}{2 \sqrt{5}} & \frac{\sqrt{\frac{3}{5}}}{2} & -\frac{1}{\sqrt{10}} & -\frac{1}{\sqrt{5}} & 0 & 0 \\
 \frac{1}{\sqrt{30}} & \frac{1}{\sqrt{30}} & -\frac{1}{2 \sqrt{3}} & -\frac{1}{2 \sqrt{3}} & -\frac{1}{3 \sqrt{5}} & -\frac{1}{3 \sqrt{5}} & 0 & -\frac{1}{3 \sqrt{2}} & 0 & \frac{1}{\sqrt{30}} & 0 &
   \sqrt{\frac{3}{35}} & 0 & \frac{1}{\sqrt{21}} & -\frac{1}{2 \sqrt{5}} & -\frac{\sqrt{\frac{3}{5}}}{2} & \frac{1}{\sqrt{10}} & \frac{1}{\sqrt{5}} & 0 & 0 \\
 -\frac{1}{\sqrt{30}} & \frac{1}{\sqrt{30}} & \frac{1}{2 \sqrt{3}} & -\frac{1}{2 \sqrt{3}} & -\frac{1}{2 \sqrt{5}} & \frac{1}{2 \sqrt{5}} & \frac{1}{6} & 0 & -\frac{1}{3 \sqrt{2}} & \sqrt{\frac{2}{15}} &
   \frac{1}{\sqrt{30}} & -\frac{\sqrt{\frac{3}{35}}}{2} & -\frac{1}{2 \sqrt{5}} & -\frac{1}{2 \sqrt{21}} & -\frac{1}{2 \sqrt{5}} & \frac{1}{2 \sqrt{15}} & -\frac{1}{3 \sqrt{10}} & \frac{1}{6 \sqrt{5}} &
   \frac{1}{\sqrt{6}} & \frac{1}{2 \sqrt{3}} \\
 \frac{1}{2 \sqrt{15}} & -\frac{1}{2 \sqrt{15}} & -\frac{1}{2 \sqrt{6}} & \frac{1}{2 \sqrt{6}} & -\frac{7}{6 \sqrt{10}} & \frac{1}{3 \sqrt{10}} & \frac{1}{3 \sqrt{2}} & \frac{1}{6} & -\frac{1}{3} &
   \frac{1}{2 \sqrt{15}} & \frac{1}{\sqrt{15}} & \sqrt{\frac{3}{70}} & -\frac{1}{\sqrt{10}} & -\frac{5}{2 \sqrt{42}} & \frac{1}{2 \sqrt{10}} & -\frac{1}{2 \sqrt{30}} & \frac{1}{6 \sqrt{5}} & -\frac{1}{6
   \sqrt{10}} & -\frac{1}{2 \sqrt{3}} & -\frac{1}{2 \sqrt{6}} \\
 \frac{1}{2 \sqrt{15}} & -\frac{1}{2 \sqrt{15}} & \frac{1}{4 \sqrt{6}} & -\frac{1}{4 \sqrt{6}} & \frac{11}{12 \sqrt{10}} & -\frac{2 \sqrt{\frac{2}{5}}}{3} & -\frac{1}{3 \sqrt{2}} & \frac{1}{12} &
   \frac{1}{3} & \frac{\sqrt{\frac{5}{3}}}{4} & \frac{1}{\sqrt{15}} & 0 & -\frac{1}{\sqrt{10}} & -\frac{\sqrt{\frac{7}{6}}}{4} & \frac{1}{2 \sqrt{10}} & -\frac{1}{2 \sqrt{30}} & \frac{7}{12 \sqrt{5}} &
   -\frac{7}{12 \sqrt{10}} & \frac{1}{4 \sqrt{3}} & \frac{1}{4 \sqrt{6}} \\
 \frac{1}{2 \sqrt{5}} & -\frac{1}{2 \sqrt{5}} & -\frac{1}{4 \sqrt{2}} & \frac{1}{4 \sqrt{2}} & \frac{1}{4 \sqrt{30}} & \frac{1}{\sqrt{30}} & 0 & -\frac{1}{4 \sqrt{3}} & 0 & \frac{3}{4 \sqrt{5}} & 0 &
   -\frac{3}{\sqrt{70}} & 0 & \frac{3}{4 \sqrt{14}} & \frac{\sqrt{\frac{3}{10}}}{2} & -\frac{1}{2 \sqrt{10}} & -\frac{\sqrt{\frac{3}{5}}}{4} & \frac{\sqrt{\frac{3}{10}}}{4} & -\frac{1}{4} & \frac{3}{4
   \sqrt{2}} \\
 -\frac{1}{\sqrt{30}} & \frac{1}{\sqrt{30}} & -\frac{1}{4 \sqrt{3}} & \frac{1}{4 \sqrt{3}} & \frac{3}{4 \sqrt{5}} & \frac{1}{2 \sqrt{5}} & -\frac{1}{6} & -\frac{1}{2 \sqrt{2}} & \frac{1}{3 \sqrt{2}} &
   \frac{1}{2 \sqrt{30}} & \frac{1}{\sqrt{30}} & \frac{\sqrt{\frac{3}{35}}}{2} & -\frac{1}{2 \sqrt{5}} & -\frac{5}{4 \sqrt{21}} & -\frac{1}{2 \sqrt{5}} & \frac{1}{2 \sqrt{15}} & -\frac{7}{6 \sqrt{10}} &
   \frac{7}{12 \sqrt{5}} & -\frac{1}{2 \sqrt{6}} & -\frac{1}{4 \sqrt{3}} \\
 -\frac{1}{\sqrt{10}} & \frac{1}{\sqrt{10}} & -\frac{1}{4} & \frac{1}{4} & -\frac{1}{4 \sqrt{15}} & -\frac{1}{\sqrt{15}} & 0 & \frac{1}{2 \sqrt{6}} & 0 & \frac{1}{2 \sqrt{10}} & 0 & -\frac{1}{\sqrt{35}} &
   0 & \frac{1}{4 \sqrt{7}} & -\frac{\sqrt{\frac{3}{5}}}{2} & \frac{1}{2 \sqrt{5}} & \frac{\sqrt{\frac{3}{10}}}{2} & -\frac{\sqrt{\frac{3}{5}}}{4} & -\frac{1}{2 \sqrt{2}} & \frac{1}{4} \\
 \frac{1}{2 \sqrt{15}} & -\frac{1}{2 \sqrt{15}} & -\frac{1}{2 \sqrt{6}} & \frac{1}{2 \sqrt{6}} & -\frac{7}{6 \sqrt{10}} & \frac{1}{3 \sqrt{10}} & -\frac{1}{3 \sqrt{2}} & \frac{1}{6} & \frac{1}{3} &
   \frac{1}{2 \sqrt{15}} & -\frac{1}{\sqrt{15}} & \sqrt{\frac{3}{70}} & \frac{1}{\sqrt{10}} & -\frac{5}{2 \sqrt{42}} & -\frac{1}{2 \sqrt{10}} & \frac{1}{2 \sqrt{30}} & -\frac{1}{6 \sqrt{5}} & \frac{1}{6
   \sqrt{10}} & \frac{1}{2 \sqrt{3}} & \frac{1}{2 \sqrt{6}} \\
 -\frac{1}{\sqrt{30}} & \frac{1}{\sqrt{30}} & \frac{1}{2 \sqrt{3}} & -\frac{1}{2 \sqrt{3}} & -\frac{1}{2 \sqrt{5}} & \frac{1}{2 \sqrt{5}} & -\frac{1}{6} & 0 & \frac{1}{3 \sqrt{2}} & \sqrt{\frac{2}{15}} &
   -\frac{1}{\sqrt{30}} & -\frac{\sqrt{\frac{3}{35}}}{2} & \frac{1}{2 \sqrt{5}} & -\frac{1}{2 \sqrt{21}} & \frac{1}{2 \sqrt{5}} & -\frac{1}{2 \sqrt{15}} & \frac{1}{3 \sqrt{10}} & -\frac{1}{6 \sqrt{5}} &
   -\frac{1}{\sqrt{6}} & -\frac{1}{2 \sqrt{3}} \\
 -\frac{1}{\sqrt{30}} & \frac{1}{\sqrt{30}} & -\frac{1}{4 \sqrt{3}} & \frac{1}{4 \sqrt{3}} & \frac{3}{4 \sqrt{5}} & \frac{1}{2 \sqrt{5}} & \frac{1}{6} & -\frac{1}{2 \sqrt{2}} & -\frac{1}{3 \sqrt{2}} &
   \frac{1}{2 \sqrt{30}} & -\frac{1}{\sqrt{30}} & \frac{\sqrt{\frac{3}{35}}}{2} & \frac{1}{2 \sqrt{5}} & -\frac{5}{4 \sqrt{21}} & \frac{1}{2 \sqrt{5}} & -\frac{1}{2 \sqrt{15}} & \frac{7}{6 \sqrt{10}} &
   -\frac{7}{12 \sqrt{5}} & \frac{1}{2 \sqrt{6}} & \frac{1}{4 \sqrt{3}} \\
 \frac{1}{2 \sqrt{15}} & -\frac{1}{2 \sqrt{15}} & \frac{1}{4 \sqrt{6}} & -\frac{1}{4 \sqrt{6}} & \frac{11}{12 \sqrt{10}} & -\frac{2 \sqrt{\frac{2}{5}}}{3} & \frac{1}{3 \sqrt{2}} & \frac{1}{12} &
   -\frac{1}{3} & \frac{\sqrt{\frac{5}{3}}}{4} & -\frac{1}{\sqrt{15}} & 0 & \frac{1}{\sqrt{10}} & -\frac{\sqrt{\frac{7}{6}}}{4} & -\frac{1}{2 \sqrt{10}} & \frac{1}{2 \sqrt{30}} & -\frac{7}{12 \sqrt{5}} &
   \frac{7}{12 \sqrt{10}} & -\frac{1}{4 \sqrt{3}} & -\frac{1}{4 \sqrt{6}} \\
 -\frac{1}{2 \sqrt{5}} & \frac{1}{2 \sqrt{5}} & \frac{1}{4 \sqrt{2}} & -\frac{1}{4 \sqrt{2}} & -\frac{1}{4 \sqrt{30}} & -\frac{1}{\sqrt{30}} & 0 & \frac{1}{4 \sqrt{3}} & 0 & -\frac{3}{4 \sqrt{5}} & 0 &
   \frac{3}{\sqrt{70}} & 0 & -\frac{3}{4 \sqrt{14}} & \frac{\sqrt{\frac{3}{10}}}{2} & -\frac{1}{2 \sqrt{10}} & -\frac{\sqrt{\frac{3}{5}}}{4} & \frac{\sqrt{\frac{3}{10}}}{4} & -\frac{1}{4} & \frac{3}{4
   \sqrt{2}} \\
 \frac{1}{\sqrt{10}} & -\frac{1}{\sqrt{10}} & \frac{1}{4} & -\frac{1}{4} & \frac{1}{4 \sqrt{15}} & \frac{1}{\sqrt{15}} & 0 & -\frac{1}{2 \sqrt{6}} & 0 & -\frac{1}{2 \sqrt{10}} & 0 & \frac{1}{\sqrt{35}} & 0
   & -\frac{1}{4 \sqrt{7}} & -\frac{\sqrt{\frac{3}{5}}}{2} & \frac{1}{2 \sqrt{5}} & \frac{\sqrt{\frac{3}{10}}}{2} & -\frac{\sqrt{\frac{3}{5}}}{4} & -\frac{1}{2 \sqrt{2}} & \frac{1}{4} \\
\end{array}
\right)
\left(
\begin{array}{c}
\langle 8 \parallel \mathbf{15_{1}} \parallel \bar{3}\\
\langle 8 \parallel \mathbf{6}_{\mathbf{1}} \parallel \bar{3}\\
\langle 10 \parallel \mathbf{15_{1}} \parallel \bar{3}\\
\langle 10 \parallel \mathbf{15^{'}_{1}} \parallel \bar{3}\\
\langle 27 \parallel \mathbf{15_{1}} \parallel \bar{3}\\
\langle 27 \parallel \mathbf{24_{1}} \parallel \bar{3}\\
\langle 27 \parallel \mathbf{24_{2}} \parallel \bar{3}\\
\langle 27 \parallel \mathbf{42_{1}} \parallel \bar{3}\\
\langle 27 \parallel \mathbf{42_{2}} \parallel \bar{3}\\
\langle 35 \parallel \mathbf{42_{1}} \parallel \bar{3}\\
\langle 35 \parallel \mathbf{42_{2}} \parallel \bar{3}\\
\langle 35 \parallel \mathbf{48_{1}} \parallel \bar{3}\\
\langle 35 \parallel \mathbf{48_{2}} \parallel \bar{3}\\
\langle 35 \parallel \mathbf{15^{'}_{1}} \parallel \bar{3}\\
\langle 8 \parallel \mathbf{3_{0}} \parallel \bar{3}\\
\langle 8 \parallel \mathbf{15_{0}} \parallel \bar{3}\\
\langle 27 \parallel \mathbf{15_{0}} \parallel \bar{3}\\
\langle 27 \parallel \mathbf{42_{0}} \parallel \bar{3}\\
\langle 10 \parallel \mathbf{15_{0}} \parallel \bar{3}\\
\langle 35 \parallel \mathbf{42_{0}} \parallel \bar{3}\\
\end{array}
\right)
$
	}
$
	}
\end{equation}
$SU(3)$-decomposition of $\Delta S=0$ processes for a generic Hamiltonian without any dynamical assumptions:
\begin{equation}
	\centering
\rotatebox{90}{
$
\resizebox{0.9\vsize}{!}{%
$
\left(
\begin{array}{c}
\mathcal{A}(\Lambda_{b}^{0}\to  \Delta^{+} \pi^{-})\\
\mathcal{A}(\Lambda_{b}^{0}\to  \Delta^{0} \pi^{0})\\
\mathcal{A}(\Lambda_{b}^{0}\to  \Delta^{0} \eta_{8})\\
\mathcal{A}(\Lambda_{b}^{0}\to  \Delta^{-}\pi^{+})\\
\mathcal{A}(\Lambda_{b}^{0}\to  \Sigma^{'0}K^{0})\\
\mathcal{A}(\Lambda_{b}^{0}\to  \Sigma^{'-}K^{+})\\
\mathcal{A}(\Xi_{b}^{0}\to \Delta^{+} K^{-})\\
\mathcal{A}(\Xi_{b}^{0}\to  \Delta^{0} \overline{K}^{0})\\
\mathcal{A}(\Xi_{b}^{0}\to  \Sigma^{'+}\pi^{-})\\
\mathcal{A}(\Xi_{b}^{0}\to  \Sigma^{'0}\pi^{0})\\
\mathcal{A}(\Xi_{b}^{0}\to  \Sigma^{'0} \eta_{8})\\
\mathcal{A}(\Xi_{b}^{0}\to  \Sigma^{'-}\pi^{+})\\
\mathcal{A}(\Xi_{b}^{0}\to \Xi^{'0} K^{0})\\
\mathcal{A}(\Xi_{b}^{0}\to  \Xi^{'-}K^{+})\\
\mathcal{A}(\Xi_{b}^{-}\to  \Delta^{'0}K^{-})\\
\mathcal{A}(\Xi_{b}^{-}\to \Delta^{'-}\overline{K}^{0})\\
\mathcal{A}(\Xi_{b}^{-}\to \Sigma^{'0} \pi^{-})\\
\mathcal{A}(\Xi_{b}^{-}\to \Sigma^{'-} \pi^{0})\\
\mathcal{A}(\Xi_{b}^{-}\to \Sigma^{'-} \eta_{8})\\
\mathcal{A}(\Xi_{b}^{-}\to  \Sigma^{'-} K^{0})\\
\end{array}
\right)
=\left(
\begin{array}{cccccccccccccccccccc}
-\frac{1}{2 \sqrt{5}} & -\frac{1}{2 \sqrt{5}} & -\frac{1}{\sqrt{30}} & \frac{1}{2} & \frac{1}{2 \sqrt{3}} & -\frac{\sqrt{\frac{2}{15}}}{3} & -\frac{\sqrt{\frac{5}{3}}}{6} & -\frac{1}{3 \sqrt{30}} &
   -\frac{\sqrt{\frac{2}{3}}}{3} & -\frac{\sqrt{\frac{2}{15}}}{3} & -\frac{\sqrt{\frac{5}{3}}}{6} & \frac{1}{10} & \frac{\sqrt{6}}{5} & \frac{\sqrt{\frac{2}{7}}}{5} & \frac{\sqrt{\frac{3}{2}}}{5} &
   \frac{1}{2 \sqrt{21}} & 0 & 0 & 0 & 0 \\
 \frac{1}{\sqrt{10}} & \frac{1}{\sqrt{10}} & \frac{1}{\sqrt{15}} & -\frac{1}{4 \sqrt{2}} & -\frac{1}{4 \sqrt{6}} & \frac{2}{3 \sqrt{15}} & \frac{\sqrt{\frac{5}{6}}}{12} & \frac{1}{3 \sqrt{15}} & \frac{1}{6
   \sqrt{3}} & \frac{2}{3 \sqrt{15}} & \frac{\sqrt{\frac{5}{6}}}{12} & -\frac{1}{20 \sqrt{2}} & \frac{2 \sqrt{3}}{5} & -\frac{1}{10 \sqrt{7}} & \frac{\sqrt{3}}{5} & -\frac{1}{4 \sqrt{42}} & 0 & 0 & 0 & 0
   \\
 0 & 0 & 0 & \frac{\sqrt{\frac{3}{2}}}{4} & \frac{1}{4 \sqrt{2}} & 0 & \frac{\sqrt{\frac{5}{2}}}{4} & 0 & \frac{1}{2} & 0 & \frac{\sqrt{\frac{5}{2}}}{4} & \frac{\sqrt{\frac{3}{2}}}{4} & 0 &
   \frac{\sqrt{\frac{3}{7}}}{2} & 0 & \frac{5}{4 \sqrt{14}} & 0 & 0 & 0 & 0 \\
 -\frac{\sqrt{\frac{3}{5}}}{2} & -\frac{\sqrt{\frac{3}{5}}}{2} & -\frac{1}{\sqrt{10}} & -\frac{\sqrt{3}}{4} & -\frac{1}{4} & -\frac{\sqrt{\frac{2}{5}}}{3} & \frac{\sqrt{5}}{12} & -\frac{1}{3 \sqrt{10}} &
   \frac{1}{3 \sqrt{2}} & -\frac{\sqrt{\frac{2}{5}}}{3} & \frac{\sqrt{5}}{12} & -\frac{\sqrt{3}}{20} & \frac{\sqrt{2}}{5} & -\frac{\sqrt{\frac{3}{14}}}{5} & \frac{1}{5 \sqrt{2}} & -\frac{1}{4 \sqrt{7}} & 0
   & 0 & 0 & 0 \\
 \frac{1}{2 \sqrt{10}} & \frac{1}{2 \sqrt{10}} & \frac{1}{2 \sqrt{15}} & -\frac{1}{2 \sqrt{2}} & -\frac{1}{2 \sqrt{6}} & -\frac{4}{3 \sqrt{15}} & -\frac{\sqrt{\frac{5}{6}}}{6} & -\frac{2}{3 \sqrt{15}} &
   -\frac{1}{3 \sqrt{3}} & -\frac{4}{3 \sqrt{15}} & -\frac{\sqrt{\frac{5}{6}}}{6} & \frac{1}{2 \sqrt{2}} & 0 & \frac{1}{\sqrt{7}} & 0 & \frac{5}{2 \sqrt{42}} & 0 & 0 & 0 & 0 \\
 -\frac{1}{2 \sqrt{5}} & -\frac{1}{2 \sqrt{5}} & -\frac{1}{\sqrt{30}} & -\frac{1}{4} & -\frac{1}{4 \sqrt{3}} & \frac{4 \sqrt{\frac{2}{15}}}{3} & -\frac{\sqrt{\frac{5}{3}}}{12} & \frac{2
   \sqrt{\frac{2}{15}}}{3} & -\frac{1}{3 \sqrt{6}} & \frac{4 \sqrt{\frac{2}{15}}}{3} & -\frac{\sqrt{\frac{5}{3}}}{12} & \frac{1}{4} & 0 & \frac{1}{\sqrt{14}} & 0 & \frac{5}{4 \sqrt{21}} & 0 & 0 & 0 & 0 \\
 -\frac{1}{2 \sqrt{5}} & \frac{1}{6 \sqrt{5}} & \frac{1}{\sqrt{30}} & \frac{1}{6} & -\frac{1}{2 \sqrt{3}} & -\frac{1}{3 \sqrt{30}} & -\frac{7}{6 \sqrt{15}} & \frac{1}{3 \sqrt{30}} &
   \frac{\sqrt{\frac{2}{3}}}{3} & \frac{1}{6 \sqrt{30}} & -\frac{1}{6 \sqrt{15}} & \frac{7}{30} & \frac{\sqrt{\frac{3}{2}}}{10} & -\frac{\sqrt{\frac{2}{7}}}{5} & -\frac{\sqrt{\frac{3}{2}}}{5} & -\frac{1}{2
   \sqrt{21}} & -\frac{2 \sqrt{\frac{2}{5}}}{3} & \frac{1}{3 \sqrt{2}} & -\frac{3}{2 \sqrt{10}} & \frac{1}{6 \sqrt{2}} \\
 \frac{1}{2 \sqrt{5}} & -\frac{1}{6 \sqrt{5}} & -\frac{1}{\sqrt{30}} & -\frac{1}{6} & \frac{1}{2 \sqrt{3}} & \frac{1}{3 \sqrt{30}} & -\frac{4}{3 \sqrt{15}} & -\frac{1}{3 \sqrt{30}} & \frac{1}{3 \sqrt{6}} &
   -\frac{1}{6 \sqrt{30}} & \frac{2}{3 \sqrt{15}} & \frac{1}{15} & \frac{\sqrt{\frac{3}{2}}}{10} & \frac{3}{5 \sqrt{14}} & -\frac{\sqrt{\frac{3}{2}}}{5} & -\frac{1}{\sqrt{21}} & \frac{2
   \sqrt{\frac{2}{5}}}{3} & -\frac{1}{3 \sqrt{2}} & -\frac{3}{2 \sqrt{10}} & -\frac{1}{6 \sqrt{2}} \\
 0 & \frac{2}{3 \sqrt{5}} & -\sqrt{\frac{2}{15}} & \frac{1}{6} & -\frac{1}{2 \sqrt{3}} & \frac{1}{3 \sqrt{30}} & \frac{1}{6 \sqrt{15}} & -\frac{7}{3 \sqrt{30}} & -\frac{\sqrt{\frac{2}{3}}}{3} &
   \frac{\sqrt{\frac{5}{6}}}{6} & \frac{7}{6 \sqrt{15}} & \frac{7}{30} & \frac{\sqrt{\frac{3}{2}}}{10} & -\frac{\sqrt{\frac{2}{7}}}{5} & -\frac{\sqrt{\frac{3}{2}}}{5} & -\frac{1}{2 \sqrt{21}} &
   \frac{\sqrt{\frac{2}{5}}}{3} & \frac{1}{3 \sqrt{2}} & \frac{1}{2 \sqrt{10}} & \frac{1}{6 \sqrt{2}} \\
 \frac{1}{4 \sqrt{5}} & -\frac{3}{4 \sqrt{5}} & \frac{\sqrt{\frac{3}{10}}}{2} & 0 & 0 & \frac{\sqrt{\frac{2}{15}}}{3} & \frac{\sqrt{\frac{5}{3}}}{6} & \frac{2 \sqrt{\frac{2}{15}}}{3} & -\frac{1}{3
   \sqrt{6}} & -\frac{2 \sqrt{\frac{2}{15}}}{3} & -\frac{1}{6 \sqrt{15}} & \frac{3}{10} & \frac{\sqrt{\frac{3}{2}}}{5} & \frac{1}{5 \sqrt{14}} & -\frac{\sqrt{6}}{5} & -\frac{\sqrt{\frac{3}{7}}}{2} & 0 & 0
   & \frac{1}{\sqrt{10}} & 0 \\
 \frac{\sqrt{\frac{3}{5}}}{4} & -\frac{1}{4 \sqrt{15}} & -\frac{1}{2 \sqrt{10}} & 0 & 0 & -\frac{\sqrt{\frac{2}{5}}}{3} & \frac{1}{6 \sqrt{5}} & \frac{\sqrt{\frac{2}{5}}}{3} & \frac{1}{3 \sqrt{2}} &
   \frac{1}{3 \sqrt{10}} & -\frac{\sqrt{5}}{6} & \frac{1}{2 \sqrt{3}} & 0 & -\sqrt{\frac{3}{14}} & 0 & \frac{1}{2 \sqrt{7}} & \sqrt{\frac{2}{15}} & 0 & 0 & \frac{1}{\sqrt{6}} \\
 -\frac{1}{2 \sqrt{5}} & \frac{\sqrt{5}}{6} & -\frac{1}{\sqrt{30}} & -\frac{1}{6} & \frac{1}{2 \sqrt{3}} & -\frac{\sqrt{\frac{5}{6}}}{3} & \frac{2}{3 \sqrt{15}} & -\frac{1}{3 \sqrt{30}} & \frac{1}{3
   \sqrt{6}} & \frac{11}{6 \sqrt{30}} & -\frac{4}{3 \sqrt{15}} & \frac{1}{15} & \frac{\sqrt{\frac{3}{2}}}{10} & \frac{3}{5 \sqrt{14}} & -\frac{\sqrt{\frac{3}{2}}}{5} & -\frac{1}{\sqrt{21}} &
   -\frac{\sqrt{\frac{2}{5}}}{3} & -\frac{1}{3 \sqrt{2}} & \frac{1}{2 \sqrt{10}} & -\frac{1}{6 \sqrt{2}} \\
 0 & -\frac{2}{3 \sqrt{5}} & \sqrt{\frac{2}{15}} & -\frac{1}{6} & \frac{1}{2 \sqrt{3}} & -\frac{1}{3 \sqrt{30}} & -\frac{1}{6 \sqrt{15}} & -\frac{4 \sqrt{\frac{2}{15}}}{3} & -\frac{1}{3 \sqrt{6}} &
   \frac{\sqrt{\frac{5}{6}}}{3} & \frac{\sqrt{\frac{5}{3}}}{6} & \frac{1}{6} & 0 & -\frac{1}{\sqrt{14}} & 0 & \frac{1}{2 \sqrt{21}} & -\frac{\sqrt{\frac{2}{5}}}{3} & -\frac{1}{3 \sqrt{2}} & 0 & \frac{1}{3
   \sqrt{2}} \\
 -\frac{1}{2 \sqrt{5}} & \frac{\sqrt{5}}{6} & -\frac{1}{\sqrt{30}} & -\frac{1}{6} & \frac{1}{2 \sqrt{3}} & \frac{\sqrt{\frac{5}{6}}}{3} & -\frac{1}{6 \sqrt{15}} & \frac{2 \sqrt{\frac{2}{15}}}{3} &
   -\frac{1}{3 \sqrt{6}} & -\frac{7}{3 \sqrt{30}} & \frac{\sqrt{\frac{5}{3}}}{6} & \frac{1}{6} & 0 & -\frac{1}{\sqrt{14}} & 0 & \frac{1}{2 \sqrt{21}} & -\frac{\sqrt{\frac{2}{5}}}{3} & -\frac{1}{3 \sqrt{2}}
   & 0 & \frac{1}{3 \sqrt{2}} \\
 -\frac{1}{2 \sqrt{5}} & \frac{1}{6 \sqrt{5}} & \frac{1}{\sqrt{30}} & -\frac{1}{12} & \frac{1}{4 \sqrt{3}} & -\frac{1}{3 \sqrt{30}} & -\frac{23}{12 \sqrt{15}} & \frac{1}{3 \sqrt{30}} &
   \frac{\sqrt{\frac{2}{3}}}{3} & \frac{1}{6 \sqrt{30}} & \frac{7}{12 \sqrt{15}} & \frac{11}{60} & -\frac{\sqrt{\frac{3}{2}}}{10} & \frac{\sqrt{\frac{2}{7}}}{5} & \frac{\sqrt{\frac{3}{2}}}{5} & -\frac{5}{4
   \sqrt{21}} & \frac{\sqrt{\frac{2}{5}}}{3} & \frac{1}{3 \sqrt{2}} & \frac{3}{2 \sqrt{10}} & \frac{1}{6 \sqrt{2}} \\
 \frac{\sqrt{\frac{3}{5}}}{2} & -\frac{1}{2 \sqrt{15}} & -\frac{1}{\sqrt{10}} & \frac{1}{4 \sqrt{3}} & -\frac{1}{4} & \frac{1}{3 \sqrt{10}} & -\frac{7}{12 \sqrt{5}} & -\frac{1}{3 \sqrt{10}} & \frac{1}{3
   \sqrt{2}} & -\frac{1}{6 \sqrt{10}} & -\frac{1}{12 \sqrt{5}} & \frac{7}{20 \sqrt{3}} & -\frac{1}{10 \sqrt{2}} & -\frac{\sqrt{\frac{3}{14}}}{5} & \frac{1}{5 \sqrt{2}} & -\frac{1}{4 \sqrt{7}} &
   -\sqrt{\frac{2}{15}} & -\frac{1}{\sqrt{6}} & \frac{\sqrt{\frac{3}{10}}}{2} & -\frac{1}{2 \sqrt{6}} \\
 \frac{1}{2 \sqrt{10}} & -\frac{1}{6 \sqrt{10}} & -\frac{1}{2 \sqrt{15}} & -\frac{1}{6 \sqrt{2}} & \frac{1}{2 \sqrt{6}} & \frac{1}{\sqrt{15}} & \frac{\sqrt{\frac{3}{10}}}{2} & -\frac{1}{\sqrt{15}} & 0 &
   -\frac{1}{2 \sqrt{15}} & -\frac{\sqrt{\frac{3}{10}}}{2} & \frac{11}{30 \sqrt{2}} & -\frac{\sqrt{3}}{10} & \frac{2}{5 \sqrt{7}} & \frac{\sqrt{3}}{5} & -\frac{5}{2 \sqrt{42}} & -\frac{1}{3 \sqrt{5}} &
   \frac{1}{3} & -\frac{1}{2 \sqrt{5}} & \frac{1}{6} \\
 -\frac{1}{2 \sqrt{10}} & \frac{1}{6 \sqrt{10}} & \frac{1}{2 \sqrt{15}} & \frac{1}{6 \sqrt{2}} & -\frac{1}{2 \sqrt{6}} & -\frac{1}{\sqrt{15}} & \frac{1}{\sqrt{30}} & \frac{1}{\sqrt{15}} & -\frac{1}{2
   \sqrt{3}} & \frac{1}{2 \sqrt{15}} & \frac{1}{\sqrt{30}} & \frac{4 \sqrt{2}}{15} & -\frac{\sqrt{3}}{10} & -\frac{1}{10 \sqrt{7}} & \frac{\sqrt{3}}{5} & -\sqrt{\frac{2}{21}} & \frac{1}{3 \sqrt{5}} &
   -\frac{1}{3} & -\frac{1}{2 \sqrt{5}} & -\frac{1}{6} \\
 \frac{\sqrt{\frac{3}{10}}}{2} & -\frac{1}{2 \sqrt{30}} & -\frac{1}{2 \sqrt{5}} & 0 & 0 & -\frac{2}{3 \sqrt{5}} & -\frac{1}{6 \sqrt{10}} & \frac{2}{3 \sqrt{5}} & -\frac{1}{6} & \frac{1}{3 \sqrt{5}} &
   \frac{\sqrt{\frac{5}{2}}}{6} & -\frac{1}{2 \sqrt{6}} & 0 & \frac{\sqrt{\frac{3}{7}}}{2} & 0 & -\frac{1}{2 \sqrt{14}} & -\frac{1}{\sqrt{15}} & 0 & 0 & \frac{1}{\sqrt{3}} \\
 -\frac{1}{2 \sqrt{5}} & \frac{1}{6 \sqrt{5}} & \frac{1}{\sqrt{30}} & \frac{1}{6} & -\frac{1}{2 \sqrt{3}} & \frac{2 \sqrt{\frac{2}{15}}}{3} & \frac{1}{6 \sqrt{15}} & -\frac{2 \sqrt{\frac{2}{15}}}{3} &
   \frac{1}{3 \sqrt{6}} & -\frac{\sqrt{\frac{2}{15}}}{3} & -\frac{\sqrt{\frac{5}{3}}}{6} & -\frac{1}{6} & 0 & \frac{1}{\sqrt{14}} & 0 & -\frac{1}{2 \sqrt{21}} & \frac{\sqrt{\frac{2}{5}}}{3} &
   -\frac{\sqrt{2}}{3} & 0 & \frac{\sqrt{2}}{3} \\
\end{array}
\right)
\left(
\begin{array}{c}
\langle 8 \parallel \mathbf{3_{1/2}} \parallel \bar{3}\rangle\\
\langle 8 \parallel \mathbf{15_{1/2}} \parallel \bar{3}\rangle\\
\langle 8 \parallel \overline{\mathbf{6}_{1/2}} \parallel \bar{3}\rangle\\
\langle 10 \parallel \mathbf{15_{3/2}} \parallel \bar{3}\rangle\\
\langle 10 \parallel \mathbf{15^{'}_{3/2}} \parallel \bar{3}\rangle\\
\langle 27 \parallel \mathbf{15_{1/2}} \parallel \bar{3}\rangle\\
\langle 27 \parallel \mathbf{15_{3/2}} \parallel \bar{3}\rangle\\
\langle 27 \parallel \mathbf{24_{1/2}} \parallel \bar{3}\rangle\\
\langle 27 \parallel \mathbf{24_{3/2}} \parallel \bar{3}\rangle\\
\langle 27 \parallel \mathbf{42_{1/2}} \parallel \bar{3}\rangle\\
\langle 27 \parallel \mathbf{42_{3/2}} \parallel \bar{3}\rangle\\
\langle 35 \parallel \mathbf{42_{3/2}} \parallel \bar{3}\rangle\\
\langle 35 \parallel \mathbf{42_{5/2}} \parallel \bar{3}\rangle\\
\langle 35 \parallel \mathbf{48_{3/2}} \parallel \bar{3}\rangle\\
\langle 35 \parallel \mathbf{48_{5/2}} \parallel \bar{3}\rangle\\
\langle 35 \parallel \mathbf{15^{'}_{3/2}} \parallel \bar{3}\rangle\\
\langle 8 \parallel \mathbf{15_{3/2}} \parallel \bar{3}\rangle\\
\langle 10 \parallel \mathbf{15_{1/2}} \parallel \bar{3}\rangle\\
\langle 27 \parallel \mathbf{42_{5/2}} \parallel \bar{3}\rangle\\
\langle 35 \parallel \mathbf{42_{1/2}} \parallel \bar{3}\rangle\\
\end{array}
\right)
$
	}
$
	}	
\end{equation} 

\section{}
\label{sec:App2}	
$\mathcal{T}$ matrix for $\Delta S=0$ processes assuming the particular form of dim-6 tree Hamiltonian described in Eq.~\eqref{T Ham}: 	
\begin{equation}
	\centering
\rotatebox{90}{
$
\resizebox{0.9\vsize}{!}{%
$
\left(
\begin{array}{c}
\mathcal{A}(\Lambda_{b}^{0}\to  \Delta^{+} \pi^{-})\\
\mathcal{A}(\Lambda_{b}^{0}\to  \Delta^{0} \pi^{0})\\
\mathcal{A}(\Lambda_{b}^{0}\to  \Delta^{0} \eta_{8})\\
\mathcal{A}(\Lambda_{b}^{0}\to  \Delta^{-}\pi^{+})\\
\mathcal{A}(\Lambda_{b}^{0}\to  \Sigma^{'0}K^{0})\\
\mathcal{A}(\Lambda_{b}^{0}\to  \Sigma^{'-}K^{+})\\
\mathcal{A}(\Xi_{b}^{0}\to \Delta^{+} K^{-})\\
\mathcal{A}(\Xi_{b}^{0}\to  \Delta^{0} \overline{K}^{0})\\
\mathcal{A}(\Xi_{b}^{0}\to  \Sigma^{'+}\pi^{-})\\
\mathcal{A}(\Xi_{b}^{0}\to  \Sigma^{'0}\pi^{0})\\
\mathcal{A}(\Xi_{b}^{0}\to  \Sigma^{'0} \eta_{8})\\
\mathcal{A}(\Xi_{b}^{0}\to  \Sigma^{'-}\pi^{+})\\
\mathcal{A}(\Xi_{b}^{0}\to \Xi^{'0} K^{0})\\
\mathcal{A}(\Xi_{b}^{0}\to  \Xi^{'-}K^{+})\\
\mathcal{A}(\Xi_{b}^{-}\to  \Delta^{'0}K^{-})\\
\mathcal{A}(\Xi_{b}^{-}\to \Delta^{'-}\overline{K}^{0})\\
\mathcal{A}(\Xi_{b}^{-}\to \Sigma^{'0} \pi^{-})\\
\mathcal{A}(\Xi_{b}^{-}\to \Sigma^{'-} \pi^{0})\\
\mathcal{A}(\Xi_{b}^{-}\to \Sigma^{'-} \eta_{8})\\
\mathcal{A}(\Xi_{b}^{-}\to  \Sigma^{'-} K^{0})\\
\end{array}
\right)
=\left(
\begin{array}{ccccc}
-\frac{C^{+}_{1,2}}{3 \sqrt{2}} & \frac{C^{+}_{1,2}}{5} & \frac{C^{+}_{1,2}}{30} & \frac{C^{-}_{1,2}}{3 \sqrt{10}} & \frac{C^{+}_{1,2}}{2 \sqrt{15}}-\frac{C^{-}_{1,2}}{\sqrt{30}} \\
 \frac{C^{+}_{1,2}}{12} & -\frac{3 C^{+}_{1,2}}{20 \sqrt{2}} & -\frac{C^{+}_{1,2}}{15 \sqrt{2}} & -\frac{C^{-}_{1,2}}{3 \sqrt{5}} & \frac{C^{-}_{1,2}}{\sqrt{15}}-\frac{C^{+}_{1,2}}{\sqrt{30}} \\
 -\frac{C^{+}_{1,2}}{4 \sqrt{3}} & -\frac{1}{4} \sqrt{\frac{3}{2}} C^{+}_{1,2} & 0 & 0 & 0 \\
 \frac{C^{+}_{1,2}}{2 \sqrt{6}} & -\frac{\sqrt{3} C^{+}_{1,2}}{20} & \frac{C^{+}_{1,2}}{10 \sqrt{3}} & \frac{C^{-}_{1,2}}{\sqrt{30}} & \frac{C^{+}_{1,2}}{2 \sqrt{5}}-\frac{C^{-}_{1,2}}{\sqrt{10}} \\
 \frac{C^{+}_{1,2}}{6} & \frac{3 C^{+}_{1,2}}{10 \sqrt{2}} & -\frac{C^{+}_{1,2}}{30 \sqrt{2}} & -\frac{C^{-}_{1,2}}{6 \sqrt{5}} & \frac{C^{-}_{1,2}}{2 \sqrt{15}}-\frac{C^{+}_{1,2}}{2 \sqrt{30}} \\
 \frac{C^{+}_{1,2}}{6 \sqrt{2}} & -\frac{C^{+}_{1,2}}{20} & \frac{C^{+}_{1,2}}{30} & \frac{C^{-}_{1,2}}{3 \sqrt{10}} & \frac{C^{+}_{1,2}}{2 \sqrt{15}}-\frac{C^{-}_{1,2}}{\sqrt{30}} \\
 -\frac{C^{+}_{1,2}}{6 \sqrt{2}} & \frac{C^{+}_{1,2}}{4} & \frac{C^{+}_{1,2}}{6} & -\frac{C^{-}_{1,2}}{3 \sqrt{10}} & \frac{C^{+}_{1,2}}{2 \sqrt{15}}-\frac{C^{-}_{1,2}}{\sqrt{30}} \\
 \frac{C^{+}_{1,2}}{6 \sqrt{2}} & \frac{C^{+}_{1,2}}{4} & -\frac{C^{+}_{1,2}}{6} & \frac{C^{-}_{1,2}}{3 \sqrt{10}} & \frac{C^{-}_{1,2}}{\sqrt{30}}-\frac{C^{+}_{1,2}}{2 \sqrt{15}} \\
 -\frac{C^{+}_{1,2}}{6 \sqrt{2}} & -\frac{C^{+}_{1,2}}{20} & -\frac{2 C^{+}_{1,2}}{15} & \frac{1}{3} \sqrt{\frac{2}{5}} C^{-}_{1,2} & 0 \\
 0 & -\frac{C^{+}_{1,2}}{5} & \frac{C^{+}_{1,2}}{20} & -\frac{C^{-}_{1,2}}{2 \sqrt{10}} & \frac{C^{-}_{1,2}}{2 \sqrt{30}}-\frac{C^{+}_{1,2}}{4 \sqrt{15}} \\
 0 & 0 & -\frac{C^{+}_{1,2}}{4 \sqrt{3}} & \frac{C^{-}_{1,2}}{2 \sqrt{30}} & \frac{C^{-}_{1,2}}{2 \sqrt{10}}-\frac{C^{+}_{1,2}}{4 \sqrt{5}} \\
 \frac{C^{+}_{1,2}}{6 \sqrt{2}} & -\frac{C^{+}_{1,2}}{20} & \frac{C^{+}_{1,2}}{30} & \frac{C^{-}_{1,2}}{3 \sqrt{10}} & \frac{C^{+}_{1,2}}{2 \sqrt{15}}-\frac{C^{-}_{1,2}}{\sqrt{30}} \\
 \frac{C^{+}_{1,2}}{6 \sqrt{2}} & \frac{C^{+}_{1,2}}{20} & \frac{2 C^{+}_{1,2}}{15} & -\frac{1}{3} \sqrt{\frac{2}{5}} C^{-}_{1,2} & 0 \\
 \frac{C^{+}_{1,2}}{6 \sqrt{2}} & -\frac{C^{+}_{1,2}}{20} & \frac{C^{+}_{1,2}}{30} & \frac{C^{-}_{1,2}}{3 \sqrt{10}} & \frac{C^{+}_{1,2}}{2 \sqrt{15}}-\frac{C^{-}_{1,2}}{\sqrt{30}} \\
 0 & \frac{2 C^{+}_{1,2}}{5} & -\frac{C^{+}_{1,2}}{10} & -\frac{C^{-}_{1,2}}{3 \sqrt{10}} & \frac{C^{+}_{1,2}}{2 \sqrt{15}}-\frac{C^{-}_{1,2}}{\sqrt{30}} \\
 0 & \frac{\sqrt{3} C^{+}_{1,2}}{10} & \frac{\sqrt{3} C^{+}_{1,2}}{10} & \frac{C^{-}_{1,2}}{\sqrt{30}} & \frac{C^{-}_{1,2}}{\sqrt{10}}-\frac{C^{+}_{1,2}}{2 \sqrt{5}} \\
 0 & -\frac{\sqrt{2} C^{+}_{1,2}}{5} & \frac{C^{+}_{1,2}}{10 \sqrt{2}} & \frac{C^{-}_{1,2}}{6 \sqrt{5}} & \frac{C^{-}_{1,2}}{2 \sqrt{15}}-\frac{C^{+}_{1,2}}{2 \sqrt{30}} \\
 0 & -\frac{C^{+}_{1,2}}{10 \sqrt{2}} & -\frac{C^{+}_{1,2}}{10 \sqrt{2}} & -\frac{C^{-}_{1,2}}{6 \sqrt{5}} & \frac{C^{+}_{1,2}}{2 \sqrt{30}}-\frac{C^{-}_{1,2}}{2 \sqrt{15}} \\
 0 & \frac{1}{10} \sqrt{\frac{3}{2}} C^{+}_{1,2} & \frac{1}{10} \sqrt{\frac{3}{2}} C^{+}_{1,2} & \frac{C^{-}_{1,2}}{2 \sqrt{15}} & \frac{C^{-}_{1,2}}{2 \sqrt{5}}-\frac{C^{+}_{1,2}}{2 \sqrt{10}} \\
 0 & -\frac{C^{+}_{1,2}}{10} & -\frac{C^{+}_{1,2}}{10} & -\frac{C^{-}_{1,2}}{3 \sqrt{10}} & \frac{C^{+}_{1,2}}{2 \sqrt{15}}-\frac{C^{-}_{1,2}}{\sqrt{30}} \\
\end{array}
\right)
\left(
\begin{array}{c}
\langle 10 \parallel \mathbf{15} \parallel \bar{3}\rangle\\
\langle 27 \parallel \mathbf{15} \parallel \bar{3}\rangle\\
\langle 8 \parallel \mathbf{15} \parallel \bar{3}\rangle\\
\langle 8 \parallel \overline{\mathbf{6}} \parallel \bar{3}\rangle\\
\langle 8 \parallel \mathbf{3} \parallel \bar{3}\rangle\\
\end{array}
\right)
$
	}
$
	}
\end{equation}

$\mathcal{T}$ matrix for $\Delta S=-1$ processes assuming the particular form of dim-6 tree Hamiltonian described in Eq.~\eqref{T Ham}:  	

\begin{equation}
\label{Tab:A3}
	\centering
\rotatebox{90}{
$
\resizebox{0.9\vsize}{!}{%
$
\left(
\begin{array}{c}
\mathcal{A}(\Lambda_{b}^{0}\to  \Delta^{+}K^{-})\\
\mathcal{A}(\Lambda_{b}^{0}\to  \Delta^{0}\bar{K}^{0})\\
\mathcal{A}(\Lambda_{b}^{0}\to  \Sigma^{'+}\pi^{-})\\
\mathcal{A}(\Lambda_{b}^{0}\to  \Sigma^{'0}\pi^{0})\\
\mathcal{A}(\Lambda_{b}^{0}\to  \Sigma^{'0}\eta_{8})\\
\mathcal{A}(\Lambda_{b}^{0}\to  \Sigma^{'-}\pi^{+})\\
\mathcal{A}(\Lambda_{b}^{0}\to  \Xi^{'0}K^{0})\\
\mathcal{A}(\Lambda_{b}^{0}\to  \Xi^{'-}K^{+})\\
\mathcal{A}(\Xi_{b}^{0}\to  \Sigma^{'+}K^{-})\\
\mathcal{A}(\Xi_{b}^{0}\to  \Sigma^{'0} \bar{K}^{0})\\
\mathcal{A}(\Xi_{b}^{0}\to  \Xi^{'0} \pi^{0})\\
\mathcal{A}(\Xi_{b}^{0}\to  \Xi^{'0}\eta_{8})\\
\mathcal{A}(\Xi_{b}^{0}\to  \Xi^{'-}\pi^{+})\\
\mathcal{A}(\Xi_{b}^{0}\to  \Omega^{-}K^{+})\\
\mathcal{A}(\Xi_{b}^{-}\to  \Sigma^{'0}K^{-})\\
\mathcal{A}(\Xi_{b}^{-}\to  \Sigma^{'-}\bar{K}^{0})\\
\mathcal{A}(\Xi_{b}^{-}\to  \Xi^{'0}\pi^{-})\\
\mathcal{A}(\Xi_{b}^{-}\to  \Xi^{'-} \pi^{0})\\
\mathcal{A}(\Xi_{b}^{-}\to  \Xi^{'-}\eta_{8})\\
\mathcal{A}(\Xi_{b}^{-}\to  \Omega^{-} K^{0})\\
\end{array}
\right)
=\left(
\begin{array}{ccccc}
-\frac{C^{+}_{1,2}}{6 \sqrt{2}} & -\frac{C^{+}_{1,2}}{20} & -\frac{2 C^{+}_{1,2}}{15} & \frac{1}{3} \sqrt{\frac{2}{5}} C^{-}_{1,2} & 0 \\
 \frac{C^{+}_{1,2}}{6 \sqrt{2}} & \frac{C^{+}_{1,2}}{20} & \frac{2 C^{+}_{1,2}}{15} & -\frac{1}{3} \sqrt{\frac{2}{5}} C^{-}_{1,2} & 0 \\
 -\frac{C^{+}_{1,2}}{6 \sqrt{2}} & \frac{C^{+}_{1,2}}{4} & \frac{C^{+}_{1,2}}{6} & -\frac{C^{-}_{1,2}}{3 \sqrt{10}} & \frac{C^{+}_{1,2}}{2 \sqrt{15}}-\frac{C^{-}_{1,2}}{\sqrt{30}} \\
 0 & -\frac{C^{+}_{1,2}}{10} & -\frac{C^{+}_{1,2}}{10} & 0 & \frac{C^{-}_{1,2}}{\sqrt{30}}-\frac{C^{+}_{1,2}}{2 \sqrt{15}} \\
 0 & -\frac{\sqrt{3} C^{+}_{1,2}}{10} & \frac{C^{+}_{1,2}}{5 \sqrt{3}} & -\frac{C^{-}_{1,2}}{\sqrt{30}} & 0 \\
 \frac{C^{+}_{1,2}}{6 \sqrt{2}} & -\frac{C^{+}_{1,2}}{20} & \frac{C^{+}_{1,2}}{30} & \frac{C^{-}_{1,2}}{3 \sqrt{10}} & \frac{C^{+}_{1,2}}{2 \sqrt{15}}-\frac{C^{-}_{1,2}}{\sqrt{30}} \\
 \frac{C^{+}_{1,2}}{6 \sqrt{2}} & \frac{C^{+}_{1,2}}{4} & -\frac{C^{+}_{1,2}}{6} & \frac{C^{-}_{1,2}}{3 \sqrt{10}} & \frac{C^{-}_{1,2}}{\sqrt{30}}-\frac{C^{+}_{1,2}}{2 \sqrt{15}} \\
 \frac{C^{+}_{1,2}}{6 \sqrt{2}} & -\frac{C^{+}_{1,2}}{20} & \frac{C^{+}_{1,2}}{30} & \frac{C^{-}_{1,2}}{3 \sqrt{10}} & \frac{C^{+}_{1,2}}{2 \sqrt{15}}-\frac{C^{-}_{1,2}}{\sqrt{30}} \\
 -\frac{C^{+}_{1,2}}{3 \sqrt{2}} & \frac{C^{+}_{1,2}}{5} & \frac{C^{+}_{1,2}}{30} & \frac{C^{-}_{1,2}}{3 \sqrt{10}} & \frac{C^{+}_{1,2}}{2 \sqrt{15}}-\frac{C^{-}_{1,2}}{\sqrt{30}} \\
 \frac{C^{+}_{1,2}}{6} & \frac{3 C^{+}_{1,2}}{10 \sqrt{2}} & -\frac{C^{+}_{1,2}}{30 \sqrt{2}} & -\frac{C^{-}_{1,2}}{6 \sqrt{5}} & \frac{C^{-}_{1,2}}{2 \sqrt{15}}-\frac{C^{+}_{1,2}}{2 \sqrt{30}} \\
 -\frac{C^{+}_{1,2}}{12} & -\frac{9 C^{+}_{1,2}}{20 \sqrt{2}} & -\frac{C^{+}_{1,2}}{30 \sqrt{2}} & -\frac{C^{-}_{1,2}}{6 \sqrt{5}} & \frac{C^{-}_{1,2}}{2 \sqrt{15}}-\frac{C^{+}_{1,2}}{2 \sqrt{30}} \\
 \frac{C^{+}_{1,2}}{4 \sqrt{3}} & \frac{1}{20} \sqrt{\frac{3}{2}} C^{+}_{1,2} & -\frac{C^{+}_{1,2}}{10 \sqrt{6}} & -\frac{C^{-}_{1,2}}{2 \sqrt{15}} & \frac{C^{-}_{1,2}}{2 \sqrt{5}}-\frac{C^{+}_{1,2}}{2
   \sqrt{10}} \\
 \frac{C^{+}_{1,2}}{6 \sqrt{2}} & -\frac{C^{+}_{1,2}}{20} & \frac{C^{+}_{1,2}}{30} & \frac{C^{-}_{1,2}}{3 \sqrt{10}} & \frac{C^{+}_{1,2}}{2 \sqrt{15}}-\frac{C^{-}_{1,2}}{\sqrt{30}} \\
 \frac{C^{+}_{1,2}}{2 \sqrt{6}} & -\frac{\sqrt{3} C^{+}_{1,2}}{20} & \frac{C^{+}_{1,2}}{10 \sqrt{3}} & \frac{C^{-}_{1,2}}{\sqrt{30}} & \frac{C^{+}_{1,2}}{2 \sqrt{5}}-\frac{C^{-}_{1,2}}{\sqrt{10}} \\
 0 & \frac{\sqrt{2} C^{+}_{1,2}}{5} & -\frac{C^{+}_{1,2}}{10 \sqrt{2}} & -\frac{C^{-}_{1,2}}{6 \sqrt{5}} & \frac{C^{+}_{1,2}}{2 \sqrt{30}}-\frac{C^{-}_{1,2}}{2 \sqrt{15}} \\
 0 & \frac{C^{+}_{1,2}}{10} & \frac{C^{+}_{1,2}}{10} & \frac{C^{-}_{1,2}}{3 \sqrt{10}} & \frac{C^{-}_{1,2}}{\sqrt{30}}-\frac{C^{+}_{1,2}}{2 \sqrt{15}} \\
 0 & -\frac{2 C^{+}_{1,2}}{5} & \frac{C^{+}_{1,2}}{10} & \frac{C^{-}_{1,2}}{3 \sqrt{10}} & \frac{C^{-}_{1,2}}{\sqrt{30}}-\frac{C^{+}_{1,2}}{2 \sqrt{15}} \\
 0 & -\frac{C^{+}_{1,2}}{10 \sqrt{2}} & -\frac{C^{+}_{1,2}}{10 \sqrt{2}} & -\frac{C^{-}_{1,2}}{6 \sqrt{5}} & \frac{C^{+}_{1,2}}{2 \sqrt{30}}-\frac{C^{-}_{1,2}}{2 \sqrt{15}} \\
 0 & \frac{1}{10} \sqrt{\frac{3}{2}} C^{+}_{1,2} & \frac{1}{10} \sqrt{\frac{3}{2}} C^{+}_{1,2} & \frac{C^{-}_{1,2}}{2 \sqrt{15}} & \frac{C^{-}_{1,2}}{2 \sqrt{5}}-\frac{C^{+}_{1,2}}{2 \sqrt{10}} \\
 0 & -\frac{\sqrt{3} C^{+}_{1,2}}{10} & -\frac{\sqrt{3} C^{+}_{1,2}}{10} & -\frac{C^{-}_{1,2}}{\sqrt{30}} & \frac{C^{+}_{1,2}}{2 \sqrt{5}}-\frac{C^{-}_{1,2}}{\sqrt{10}} \\
\end{array}
\right)
\left(
\begin{array}{c}
\langle 10 \parallel \mathbf{15} \parallel \bar{3}\rangle\\
\langle 27 \parallel \mathbf{15} \parallel \bar{3}\rangle\\
\langle 8 \parallel \mathbf{15} \parallel \bar{3}\rangle\\
\langle 8 \parallel \overline{\mathbf{6}} \parallel \bar{3}\rangle\\
\langle 8 \parallel \mathbf{3} \parallel \bar{3}\rangle\\
\end{array}
\right)
$
	}
$
	}
\end{equation}
\newpage
$\mathcal{P}$ matrix for $\Delta S=0$ processes assuming the particular form of dim-6 gluonic and Electroweak penguin Hamiltonian described in Eq.~\eqref{P g Ham} and Eq.~\eqref{P EW Ham}: 
\begin{equation}
\label{PEW S0 tab}
\centering
\rotatebox{90}{
	$
	\resizebox{0.9\vsize}{!}{%
		$
		\left(
		\begin{array}{c}
		\mathcal{A}(\Lambda_{b}^{0}\to  \Delta^{+} \pi^{-})\\
		\mathcal{A}(\Lambda_{b}^{0}\to  \Delta^{0} \pi^{0})\\
		\mathcal{A}(\Lambda_{b}^{0}\to  \Delta^{0} \eta_{8})\\
		\mathcal{A}(\Lambda_{b}^{0}\to  \Delta^{-}\pi^{+})\\
		\mathcal{A}(\Lambda_{b}^{0}\to  \Sigma^{'0}K^{0})\\
		\mathcal{A}(\Lambda_{b}^{0}\to  \Sigma^{'-}K^{+})\\
		\mathcal{A}(\Xi_{b}^{0}\to \Delta^{+} K^{-})\\
		\mathcal{A}(\Xi_{b}^{0}\to  \Delta^{0} \overline{K}^{0})\\
		\mathcal{A}(\Xi_{b}^{0}\to  \Sigma^{'+}\pi^{-})\\
		\mathcal{A}(\Xi_{b}^{0}\to  \Sigma^{'0}\pi^{0})\\
		\mathcal{A}(\Xi_{b}^{0}\to  \Sigma^{'0} \eta_{8})\\
		\mathcal{A}(\Xi_{b}^{0}\to  \Sigma^{'-}\pi^{+})\\
		\mathcal{A}(\Xi_{b}^{0}\to \Xi^{'0} K^{0})\\
		\mathcal{A}(\Xi_{b}^{0}\to  \Xi^{'-}K^{+})\\
		\mathcal{A}(\Xi_{b}^{-}\to  \Delta^{'0}K^{-})\\
		\mathcal{A}(\Xi_{b}^{-}\to \Delta^{'-}\overline{K}^{0})\\
		\mathcal{A}(\Xi_{b}^{-}\to \Sigma^{'0} \pi^{-})\\
		\mathcal{A}(\Xi_{b}^{-}\to \Sigma^{'-} \pi^{0})\\
		\mathcal{A}(\Xi_{b}^{-}\to \Sigma^{'-} \eta_{8})\\
		\mathcal{A}(\Xi_{b}^{-}\to  \Xi^{'-} K^{0})\\
		\end{array}
		\right)
		=
		\left(
		\begin{array}{ccccc}
		\frac{C^{+}_{9,10}}{2 \sqrt{2}} & -\frac{3 C^{+}_{9,10}}{10} & 
		-\frac{C^{+}_{9,10}}{20} & -\frac{C^{-}_{9,10}}{2 \sqrt{10}} & 
		-\sqrt{\frac{2}{15}} D \\
		-\frac{C^{+}_{9,10}}{8} & \frac{9 C^{+}_{9,10}}{40 \sqrt{2}} & 
		\frac{C^{+}_{9,10}}{10 \sqrt{2}} & \frac{C^{-}_{9,10}}{2 \sqrt{5}} & 
		\frac{2 D}{\sqrt{15}} \\
		\frac{\sqrt{3} C^{+}_{9,10}}{8} & \frac{3}{8} \sqrt{\frac{3}{2}} 
		C^{+}_{9,10} & 0 & 0 & 0 \\
		-\frac{1}{4} \sqrt{\frac{3}{2}} C^{+}_{9,10} & \frac{3 \sqrt{3} 
		C^{+}_{9,10}}{40} & -\frac{\sqrt{3} C^{+}_{9,10}}{20} & -\frac{1}{2} 
		\sqrt{\frac{3}{10}} C^{-}_{9,10} & -\sqrt{\frac{2}{5}} D \\
		-\frac{C^{+}_{9,10}}{4} & -\frac{9 C^{+}_{9,10}}{20 \sqrt{2}} & 
		\frac{C^{+}_{9,10}}{20 \sqrt{2}} & \frac{C^{-}_{9,10}}{4 \sqrt{5}} & 
		\frac{D}{\sqrt{15}} \\
		-\frac{C^{+}_{9,10}}{4 \sqrt{2}} & \frac{3 C^{+}_{9,10}}{40} & 
		-\frac{C^{+}_{9,10}}{20} & -\frac{C^{-}_{9,10}}{2 \sqrt{10}} & 
		-\sqrt{\frac{2}{15}} D \\
		\frac{C^{+}_{9,10}}{4 \sqrt{2}} & -\frac{3 C^{+}_{9,10}}{8} & 
		-\frac{C^{+}_{9,10}}{4} & \frac{C^{-}_{9,10}}{2 \sqrt{10}} & 
		-\sqrt{\frac{2}{15}} D \\
		-\frac{C^{+}_{9,10}}{4 \sqrt{2}} & -\frac{3 C^{+}_{9,10}}{8} & 
		\frac{C^{+}_{9,10}}{4} & -\frac{C^{-}_{9,10}}{2 \sqrt{10}} & 
		\sqrt{\frac{2}{15}} D \\
		\frac{C^{+}_{9,10}}{4 \sqrt{2}} & \frac{3 C^{+}_{9,10}}{40} & 
		\frac{C^{+}_{9,10}}{5} & -\frac{C^{-}_{9,10}}{\sqrt{10}} & 0 \\
		0 & \frac{3 C^{+}_{9,10}}{10} & -\frac{3 C^{+}_{9,10}}{40} & \frac{3 
		C^{-}_{9,10}}{4 \sqrt{10}} & \frac{D}{\sqrt{30}} \\
		0 & 0 & \frac{\sqrt{3} C^{+}_{9,10}}{8} & -\frac{1}{4} 
		\sqrt{\frac{3}{10}} C^{-}_{9,10} & \frac{D}{\sqrt{10}} \\
		-\frac{C^{+}_{9,10}}{4 \sqrt{2}} & \frac{3 C^{+}_{9,10}}{40} & 
		-\frac{C^{+}_{9,10}}{20} & -\frac{C^{-}_{9,10}}{2 \sqrt{10}} & 
		-\sqrt{\frac{2}{15}} D \\
		-\frac{C^{+}_{9,10}}{4 \sqrt{2}} & -\frac{3 C^{+}_{9,10}}{40} & 
		-\frac{C^{+}_{9,10}}{5} & \frac{C^{-}_{9,10}}{\sqrt{10}} & 0 \\
		-\frac{C^{+}_{9,10}}{4 \sqrt{2}} & \frac{3 C^{+}_{9,10}}{40} & 
		-\frac{C^{+}_{9,10}}{20} & -\frac{C^{-}_{9,10}}{2 \sqrt{10}} & 
		-\sqrt{\frac{2}{15}} D \\
		0 & -\frac{3 C^{+}_{9,10}}{5} & \frac{3 C^{+}_{9,10}}{20} & 
		\frac{C^{-}_{9,10}}{2 \sqrt{10}} & -\sqrt{\frac{2}{15}} D \\
		0 & -\frac{3 \sqrt{3} C^{+}_{9,10}}{20} & -\frac{3 \sqrt{3} 
		C^{+}_{9,10}}{20} & -\frac{1}{2} \sqrt{\frac{3}{10}} C^{-}_{9,10} & 
		\sqrt{\frac{2}{5}} D \\
		0 & \frac{3 C^{+}_{9,10}}{5 \sqrt{2}} & -\frac{3 C^{+}_{9,10}}{20 
		\sqrt{2}} & -\frac{C^{-}_{9,10}}{4 \sqrt{5}} & \frac{D}{\sqrt{15}} \\
		0 & \frac{3 C^{+}_{9,10}}{20 \sqrt{2}} & \frac{3 C^{+}_{9,10}}{20 
		\sqrt{2}} & \frac{C^{-}_{9,10}}{4 \sqrt{5}} & -\frac{D}{\sqrt{15}} \\
		0 & -\frac{3}{20} \sqrt{\frac{3}{2}} C^{+}_{9,10} & -\frac{3}{20} 
		\sqrt{\frac{3}{2}} C^{+}_{9,10} & -\frac{1}{4} \sqrt{\frac{3}{5}} 
		C^{-}_{9,10} & \frac{D}{\sqrt{5}} \\
		0 & \frac{3 C^{+}_{9,10}}{20} & \frac{3 C^{+}_{9,10}}{20} & 
		\frac{C^{-}_{9,10}}{2 \sqrt{10}} & -\sqrt{\frac{2}{15}} D \\
		\end{array}
		\right)
		\left(
		\begin{array}{c}
		\langle 10 \parallel \mathbf{15} \parallel \bar{3}\rangle\\
		\langle 27 \parallel \mathbf{15} \parallel \bar{3}\rangle\\
		\langle 8 \parallel \mathbf{15} \parallel \bar{3}\rangle\\
		\langle 8 \parallel \overline{\mathbf{6}} \parallel \bar{3}\rangle\\
		\langle 8 \parallel \mathbf{3} \parallel \bar{3}\rangle\\
		\end{array}
		\right)
		$
	}
	$
}
\end{equation}
\newpage
$\mathcal{P}$ matrix for $\Delta S=-1$ processes assuming the particular form of dim-6 gluonic and Electroweak penguin Hamiltonian described in Eq~\eqref{P g Ham} and Eq~\eqref{P EW Ham}:
\begin{equation}
\label{PEW S1 Tab}
	\centering
\rotatebox{90}{
$
\resizebox{0.9\vsize}{!}{%
$
\left(
\begin{array}{c}
\mathcal{A}(\Lambda_{b}^{0}\to  \Delta^{+}K^{-})\\
\mathcal{A}(\Lambda_{b}^{0}\to  \Delta^{0}\bar{K}^{0})\\
\mathcal{A}(\Lambda_{b}^{0}\to  \Sigma^{'+}\pi^{-})\\
\mathcal{A}(\Lambda_{b}^{0}\to  \Sigma^{'0}\pi^{0})\\
\mathcal{A}(\Lambda_{b}^{0}\to  \Sigma^{'0}\eta_{8})\\
\mathcal{A}(\Lambda_{b}^{0}\to  \Sigma^{'-}\pi^{+})\\
\mathcal{A}(\Lambda_{b}^{0}\to  \Xi^{'0}K^{0})\\
\mathcal{A}(\Lambda_{b}^{0}\to  \Xi^{'-}K^{+})\\
\mathcal{A}(\Xi_{b}^{0}\to  \Sigma^{'+}K^{-})\\
\mathcal{A}(\Xi_{b}^{0}\to  \Sigma^{'0} \bar{K}^{0})\\
\mathcal{A}(\Xi_{b}^{0}\to  \Xi^{'0} \pi^{0})\\
\mathcal{A}(\Xi_{b}^{0}\to  \Xi^{'0}\eta_{8})\\
\mathcal{A}(\Xi_{b}^{0}\to  \Xi^{'-}\pi^{+})\\
\mathcal{A}(\Xi_{b}^{0}\to  \Omega^{-}K^{+})\\
\mathcal{A}(\Xi_{b}^{-}\to  \Sigma^{'0}K^{-})\\
\mathcal{A}(\Xi_{b}^{-}\to  \Sigma^{'-}\bar{K}^{0})\\
\mathcal{A}(\Xi_{b}^{-}\to  \Xi^{'0}\pi^{-})\\
\mathcal{A}(\Xi_{b}^{-}\to  \Xi^{'-} \pi^{0})\\
\mathcal{A}(\Xi_{b}^{-}\to  \Xi^{'-}\eta_{8})\\
\mathcal{A}(\Xi_{b}^{-}\to  \Omega^{-} K^{0})\\
\end{array}
\right)
=
\left(
\begin{array}{ccccc}
\frac{C^{+}_{9,10}}{4 \sqrt{2}} & \frac{3 C^{+}_{9,10}}{40} & 
\frac{C^{+}_{9,10}}{5} & -\frac{C^{-}_{9,10}}{\sqrt{10}} & 0 \\
-\frac{C^{+}_{9,10}}{4 \sqrt{2}} & -\frac{3 C^{+}_{9,10}}{40} & 
-\frac{C^{+}_{9,10}}{5} & \frac{C^{-}_{9,10}}{\sqrt{10}} & 0 \\
\frac{C^{+}_{9,10}}{4 \sqrt{2}} & -\frac{3 C^{+}_{9,10}}{8} & 
-\frac{C^{+}_{9,10}}{4} & \frac{C^{-}_{9,10}}{2 \sqrt{10}} & 
-\sqrt{\frac{2}{15}} D \\
0 & \frac{3 C^{+}_{9,10}}{20} & \frac{3 C^{+}_{9,10}}{20} & 0 & 
\sqrt{\frac{2}{15}} D \\
0 & \frac{3 \sqrt{3} C^{+}_{9,10}}{20} & -\frac{\sqrt{3} C^{+}_{9,10}}{10} & 
\frac{1}{2} \sqrt{\frac{3}{10}} C^{-}_{9,10} & 0 \\
-\frac{C^{+}_{9,10}}{4 \sqrt{2}} & \frac{3 C^{+}_{9,10}}{40} & 
-\frac{C^{+}_{9,10}}{20} & -\frac{C^{-}_{9,10}}{2 \sqrt{10}} & 
-\sqrt{\frac{2}{15}} D \\
-\frac{C^{+}_{9,10}}{4 \sqrt{2}} & -\frac{3 C^{+}_{9,10}}{8} & 
\frac{C^{+}_{9,10}}{4} & -\frac{C^{-}_{9,10}}{2 \sqrt{10}} & 
\sqrt{\frac{2}{15}} D \\
-\frac{C^{+}_{9,10}}{4 \sqrt{2}} & \frac{3 C^{+}_{9,10}}{40} & 
-\frac{C^{+}_{9,10}}{20} & -\frac{C^{-}_{9,10}}{2 \sqrt{10}} & 
-\sqrt{\frac{2}{15}} D \\
\frac{C^{+}_{9,10}}{2 \sqrt{2}} & -\frac{3 C^{+}_{9,10}}{10} & 
-\frac{C^{+}_{9,10}}{20} & -\frac{C^{-}_{9,10}}{2 \sqrt{10}} & 
-\sqrt{\frac{2}{15}} D \\
-\frac{C^{+}_{9,10}}{4} & -\frac{9 C^{+}_{9,10}}{20 \sqrt{2}} & 
\frac{C^{+}_{9,10}}{20 \sqrt{2}} & \frac{C^{-}_{9,10}}{4 \sqrt{5}} & 
\frac{D}{\sqrt{15}} \\
\frac{C^{+}_{9,10}}{8} & \frac{27 C^{+}_{9,10}}{40 \sqrt{2}} & 
\frac{C^{+}_{9,10}}{20 \sqrt{2}} & \frac{C^{-}_{9,10}}{4 \sqrt{5}} & 
\frac{D}{\sqrt{15}} \\
-\frac{\sqrt{3} C^{+}_{9,10}}{8} & -\frac{3}{40} \sqrt{\frac{3}{2}} 
C^{+}_{9,10} & \frac{1}{20} \sqrt{\frac{3}{2}} C^{+}_{9,10} & \frac{1}{4} 
\sqrt{\frac{3}{5}} C^{-}_{9,10} & \frac{D}{\sqrt{5}} \\
-\frac{C^{+}_{9,10}}{4 \sqrt{2}} & \frac{3 C^{+}_{9,10}}{40} & 
-\frac{C^{+}_{9,10}}{20} & -\frac{C^{-}_{9,10}}{2 \sqrt{10}} & 
-\sqrt{\frac{2}{15}} D \\
-\frac{1}{4} \sqrt{\frac{3}{2}} C^{+}_{9,10} & \frac{3 \sqrt{3} 
C^{+}_{9,10}}{40} & -\frac{\sqrt{3} C^{+}_{9,10}}{20} & -\frac{1}{2} 
\sqrt{\frac{3}{10}} C^{-}_{9,10} & -\sqrt{\frac{2}{5}} D \\
0 & -\frac{3 C^{+}_{9,10}}{5 \sqrt{2}} & \frac{3 C^{+}_{9,10}}{20 \sqrt{2}} & 
\frac{C^{-}_{9,10}}{4 \sqrt{5}} & -\frac{D}{\sqrt{15}} \\
0 & -\frac{3 C^{+}_{9,10}}{20} & -\frac{3 C^{+}_{9,10}}{20} & 
-\frac{C^{-}_{9,10}}{2 \sqrt{10}} & \sqrt{\frac{2}{15}} D \\
0 & \frac{3 C^{+}_{9,10}}{5} & -\frac{3 C^{+}_{9,10}}{20} & 
-\frac{C^{-}_{9,10}}{2 \sqrt{10}} & \sqrt{\frac{2}{15}} D \\
0 & \frac{3 C^{+}_{9,10}}{20 \sqrt{2}} & \frac{3 C^{+}_{9,10}}{20 \sqrt{2}} & 
\frac{C^{-}_{9,10}}{4 \sqrt{5}} & -\frac{D}{\sqrt{15}} \\
0 & -\frac{3}{20} \sqrt{\frac{3}{2}} C^{+}_{9,10} & -\frac{3}{20} 
\sqrt{\frac{3}{2}} C^{+}_{9,10} & -\frac{1}{4} \sqrt{\frac{3}{5}} C^{-}_{9,10} 
& \frac{D}{\sqrt{5}} \\
0 & \frac{3 \sqrt{3} C^{+}_{9,10}}{20} & \frac{3 \sqrt{3} C^{+}_{9,10}}{20} & 
\frac{1}{2} \sqrt{\frac{3}{10}} C^{-}_{9,10} & -\sqrt{\frac{2}{5}} D \\
\end{array}
\right)
\left(
\begin{array}{c}
\langle 10 \parallel \mathbf{15} \parallel \bar{3}\rangle\\
\langle 27 \parallel \mathbf{15} \parallel \bar{3}\rangle\\
\langle 8 \parallel \mathbf{15} \parallel \bar{3}\rangle\\
\langle 8 \parallel \overline{\mathbf{6}} \parallel \bar{3}\rangle\\
\langle 8 \parallel \mathbf{3} \parallel \bar{3}\rangle\\
\end{array}
\right)
$
	}
$
	}
\end{equation}
\pagebreak
\twocolumngrid
The $SU(3)$-decomposition of decay amplitudes without
assuming $SU(3)$-breaking for $\overline{\mathbf{3}}\to\mathbf{10}\otimes\mathbf{1}$ decay modes are given below:

$\Delta S=0$ processes:
\begin{align}
&\mathcal{A}(\Lambda_{b}^{0}\to  \Delta^{0} 
\eta_{1})=-\frac{C^{+}_{1,2}}{\sqrt{6}}\,(\lambda_{u}^{d}-\frac{3}{2}\kappa\lambda_{t}^{d})\,\langle
 10 \parallel \mathbf{15} \parallel \bar{3}\rangle\nonumber\\
&\mathcal{A}(\Xi_{b}^{-}\to \Sigma^{'-} \eta_{1})=0
\end{align}
$\Delta S=-1$ processes:
\begin{align}
&\mathcal{A}(\Lambda_{b}^{0}\to  \Sigma^{0} 
\eta_{1})=-\frac{C^{+}_{1,2}}{2\sqrt{3}}\,(\lambda_{u}^{s}-\frac{3}{2}\kappa\lambda_{t}^{s})\,\langle
 10 \parallel \mathbf{15} \parallel \bar{3}\rangle\nonumber\\
&\mathcal{A}(\Xi_{b}^{-}\to \Xi^{'-} \eta_{1})=0
\end{align}  
We have used the following shorthand notation to express the tree 
($\mathcal{T}$) and penguin ($\mathcal{P}$) matrices in a convenient form:
\begin{gather*}
(C_{10} \pm C_{9})= C_{9,10}^{\pm},\\[2ex]
(C_{1} \pm C_{2})= C_{1,2}^{\pm},\\[2ex]
(C_{3}\pm C_{4}) = C_{3,4}^{\pm},\\[2ex]
(C_{5}\pm C_{6}) = C_{5,6}^{\pm},
\end{gather*}
and $D$ is given by the particular combination of Wilson coefficients:
\begin{align}
D&=-\frac{1}{4 \sqrt{2}}\Big(C_{9,10}^{+}+\sqrt{2} C_{9,10}^{-}\Big)
\nonumber\\
&\quad + 
\Big\{\sqrt{2} C_{3,4}^{+}-C_{3,4}^{-} 
+\sqrt{2}C_{5,6}^{+}-C_{5,6}^{-}\Big\}.\qquad
\end{align}

\section{}
\label{sec:App3}
The flavor flow in five topologies~\cite{He:2018joe} given in Figure~\ref{diagram1} is described below;
\begin{align*}
E_{1}&:B_{[ij]}D^{\{kmj\}}M^{l}_{k}H^{i}_{lm}\nonumber\\
E_{2}&:B_{[ij]}D^{\{kmj\}}M^{l}_{k}H^{i}_{ml}\nonumber\\
P_{u}&:B_{[ij]}D^{\{kjl\}}M^{i}_{k}H^{m}_{ml}\nonumber\\
T&:B_{[ij]}D^{\{lmj\}}M^{i}_{k}H^{k}_{ml}\nonumber\\
E_{3}&:B_{[ij]}D^{\{klm\}}M^{j}_{m}H^{i}_{kl}\nonumber
\end{align*}
where $B_{[ij]}$, $D^{\{klm\}}$, $M^{x}_{y}$ are flavor wavefunctions of the 
initial anti-triplet ($\overline{\mathbf{3}}$) $b$-baryon, the final state 
decuplet baryon ($\mathbf{10}$) and the octet meson ($\mathbf{8}$) 
respectively. The $H^{i}_{ab}$ mediates the quark transition and for $\Delta 
S=-1$, $\Delta S=0$ processes, the only non-zero contribution come from the 
elements $H^{1}_{13}=1$ and $H^{1}_{12}=1$ respectively. According to the phase 
convention chosen for quark ($q_{i}$) and anti-quark ($\overline{q_{i}}$) 
flavor states in Eq.~\eqref{eqn:q-antiq};
the meson wave-functions are given as, 
\begin{gather*}
K^{+}=u\overline{s}, \qquad K^{-}=-s\overline{u}, \qquad K^{0}= d\overline{s}, 
\qquad  \overline{K}^{0}=s\overline{d},\\
\pi^{+}=u\overline{d},\qquad \pi^{-}=-d\overline{u},\qquad 
\pi^{0}=\frac{1}{\sqrt{2}}(d\overline{d}-u\overline{u}),\\
\eta_{8}=-\frac{1}{\sqrt{6}}(u\overline{u}+d\overline{d}-2s\overline{s}),\quad
\eta_{1}=-\frac{1}{\sqrt{3}}(u\overline{u}+d\overline{d}+s\overline{s}).
\end{gather*} 
The initial $b$-baryon anti triplet $B_{[ij]}$ consists of the following states;
\begin{align*}
\Lambda_{b}^{0}&=\frac{1}{\sqrt{2}}(ud-du)\\
\Xi_{b}^{0}&=\frac{1}{\sqrt{2}}(us-su)\\
\Xi_{b}^{-}&=\frac{1}{\sqrt{2}}(ds-sd).
\end{align*} 
The decuplet baryons are have a completely symmetric flavor wave-function:
\begin{gather*}
\Delta^{++}=uuu,\qquad\Delta^{-}=ddd\\
\Delta^{+}=\frac{1}{\sqrt{3}}(uud+udu+duu)\\
\Delta^{0}=\frac{1}{\sqrt{3}}(udd+ddu+dud)\\
\Sigma^{'+}=\frac{1}{\sqrt{3}}(uus+usu+suu)\\
\Sigma^{'0}=\frac{1}{\sqrt{6}}(uds+usd+dus\\+dsu+sud+sdu)\\
\Sigma^{'-}=\frac{1}{\sqrt{3}}(dds+dsd+sdd)\\
\Xi^{'0}=\frac{1}{\sqrt{6}}(uss+sus+ssu)\\
\Xi^{'-}=\frac{1}{\sqrt{3}}(dss+ssd+sds)\\
\Omega^{-}=sss
\end{gather*}

\section{}
\label{sec:App4}
Considering only the tree operators in the effective Hamiltonian,
\begin{align}
\mathcal{H}^{\text{eff}}=C_{1}O_{1}+C_{2}O_{2}
\end{align}
that can be recast as~\cite{Korner:1970xq,Pati:1970fg,Stech:1987fa,Neubert:1991zd,Neubert:1997uc},
\begin{align}
\label{TH}
\mathcal{H}^{\text{eff}}=C^{-}O^{-}+C^{+}O^{+},
\end{align}
where $C^{-}=C_{1}-C_{2}$, $C^{+}=C_{1}+C_{2}$ and
\begin{widetext}
\begin{align}
O^{-}=\frac{1}{2}(O_{1}-O_{2})=\frac{1}{2}\Big((\overline{u}^{i}_{L}\gamma^{\mu}b^{j}_{L})
(\overline{d}^{j}_{L}\gamma_{\mu}u^{i}_{L})-(\overline{u}^{i}_{L}\gamma^{\mu}b^{i}_{L})
(\overline{d}^{j}_{L}\gamma_{\mu}u^{j}_{L})\Big) \\
O^{+}=\frac{1}{2}(O_{1}+O_{2})=\frac{1}{2}\Big((\overline{u}^{i}_{L}\gamma^{\mu}b^{j}_{L})
(\overline{d}^{j}_{L}\gamma_{\mu}u^{i}_{L})+(\overline{u}^{i}_{L}\gamma^{\mu}b^{i}_{L})
(\overline{d}^{j}_{L}\gamma_{\mu}u^{j}_{L})\Big).
\end{align}
Rewriting $O_{1}$ and $O_{2}$ we get,
\begin{align}
O_{1}=\frac{1}{4}\Big[(\overline{u}^{i})_{\alpha}\big[\gamma^{\mu}(1-\gamma_{5}) \big]_{\alpha \beta}(b^{l})_{\beta}\Big]\Big[(\overline{d}^{j})_{\sigma}\big[\gamma_{\mu}(1-\gamma_{5}) \big]_{\sigma \lambda}(u^{k})_{\lambda}\Big]\delta^{lj}\delta^{ik},\\
O_{2}=\frac{1}{4}\Big[(\overline{u}^{i})_{\alpha}\big[\gamma^{\mu}(1-\gamma_{5}) \big]_{\alpha \beta}(b^{l})_{\beta}\Big]\Big[(\overline{d}^{j})_{\sigma}\big[\gamma_{\mu}(1-\gamma_{5}) \big]_{\sigma \lambda}(u^{k})_{\lambda}\Big]\delta^{il}\delta^{jk}
\end{align} 
Using the following identity,
\begin{align*}
\frac{1}{2}\epsilon^{aij}\epsilon^{alk}=(\delta^{il}\delta^{jk}-\delta^{ik}\delta^{jl})
\end{align*}
one can write $O^{-}$ as,
\begin{align}
\label{VA form}
O^{-}=-\frac{1}{2}.\frac{1}{2}.\frac{1}{4}\epsilon^{aij}\epsilon^{alk}\Big[(\overline{u}^{i})_{\alpha}\big[\gamma^{\mu}(1-\gamma_{5}) \big]_{\alpha \beta}(b^{l})_{\beta}\Big]\Big[(\overline{d}^{j})_{\sigma}\big[\gamma_{\mu}(1-\gamma_{5}) \big]_{\sigma \lambda}(u^{k})_{\lambda}\Big]
\end{align}
Using Firez transformation~\cite{Nieves:2003in} one can recast Eq.~\eqref{VA 
form} as~\cite{Stech:1987fa},
\begin{align}
O^{-}=-\frac{1}{4}\frac{1}{2}\underbrace{\epsilon^{alk}\Big[(\overline{b}^{l})^{\mathbf{C}}(1-\gamma_{5})\,(u^{k}) \Big]}_{(bu)_{\overline{3}}}\underbrace{\epsilon^{aij}\Big[(\overline{u}^{i})(1+\gamma_{5})\,(d^{j})^{\mathbf{C}} \Big]}_{(ud)_{\overline{3}}^{\dagger}}
\end{align}
\end{widetext}
where $q^{\textbf{C}}=\mathbf{C}q^{*}$, $\mathbf{C}$ being the 
charge-conjugation operator. According to diquark 
mechanism~\cite{Stech:1987fa,Neubert:1991zd}, the diquark 
$\epsilon^{aij}\Big[(\overline{u}^{i})(1+\gamma_{5})\,(d^{j})^{\mathbf{C}} 
\Big]$ has a total spin 0 and it transforms as a $\overline{\mathbf{3}}$ and 
$\overline{\mathbf{3}}$ under $SU(3)_{F}$ and $SU(3)_{\text{color}}$ 
respectively. This diquark ending up completely in the final state baryon 
cannot produce\footnote{Assuming the baryon made of the diquark and a third 
quark transforming as a $\mathbf{3}$ and $\mathbf{3}$ under $SU(3)_{F}$ and 
$SU(3)_{\text{color}}$ the final state transforms as
$SU(3)_\text{color}:\,\overline{\mathbf{3}}\otimes\mathbf{3}=\mathbf{1}\oplus\mathbf{8}$,  
$SU(3)_{F}:\,\overline{\mathbf{3}}\otimes\mathbf{3}=\mathbf{1}\oplus\mathbf{8}$.
 Since baryons are color singlet, a $\mathbf{6}$ of $SU(3)_{\text{color}}$ 
diquark can never form a baryon.} a decuplet baryon as any quark pair inside 
the decuplet baryon transforms as $\mathbf{6}$ and $\overline{\mathbf{3}}$ 
under $SU(3)_{F}$ and $SU(3)_{\text{color}}$ with a total spin equaling 1. On 
the other hand, the operator $Q^{+}$ which is a product of 
$SU(3)_{\text{color}}$ sextet currents cannot produce a color singlet baryon. 
In diagrams $T$ and $E_{3}$, the quark pair antisymmetric in color and 
flavor~\cite{Korner:1970xq,Pati:1970fg} originating from weak interaction ends 
up in the decuplet baryon which is forbidden by the above mentioned 
argument~\cite{Kohara:1991ug,Xu:1992sw,Chau:1995gk,Geng:2019awr,Liang:2018rkl,Gutsche:2018utw,Gutsche:2018msz}.
 Since the quark pair $(ud)_{\overline{3}}$ is in isospin 0 state\footnote{Only 
allowed isospin values in a $\overline{\mathbf{3}}$ of $SU(3)_{F}$ is 0 or 
$1/2$.}, the total quark transition obeys $\Delta I=1/2$ rule. In contrast, in 
diagrams $E_{1}$ and $E_{2}$, the diquark argument is not applicable as only 
one of the quarks from weak interaction form the final state baryon while the 
other ends up in the meson.

As a consequence, the number of independent diagrams reduces to three. If we demand an equivalent description of all possible decays in terms of $SU(3)$-reduced amplitudes, some of the reduced amplitudes can no longer be independent. The relation between the remaining diagrams and the $SU(3)$-reduced amplitudes is given below,
\begin{align}
&E_{1}=\frac{\sqrt{3} \,a_{10}}{4}-\frac{2\, b_{8} }{\sqrt{15}},\\
&E_{2}=\frac{\sqrt{3} \,a_{10}}{4}+\frac{2\,b_{8}}{\sqrt{15}},\\
&P=-\frac{\sqrt{3}
   \,a_{10}}{16}+\frac{c_{8}}{\sqrt{10}}-\frac{b_{8}}{\sqrt{15}}
\end{align}
where $c_{8},\,b_{8},\,a_{8},\,a_{10},\,a_{27}$ are defined in Eq.~\eqref{alg 
amp}.

Equivalently, the following relations hold for the $SU(3)$-reduced matrix elements,
\begin{align}
&a_{27}=0 \qquad(C_{1}\neq 0,\, C_{2}\neq 0 )\\
&a_{8}=\frac{5}{8\sqrt{2}}a_{10}.
\end{align} 
The reduction in number of independent $SU(3)$-reduced amplitudes also imply 
additional amplitude relations that are enclosed in the box:
\begin{widetext}
\begin{align}
\label{USpin-1}
&\mathcal{A}(\Lambda_{b}^{0}\to \Delta^{+}\pi^{-})=\mathcal{A}(\Xi_{b}^{0}\to 
\Sigma^{'+} 
K^{-})\,\boxed{=-\sqrt{2}\mathcal{A}(\Lambda_{b}^{0}\to\Sigma^{'0}K^{0})=-\sqrt{2}\mathcal{A}(\Xi^{b}_{0}\to\Sigma^{'0}\overline{K^{0}}),}\\
\nonumber\\
\label{USpin-2}
&\mathcal{A}(\Lambda_{b}^{0}\to\Delta^{+}K^{-})=\mathcal{A}(\Xi_{b}^{0}\to\Sigma^{'+}\pi^{-})=-\mathcal{A}(\Lambda^{0}_{b}\to\Delta^{0}\overline{K^{0}})=-\mathcal{A}(\Xi_{b}^{0}\to\Xi^{'0}K^{0}),\\
\nonumber\\ 
\label{USpin-3}
&\mathcal{A}(\Lambda_{b}^{0}\to\Sigma^{'-}\pi^{+})=
\mathcal{A}(\Lambda_{b}^{0}\to\Xi^{'-}K^{+})=
\mathcal{A}(\Xi_{b}^{0}\to\Xi^{'-}\pi^{+})=
\frac{1}{\sqrt{3}}\mathcal{A}(\Xi_{b}^{0}\to\Omega^{-}K^{+})\nonumber\\
&\quad=\frac{1}{\sqrt{3}}\mathcal{A}(\Lambda_{b}^{0}\to\Delta^{+}\pi^{-})=
\mathcal{A}(\Lambda_{b}^{0}\to\Sigma^{'-}K^{+})=
\mathcal{A}(\Xi_{b}^{0}\to\Sigma^{'-}\pi^{+})=
\mathcal{A}(\Xi_{b}^{0}\to\Xi^{'-}K^{+})\nonumber\\
&\quad\boxed{=-\sqrt{2}\mathcal{A}(\Xi_{b}^{0}\to\Xi^{'0}\pi^{0})}\\
\label{USpin-4}
&\mathcal{A}(\Xi_{b}^{-}\to\Sigma^{'-}\overline{K^{0}})=
-\sqrt{2}\mathcal{A}(\Xi_{b}^{-}\to\Xi^{'-}\pi^{0})=
\sqrt{\frac{2}{3}}\mathcal{A}(\Xi_{b}^{-}\to\Xi^{'-}\eta_{8})=
-\frac{1}{\sqrt{3}}\mathcal{A}(\Xi_{b}^{-}\to\Omega^{-}K^{0})\nonumber\\
&\quad=\frac{1}{\sqrt{3}}\mathcal{A}(\Xi_{b}^{-}\to\Delta^{'-}\overline{K^{0}})=
\sqrt{\frac{2}{3}}\mathcal{A}(\Xi_{b}^{-}\to\Sigma^{'-}\eta_{8})=
-\mathcal{A}(\Xi_{b}^{-}\to\Xi^{'-}K^{0})=
-\sqrt{2}\mathcal{A}(\Xi_{b}^{-}\to\Sigma^{'-}\pi^{0})\nonumber\\
&\quad\boxed{=-\mathcal{A}(\Xi_{b}^{0}\to\Delta^{+}K^{-})=
	\mathcal{A}(\Xi_{b}^{0}\to\Delta^{0}\overline{K^{0}})=
	-\mathcal{A}(\Lambda^{0}_{b}\to\Sigma^{'+}\pi^{-})=
	\mathcal{A}(\Lambda_{b}^{0}\to\Xi^{'0}K^{0})}\\
&\quad\boxed{=-\mathcal{A}(\Xi_{b}^{-}\to\Delta^{0}K^{-})=
	\sqrt{2}\mathcal{A}(\Xi_{b}^{-}\to\Sigma^{'0}\pi^{-})=
	-\sqrt{2}\mathcal{A}(\Xi^{-}_{b}\to\Sigma^{'0}K^{-})=
	\mathcal{A}(\Xi_{b}^{-}\to\Xi^{'0}\pi^{-})}\nonumber
\end{align}
\end{widetext}
By measuring the phase-space corrected branching fractions and $\delta_{CP}$ one can test the diquark picture in bottom baryon decays.
\end{appendices}


\begin{thebibliography}{99}
	\bibitem{Aaij:2019pqz}
	R.~Aaij {\it et al.} [LHCb Collaboration],
	Phys.\ Rev.\ D {\bf 100} (2019) no.3,  031102
	\bibitem{Aaltonen:2011qt} 
	T.~Aaltonen {\it et al.} [CDF Collaboration],
	Phys.\ Rev.\ Lett.\  {\bf 106}, 181802 (2011)
	
	\bibitem{Aaij:2012as}
	R.~Aaij {\it et al.} [LHCb Collaboration],
	JHEP {\bf 1210} (2012) 037
	
	\bibitem{Aaltonen:2014vra}
	T.~A.~Aaltonen {\it et al.} [CDF Collaboration],
	Phys.\ Rev.\ Lett.\  {\bf 113} (2014) no.24,  242001
	
	\bibitem{Aaij:2016cla}
	R.~Aaij {\it et al.} [LHCb Collaboration],
	Nature Phys.\  {\bf 13} (2017) 391
	
	\bibitem{Aaij:2017pgy}
	R.~Aaij {\it et al.} [LHCb Collaboration],
	JHEP {\bf 1802} (2018) 098
	
	\bibitem{Aaij:2018lsx}
	R.~Aaij {\it et al.} [LHCb Collaboration],
	JHEP {\bf 1808} (2018) 039
	
	
	\bibitem{Aaij:2018tlk} 
	R.~Aaij {\it et al.} [LHCb Collaboration],
	Phys.\ Lett.\ B {\bf 787}, 124 (2018) 
	
	\bibitem{Aaij:2019rkf}
	R.~Aaij \textit{et al.} [LHCb],
	Eur. Phys. J. C \textbf{79}, no.9, 745 (2019)
	
	\bibitem{Aaij:2019mmy}
	R.~Aaij \textit{et al.} [LHCb],
	[arXiv:1912.10741 [hep-ex]].
	
	
	
	
	
	
	
	\bibitem{Zeppenfeld:1980ex} 
	D.~Zeppenfeld,
	Z.\ Phys.\ C {\bf 8}, 77 (1981).
	
	\bibitem{Savage:1989ub} 
	M.~J.~Savage and M.~B.~Wise,
	Phys.\ Rev.\ D {\bf 39}, 3346 (1989)
	Erratum: [Phys.\ Rev.\ D {\bf 40}, 3127 (1989)].
	
	\bibitem{Gronau:1990ka} 
	M.~Gronau and D.~London,
	Phys.\ Rev.\ Lett.\  {\bf 65}, 3381 (1990).
	
	\bibitem{Chau:1990ay} 
	L.~L.~Chau, H.~Y.~Cheng, W.~K.~Sze, H.~Yao and B.~Tseng,
	Phys.\ Rev.\ D {\bf 43}, 2176 (1991)
	Erratum: [Phys.\ Rev.\ D {\bf 58}, 019902 (1998)].
	
	\bibitem{Gronau:1994rj} 
	M.~Gronau, O.~F.~Hernandez, D.~London and J.~L.~Rosner,
	Phys.\ Rev.\ D {\bf 50}, 4529 (1994)
	
	\bibitem{Deshpande:1994pw} 
	N.~G.~Deshpande and X.~G.~He,
	Phys.\ Rev.\ Lett.\  {\bf 74}, 26 (1995)
	Erratum: [Phys.\ Rev.\ Lett.\  {\bf 74}, 4099 (1995)]
	
	\bibitem{Gronau:1995hn} 
	M.~Gronau, O.~F.~Hernandez, D.~London and J.~L.~Rosner,
	Phys.\ Rev.\ D {\bf 52}, 6374 (1995)  
	
	
	\bibitem{Gronau:1995hm} 
	M.~Gronau, O.~F.~Hernandez, D.~London and J.~L.~Rosner,
	Phys.\ Rev.\ D {\bf 52}, 6356 (1995)
	
	\bibitem{Grinstein:1996us}
	B.~Grinstein and R.~F.~Lebed,
	Phys.\ Rev.\ D {\bf 53} (1996) 6344
	
	
	\bibitem{Deshpande:1997rr} 
	N.~G.~Deshpande, B.~Dutta and S.~Oh,
	Phys.\ Lett.\ B {\bf 473}, 141 (2000)
	
	\bibitem{Deshpande:1997ar} 
	N.~G.~Deshpande, B.~Dutta and S.~Oh,
	Phys.\ Rev.\ D {\bf 57}, 5723 (1998)
	\bibitem{Gronau:1998fn}
	M.~Gronau, D.~Pirjol and T.~M.~Yan,
	Phys.\ Rev.\ D {\bf 60} (1999) 034021
	Erratum: [Phys.\ Rev.\ D {\bf 69} (2004) 119901]
	
	\bibitem{Neubert:1997wb}
	M.~Neubert,
	Phys. Lett. B \textbf{424}, 152-160 (1998)
	
	\bibitem{Atwood:1997iw}
	D.~Atwood and A.~Soni,
	Phys. Rev. D \textbf{58}, 036005 (1998)
	
	\bibitem{Buras:1997cv}
	A.~J.~Buras, R.~Fleischer and T.~Mannel,
	Nucl. Phys. B \textbf{533}, 3-24 (1998)
	
	\bibitem{Buras:1998ra}
	A.~J.~Buras and L.~Silvestrini,
	Nucl.\ Phys.\ B {\bf 569} (2000) 3
	
	\bibitem{Falk:1998wc}
	A.~F.~Falk, A.~L.~Kagan, Y.~Nir and A.~A.~Petrov,
	Phys. Rev. D \textbf{57}, 4290-4300 (1998)
	
	\bibitem{Beneke:2000ry}
	M.~Beneke, G.~Buchalla, M.~Neubert and C.~T.~Sachrajda,
	Nucl.\ Phys.\ B {\bf 591} (2000) 313
	
	\bibitem{He:2000dg}
	X.~G.~He, J.~Y.~Leou and C.~Y.~Wu,
	Phys.\ Rev.\ D {\bf 62} (2000) 114015
	
	\bibitem{Deshpande:2000jp}
	N.~G.~Deshpande, X.~G.~He and J.~Q.~Shi,
	Phys.\ Rev.\ D {\bf 62} (2000) 034018
	
	\bibitem{Gronau:2000pk} 
	M.~Gronau and D.~Pirjol,
	Phys.\ Rev.\ D {\bf 62}, 077301 (2000)
	
	\bibitem{Fu:2002nr}
	H.~K.~Fu, X.~G.~He, Y.~K.~Hsiao and J.~Q.~Shi,
	Chin. J. Phys. \textbf{41}, 601-617 (2003)  
	
	\bibitem{Paz:2002ev}
	G.~Paz,
	hep-ph/0206312.
	
	\bibitem{Dariescu:2002hw} 
	M.~A.~Dariescu, N.~G.~Deshpande, X.~G.~He and G.~Valencia,
	Phys.\ Lett.\ B {\bf 557}, 60 (2003)
	
	\bibitem{Wu:2002nz}
	Y.~L.~Wu and Y.~F.~Zhou,
	Eur.\ Phys.\ J.\ direct {\bf 5} (2003) 014
	[Eur.\ Phys.\ J.\ C {\bf 32} (2004) no.S1,  179]
	
	\bibitem{Beneke:2003zv}
	M.~Beneke and M.~Neubert,
	Nucl.\ Phys.\ B {\bf 675} (2003) 333
	
	\bibitem{Grossman:2003qp}
	Y.~Grossman, Z.~Ligeti, Y.~Nir and H.~Quinn,
	Phys.\ Rev.\ D {\bf 68} (2003) 015004
	
	\bibitem{Chiang:2003rb}
	C.~W.~Chiang, M.~Gronau and J.~L.~Rosner,
	Phys.\ Rev.\ D {\bf 68} (2003) 074012
	\bibitem{He:2004ck}
	X.~G.~He and B.~H.~McKellar,
	[arXiv:hep-ph/0410098 [hep-ph]]
	
	\bibitem{Chiang:2004nm}
	C.~W.~Chiang, M.~Gronau, J.~L.~Rosner and D.~A.~Suprun,
	Phys.\ Rev.\ D {\bf 70} (2004) 034020
	
	\bibitem{Buras:2004ub}
	A.~J.~Buras, R.~Fleischer, S.~Recksiegel and F.~Schwab,
	Nucl.\ Phys.\ B {\bf 697} (2004) 133
	
	
	\bibitem{Gronau:2006eb}
	M.~Gronau, Y.~Grossman, G.~Raz and J.~L.~Rosner,
	Phys.\ Lett.\ B {\bf 635} (2006) 207
	
	\bibitem{Ali:2007ff}
	A.~Ali, G.~Kramer, Y.~Li, C.~D.~Lu, Y.~L.~Shen, W.~Wang and Y.~M.~Wang,
	Phys.\ Rev.\ D {\bf 76} (2007) 074018
	
	\bibitem{Cheng:2011qh} 
	H.~Y.~Cheng and S.~Oh,
	JHEP {\bf 1109}, 024 (2011)	
	
	\bibitem{He:2013vta}
	X.~G.~He, S.~F.~Li and H.~H.~Lin,
	JHEP {\bf 1308} (2013) 065
	
	\bibitem{Cheng:2014rfa}
	H.~Y.~Cheng, C.~W.~Chiang and A.~L.~Kuo,
	Phys.\ Rev.\ D {\bf 91} (2015) no.1,  014011
	
	\bibitem{Grinstein:2014aza}
	B.~Grinstein, D.~Pirtskhalava, D.~Stone and P.~Uttayarat,
	Phys.\ Rev.\ D {\bf 89} (2014) no.11,  114014
	%
	
	\bibitem{Hsiao:2015iiu}
	Y.~K.~Hsiao, C.~F.~Chang and X.~G.~He,
	Phys.\ Rev.\ D {\bf 93} (2016) no.11,  114002
	
	\bibitem{He:2018php}
	X.~G.~He and W.~Wang,
	Chin.\ Phys.\ C {\bf 42} (2018) no.10,  103108
	
	\bibitem{He:2018joe}
	X.~G.~He, Y.~J.~Shi and W.~Wang,
	arXiv:1811.03480 [hep-ph].
	
	\bibitem{Wang:2020gmn}
	D.~Wang, C.~P.~Jia and F.~S.~Yu,
	arXiv:2001.09460 [hep-ph].  
	
	%
	%
	%
	%
	%
	
	\bibitem{Hinchliffe:1995hz}
	I.~Hinchliffe and T.~A.~Kaeding,
	Phys.\ Rev.\ D {\bf 54} (1996) 914
	
	\bibitem{Bhattacharya:2008ss}
	B.~Bhattacharya and J.~L.~Rosner,
	Phys.\ Rev.\ D {\bf 77} (2008) 114020
	
	
	\bibitem{Pirtskhalava:2011va}
	D.~Pirtskhalava and P.~Uttayarat,
	Phys.\ Lett.\ B {\bf 712} (2012) 81
	
	\bibitem{Feldmann:2012js}
	T.~Feldmann, S.~Nandi and A.~Soni,
	JHEP {\bf 1206} (2012) 007
	
	\bibitem{Hiller:2012xm}
	G.~Hiller, M.~Jung and S.~Schacht,
	Phys.\ Rev.\ D {\bf 87} (2013) no.1,  014024
	
	
	\bibitem{Bhattacharya:2012ah}
	B.~Bhattacharya, M.~Gronau and J.~L.~Rosner,
	Phys.\ Rev.\ D {\bf 85} (2012) 054014
	[Phys.\ Rev.\ D {\bf 85} (2012) no.7,  079901]
	
	
	\bibitem{Grossman:2012ry}
	Y.~Grossman and D.~J.~Robinson,
	JHEP {\bf 1304} (2013) 067
	
	\bibitem{Gronau:2015rda}
	M.~Gronau,
	Phys.\ Rev.\ D {\bf 91} (2015) no.7,  076007
	
	\bibitem{Muller:2015lua}
	S.~MC<ller, U.~Nierste and S.~Schacht,
	Phys.\ Rev.\ D {\bf 92} (2015) no.1,  014004
	
	\bibitem{Muller:2015rna}
	S.~MC<ller, U.~Nierste and S.~Schacht,
	Phys.\ Rev.\ Lett.\  {\bf 115} (2015) no.25,  251802
	
	\bibitem{Cheng:2016ejf}
	H.~Y.~Cheng, C.~W.~Chiang and A.~L.~Kuo,
	Phys.\ Rev.\ D {\bf 93} (2016) no.11,  114010
	
	\bibitem{Grossman:2019xcj}
	Y.~Grossman and S.~Schacht,
	JHEP \textbf{07}, 020 (2019)	
	
	
	
	
	
	
	\bibitem{Lu:2009cm}
	C.~D.~Lu, Y.~M.~Wang, H.~Zou, A.~Ali and G.~Kramer,
	Phys.\ Rev.\ D {\bf 80} (2009) 034011
	
	\bibitem{Gronau:2013mza}
	M.~Gronau and J.~L.~Rosner,
	Phys.\ Rev.\ D {\bf 89} (2014) no.3,  037501
	Erratum: [Phys.\ Rev.\ D {\bf 91} (2015) no.11,  119902]
	
	\bibitem{Hsiao:2014mua}
	Y.~K.~Hsiao and C.~Q.~Geng,
	Phys.\ Rev.\ D {\bf 91} (2015) no.11,  116007
	
	\bibitem{He:2015fwa} 
	X.~G.~He and G.~N.~Li,
	Phys.\ Lett.\ B {\bf 750}, 82 (2015)  
	
	\bibitem{He:2015fsa}
	M.~He, X.~G.~He and G.~N.~Li,
	Phys.\ Rev.\ D {\bf 92} (2015) no.3,  036010
	
	\bibitem{Zhu:2016bra}
	J.~Zhu, H.~W.~Ke and Z.~T.~Wei,
	Eur.\ Phys.\ J.\ C {\bf 76} (2016) no.5,  284
	
	\bibitem{Hsiao:2017tif}
	Y.~K.~Hsiao, Y.~Yao and C.~Q.~Geng,
	Phys.\ Rev.\ D {\bf 95} (2017) no.9,  093001
	
	
	
	
	\bibitem{Savage:1989qr}
	M.~J.~Savage and R.~P.~Springer,
	Phys.\ Rev.\ D {\bf 42} (1990) 1527.
	
	\bibitem{Pakvasa:1990if}
	S.~Pakvasa, S.~P.~Rosen and S.~F.~Tuan,
	Phys.\ Rev.\ D {\bf 42} (1990) 3746.
	
	
	\bibitem{Savage:1991wu}
	M.~J.~Savage,
	Phys.\ Lett.\ B {\bf 257} (1991) 414.
	
	\bibitem{Lu:2016ogy}
	C.~D.~LC<, W.~Wang and F.~S.~Yu,
	Phys.\ Rev.\ D {\bf 93} (2016) no.5,  056008
	
	\bibitem{Geng:2017mxn}
	C.~Q.~Geng, Y.~K.~Hsiao, C.~W.~Liu and T.~H.~Tsai,
	JHEP {\bf 1711} (2017) 147
	
	
	\bibitem{Geng:2018upx}
	C.~Q.~Geng, Y.~K.~Hsiao, C.~W.~Liu and T.~H.~Tsai,
	Phys.\ Rev.\ D {\bf 99} (2019) no.7,  073003
	
	\bibitem{Geng:2018bow}
	C.~Q.~Geng, Y.~K.~Hsiao, C.~W.~Liu and T.~H.~Tsai,
	Eur.\ Phys.\ J.\ C {\bf 78} (2018) no.7,  593
	
	\bibitem{Geng:2018plk}
	C.~Q.~Geng, Y.~K.~Hsiao, C.~W.~Liu and T.~H.~Tsai,
	Phys.\ Rev.\ D {\bf 97} (2018) no.7,  073006
	
	\bibitem{Geng:2018rse}
	C.~Q.~Geng, C.~W.~Liu and T.~H.~Tsai,
	Phys.\ Lett.\ B {\bf 790} (2019) 225
	
	\bibitem{Grossman:2018ptn}
	Y.~Grossman and S.~Schacht,
	Phys.\ Rev.\ D {\bf 99} (2019) no.3,  033005
	
	\bibitem{Hsiao:2019yur}
	Y.~K.~Hsiao, Y.~Yao and H.~J.~Zhao,
	Phys.\ Lett.\ B {\bf 792} (2019) 35
	
	\bibitem{Geng:2019xbo}
	C.~Q.~Geng, C.~W.~Liu and T.~H.~Tsai,
	Phys.\ Lett.\ B {\bf 794} (2019) 19
	
	\bibitem{Wang:2019alu}
	R.~M.~Wang, M.~Z.~Yang, H.~B.~Li and X.~D.~Cheng,
	Phys.\ Rev.\ D {\bf 100} (2019) no.7,  076008
	
	\bibitem{Wang:2019dls}
	D.~Wang,
	Eur.\ Phys.\ J.\ C {\bf 79} (2019) no.5,  429
	
	\bibitem{Jia:2019zxi}
	C.~P.~Jia, D.~Wang and F.~S.~Yu,
	arXiv:1910.00876 [hep-ph].
	
	\bibitem{Dery:2020lbc}
	A.~Dery, M.~Ghosh, Y.~Grossman and S.~Schacht,
	JHEP \textbf{03}, 165 (2020)
	
	
	%
	%
	%
	%
	%
	%
	%
	
	\bibitem{Grinstein:2004kn}
	B.~Grinstein and M.~A.~Savrov,
	hep-ph/0408346.
	\bibitem{Roy:2019cky} 
	S.~Roy, R.~Sinha and N.~G.~Deshpande,
	Phys.\ Rev.\ D {\bf 101}, no. 3, 036018 (2020)
	
	
	
	
	
	\bibitem{Shortley}
	E. U. Condon and G. H. Shortley, The Theory of Atomic Spectra, London: 
	Cambridge
	University Press, 1957.
	
	\bibitem{deSwart:1963pdg} 
	J.~J.~de Swart,
	Rev.\ Mod.\ Phys.\  {\bf 35}, 916 (1963)
	Erratum: [Rev.\ Mod.\ Phys.\  {\bf 37}, 326 (1965)].
	
	\bibitem{Samios:1974tw} 
	N.~P.~Samios, M.~Goldberg and B.~T.~Meadows,
	Rev.\ Mod.\ Phys.\  {\bf 46}, 49 (1974).
	
	\bibitem{Kaeding:1995re}
	T.~A.~Kaeding and H.~T.~Williams,
	Comput.\ Phys.\ Commun.\  {\bf 98} (1996) 398
	
	\bibitem{Kaeding:1995vq} 
	T.~A.~Kaeding,
		Atom.\ Data Nucl.\ Data Tabl.\  {\bf 61}, 233 (1995)
	
	\bibitem{Ciuchini:1993vr}
	M.~Ciuchini, E.~Franco, G.~Martinelli and L.~Reina,
	Nucl.\ Phys.\ B {\bf 415} (1994) 403
	
	\bibitem{Buchalla:1995vs}
	G.~Buchalla, A.~J.~Buras and M.~E.~Lautenbacher,
	Rev.\ Mod.\ Phys.\  {\bf 68} (1996) 1125
	
	\bibitem{Buras:1998raa}
	A.~J.~Buras,
	hep-ph/9806471.
	
	\bibitem{Egolf:2002nk}
	D.~A.~Egolf, R.~P.~Springer and J.~Urban,
	Phys.\ Rev.\ D {\bf 68} (2003) 013003
	\bibitem{Grossman:2013lya}
	Y.~Grossman, Z.~Ligeti and D.~J.~Robinson,
	JHEP {\bf 1401} (2014) 066
	\bibitem{Xu:2013dta} 
	D.~Xu, G.~N.~Li and X.~G.~He,
	Int.\ J.\ Mod.\ Phys.\ A {\bf 29}, 1450011 (2014)
	
	\bibitem{Brown:1983wd}
	T.~Brown, S.~F.~Tuan and S.~Pakvasa,
	Phys.\ Rev.\ Lett.\  {\bf 51} (1983) 1823.
	
	
	\bibitem{Donoghue:1985ww}
	J.~F.~Donoghue and S.~Pakvasa,
	Phys.\ Rev.\ Lett.\  {\bf 55} (1985) 162.
	
	\bibitem{Dunietz:1992ti}
	I.~Dunietz,
	Z.\ Phys.\ C {\bf 56} (1992) 129.
	
	
	
	
	\bibitem{Korner:1970xq} 
	J.~G.~Korner,
	Nucl.\ Phys.\ B {\bf 25}, 282 (1971).
	
	\bibitem{Pati:1970fg}
	J.~C.~Pati and C.~H.~Woo,
	Phys.\ Rev.\ D {\bf 3} (1971) 2920.
	
	
	\bibitem{Stech:1987fa}
	B.~Stech,
	Phys.\ Rev.\ D {\bf 36} (1987) 975.
	
	\bibitem{Neubert:1991zd}
	M.~Neubert and B.~Stech,
	Phys.\ Rev.\ D {\bf 44} (1991) 775.
	
	\bibitem{Neubert:1997uc} 
	M.~Neubert and B.~Stech,
	Adv.\ Ser.\ Direct.\ High Energy Phys.\  {\bf 15}, 294 (1998)       
	\bibitem{Nieves:2003in}
	J.~F.~Nieves and P.~B.~Pal,
	Am. J. Phys. \textbf{72}, 1100-1108 (2004)
	
	\bibitem{Kohara:1991ug}
	Y.~Kohara,
	Phys.\ Rev.\ D {\bf 44} (1991) 2799.
	
	\bibitem{Xu:1992sw}
	Q.~P.~Xu and A.~N.~Kamal,
	Phys.\ Rev.\ D {\bf 46} (1992) 3836.
	
	\bibitem{Chau:1995gk}
	L.~L.~Chau, H.~Y.~Cheng and B.~Tseng,
	Phys.\ Rev.\ D {\bf 54} (1996) 2132
	
	
	\bibitem{Geng:2019awr}
	C.~Q.~Geng, C.~W.~Liu, T.~H.~Tsai and Y.~Yu,
	Phys.\ Rev.\ D {\bf 99} (2019) no.11,  114022
	
	\bibitem{Liang:2018rkl} 
	W.~H.~Liang and E.~Oset,
	Eur.\ Phys.\ J.\ C {\bf 78}, no. 6, 528 (2018)
	\bibitem{Gutsche:2018utw}
	T.~Gutsche, M.~A.~Ivanov, J.~G.~KC6rner and V.~E.~Lyubovitskij,
	Phys.\ Rev.\ D {\bf 98} (2018) no.7,  074011
	
	\bibitem{Gutsche:2018msz} 
	T.~Gutsche, M.~A.~Ivanov, J.~G.~K\"orner, V.~E.~Lyubovitskij and 
	Z.~Tyulemissov,
	Phys.\ Rev.\ D {\bf 99}, no. 5, 056013 (2019)
	
	%
	%
	
	
\end{thebibliography}
\end{document}